\documentclass[aip,amsmath,amssymb,reprint,author-year]{revtex4-1}
\usepackage{natbib}

\usepackage{graphicx}
\usepackage{dcolumn}
\usepackage{bm}
\usepackage[utf8]{inputenc}
\usepackage[T1]{fontenc}
\usepackage{etoolbox}
\usepackage{graphicx}
\usepackage{amssymb}
\usepackage{epstopdf, epsfig}
\usepackage{yfonts}
\usepackage[colorlinks=true, linkcolor=cyan, citecolor=cyan, urlcolor=cyan]{hyperref}
\usepackage{fontenc}
\usepackage{upgreek}
\usepackage{mathrsfs}
\usepackage{booktabs}
\usepackage{pifont}
\usepackage{amsmath}
\usepackage{bm}
\usepackage{wasysym}
\usepackage{caption}
\usepackage{todonotes}
\usepackage{subcaption}
\usepackage{euscript}
\usepackage{natbib}
\usepackage{lipsum}
\usepackage{xcolor} 
\usepackage[document]{ragged2e}  

\usepackage{orcidlink}

\usepackage[export]{adjustbox} 
\usepackage{cleveref}
\usepackage[normalem]{ulem}

\newcommand{\pd}[2]{\frac{\partial #1}{\partial #2}}

\newcommand{\jfm}[1]{\textcolor{cyan}{#1}}
\newcommand{\x}[1]{\textcolor{black}{#1}}  
\newcommand{\xx}[1]{\textcolor{black}{#1}} 
\newcommand{\xxx}[1]{\textcolor{black}{#1}}  
\newcommand{\xt}[1]{\textcolor{black}{#1}}  
\newcommand{\xy}[1]{\textcolor{black}{#1}} 
\newcommand{\pf}[1]{\textcolor{black}{#1}} 
\newcommand{\sm}[1]{\textcolor{black}{#1}}
\newcommand{\ac}[1]{\textcolor{black}{#1}}
\newcommand{\yh}[1]{\textcolor{black}{#1}}
\newcommand{\jk}[1]{\textcolor{black}{#1}}

\makeatletter
\def\@email#1#2{%
 \endgroup
 \patchcmd{\titleblock@produce}
  {\frontmatter@RRAPformat}
  {\frontmatter@RRAPformat{\produce@RRAP{*#1\href{mailto:#2}{#2}}}\frontmatter@RRAPformat}
  {}{}
}%
\makeatother

\begin{document}

\preprint{AIP/123-QED}

\title[Energetics and Stochastics of Extreme Waves Breaking over a Symmetrical \pf{Breakwater}]{Energetics and Stochastics of Extreme Waves Breaking over a Symmetrical \pf{Breakwater}}
\author{S. Mendes\,\orcidlink{0000-0003-2395-781X}}
 \altaffiliation{School of Civil and Environmental Engineering, Nanyang Technological University, Singapore}
 \email{saulo.mendes@ntu.edu.sg}
\author{Y. He\,\orcidlink{0000-0003-0430-6472}}
 \altaffiliation{Key Laboratory for Mechanics in Fluid-Solid Coupling Systems, Institute of Mechanics, Chinese Academy of Sciences, Beijing 100190, PR China}
\author{J. Kasparian\,\orcidlink{0000-0003-2398-3882}}
 \altaffiliation{Group of Applied Physics, University of Geneva, Rue de l'École de Médecine 20, 1205 Geneva, Switzerland}
 \altaffiliation{Institute for Environmental Sciences, University of Geneva, Boulevard Carl-Vogt 66, 1205 Geneva, Switzerland}
\author{A. Chabchoub\,\orcidlink{0000-0002-5473-3029}}
 \altaffiliation{Marine Physics and Engineering Unit, Okinawa Institute of Science and Technology, Onna-son, Okinawa 904-0495, Japan}
 \altaffiliation{Department of Infrastructure Engineering, The University of Melbourne, Parkville, Victoria 3010, Australia}
  \altaffiliation{Department of Civil and Environmental Engineering, Imperial College London, London SW7 2AZ, United Kingdom}

\date{\today}

\begin{abstract}
\justifying{Rogue wave formation and enhancement over coastal areas have been documented over the last decade. This is in apparent contradiction with \pf{the observed} low rogue wave probability near \pf{the surf zone}. Existing theories and experiments describe rogue wave amplification in this regime; however, they do so without considering wave breaking processes. To address this gap, \pf{we consider fully nonlinear effects of wave breaking through the proxy of height-to-depth ratio, spatial changes on the wavenumber through the WKB approximation, and slope-corrected refraction mild-slope equations on the energetics of irregular wave fields travelling over a breakwater. W}e show that by increasing the significant wave height towards the breaking limit, the kinetic energy grows faster than the variance of the surface elevation due to nonlinearity. Thus, it decreases the kurtosis, albeit not to the point of getting sub-Gaussian statistics. We thereby resolve the apparent paradox of the occurrence probability of rogue waves increasing at the beginning of shoaling but subsequently decreasing when wave breaking becomes dominant. \pf{Motivated by these theoretical developments,} we experimentally probe inhomogeneous wave fields nearing the wave-breaking regime. \pf{We conduct unidirectional irregular wave experiments in a 30 m long wave flume, generating broad-banded waves over a symmetric submerged breakwater featuring a bottom slope of 1/5, allowing detailed characterization of spectral evolution and the persistence of elevated excess kurtosis even near the breaking limit for this steep slope. Through this analysis, we confirm} that the excess kurtosis can still be large if the bottom slope is steep\pf{, and its maximum value is at least four times larger than in other ocean processes, occurring atop the breakwater and about half of its deep water peak wavelength distance after the shoal}. As these conditions are typical near shorelines, this understanding is key to coastal areas.}
\end{abstract}

\maketitle

\section{Introduction} 

\justifying{

The forecast of extreme natural events is crucial for preventing loss of life, safeguarding infrastructure, and minimizing damage to both local and global environments. Over the past decades, climate extremes have also been intensely studied in addition to the most widely assessed natural hazards, such as earthquakes, tsunami or floods. Despite sailor stories, giant waves have been \sm{largely} overlooked throughout the nineteenth \jk{and most of the twentieth} century \citep{Liu2007}. However, since the New Year Wave of 1995 \citep{Haver2004}, rogue waves have been \jk{extensively investigated, as they are} responsible for considerable damage \citep{Faukner1997,Faukner2002,Gregersen2015}. \ac{Rogue waves and their respective loads, amplified} by abrupt inhomogeneities in the wave field, have recently been examined and better assessed \citep{Adcock2022,Adcock2023,Li2023,Trulsen2025}. Models for short-term and long-term statistics of water waves are used to define the operating envelope for ocean vessels and fixed offshore structures. Moreover, the prediction and control of rogue waves are particularly important for marine and offshore safety. Among the various mechanisms of rogue wave formation and amplification, shoaling in coastal areas presents the highest hazardous risk \jk{due to the higher population and infrastructure density} \sm{and higher amplification of probability density, see} \citet{Chabchoub2023} \sm{for a review}. \sm{Furthermore, coastal rogue waves also represent a strong majority of catalogued major events \citep{Didenkulova2023}.}

Over the past decade, the out-of-equilibrium dynamics of rogue wave occurrence in random seas subject to shoaling have been intensively investigated \citep{Trulsen2012,kashima2019aftereffect,Chabchoub2019,Adcock2020,Trulsen2020,Adcock2021b,Benoit2021,Chabchoub2021,Benoit2023}. Experiments have been carried out considering wave propagation either over a step or over relatively steep slopes\jk{. They} consistently showed \jk{a} growth of the steepness over the shoaling region\jk{. The enhanced nonlinearity significantly increases the probability of rogue wave formation}. \jk{The evolution of this wave height probability distribution function over a step has been modeled both} numerically \citep{Bolles2019} for surface elevation and theoretically \citep{Adcock2021c} for wave crests\jk{. Regarding finite slopes}, a theoretical wave height distribution for steep shoals \citep{Mendes2022}, successfully \jk{reproduced} experimental results~\citep{Raustol2014,Trulsen2020} for a variety of water depths and initial steepness values. However, these theories disregard wave breaking and are limited to relative water depths \sm{(or Ursell numbers)} within \sm{the limits of} wave theories \sm{up to} second-order \sm{in steepness}. To date, a proper analytical and continuous water wave solution for all water depths \citep{LeMehaute1976} is still missing, even for regular waves \sm{\citep{Antuono2022}}. \sm{From the perspective of the harmonic generation \ac{in a} theoretical framework \citep{Adcock2021c} or the statistical KdV method \citep{Majda2020}, the lack of a generalized solution that includes wave breaking makes it difficult to forecast wave statistics over a step. \xxx{However,} \cite{Draycott2022} dealt with the location of the breaker over a step through the harmonic generation framework in numerical \ac{smoothed-particle hydrodynamics (SPH) simulations}.}

\sm{Likewise, the stochastic energetic method applied to arbitrary slopes \citep{Mendes2023b} cannot provide a meaningful analysis of wave statistics under breaking, since the formulation for the wave energy density under the aforementioned conditions \ac{is missing}. Furthermore, the wave energy density influences not only wave statistics, but also sediment flows, wave shoaling coefficients, wave-structure interactions, and wave energy conversion, among others. To the best of our knowledge, computations of the wave energy density exist only for large wave steepness in deep-water or for smaller steepness in intermediate water and over flat bottoms \citep{Higgins1975b,Klopman1990,Ma2020,Henry2021}. Consequently, a comprehensive analysis is lacking for arbitrary steepness values up to wave breaking and for arbitrary bottom slopes.} \sm{On the other hand, the problem of rogue waves subject to shoaling over an arbitrarily steep slope requires a stronger theoretical background.} \jk{Relying on out-of-equilibrium theory of} excess kurtosis of the wave height within the theory of small wave amplitude\jk{, we \ac{recently} developed an analytical description of the effect of the slope magnitude on the extent of rogue wave amplification \citep{Mendes2022,Mendes2023}}. Such theoretical reasoning agrees with the observation that slopes not exceeding 1\% cannot affect wave statistics \citep{Swan2013}\sm{. It also explains} why the amplification of rogue wave probability quickly saturates at slopes near 50\% \citep{Adcock2020,Ma2021}. 

All these results brought \sm{an improved understanding of} the second-order theoretical analysis of the effect of shoaling on rogue wave occurrence for \sm{the respective \xx{isolated} cases of} deep, intermediate and shallow-water wave statistics. \sm{Shoaling induces rogue wave amplification in intermediate water due to second-order effects. But the \ac{corresponding} increase of wave steepness due to the same effect leads to wave breaking and sub-Gaussian statistics over a flat bottom or mild slope in shallow-water \citep{Glukhovskii1966,Xu2021,Karmpadakis2022}. In addition, wave breaking effects on wave statistics in the surf zone are further problematic because they induce a higher directional spread \citep{Herbers1999,Szczyrba2023}. This directional spread tames the excess kurtosis of the surface elevation, and thus the rogue wave amplification \citep{Karmpadakis2022,Mei2023,Adcock2023}.}

\jk{Here, we address the need \sm{for a model compatible with} the most common conditions encountered in the real ocean, including coastal regions, namely relatively steep slopes \sm{that ultimately have to include wave breaking}}. We thereby connect the two limiting conditions of a flat bottom and a step transition, \jk{and reconcile apparently contradictory behaviours \sm{in or near the regime of wave breaking} between very mild \citep{Glukhovskii1966,Xu2021} and steep slopes \citep{Trulsen2020,Adcock2021c}, \sm{leading to sub- and super-Gaussian wave statistics, respectively}. \xx{Although interpolation between deep-water small-amplitude waves and shallow-water breaking waves suggests a bell-shaped curve for excess kurtosis as a function of wave steepness, no model has yet been proposed \xy{to explain or confirm} this expectation.} In wave flume experiments, we characterize the evolution of waves over a shoal over a wide range of \sm{wave steepness} \ac{values and we particularly observe} the corresponding wave height statistics, as well as wave breaking.} While there is no proper analytical and continuous water wave solution for all depths \jk{to date}, we show that the ratio between significant wave height and water depth allows second-order wave theories to properly describe rogue wave statistics in a regime otherwise belonging to cnoidal wave theory. \sm{The present paper is organized as follows: in \jfm{section} \ref{sec:Edensity} we develop a new model for wave energy recognizing the relevance of wave breaking and \jk{sharp} changes in the water column depth and wavenumber. Then \jk{in \jfm{section}~\ref{sec:Conseq}}, we \jk{build on this novel energetics to develop a unifying method that approximates the statistical behaviour of finite-amplitude waves even up to wave breaking. Finally, in \jfm{section} \ref{sec:Exp} we detail experimental procedures \ac{for} a unidimensional wave flume \ac{designed} to represent the target conditions of wave breaking over steep slopes, and thus, \jk{we} use \ac{these measurements} to validate the theoretical development \ac{of} previous sections} }
}

\section{Energy Density of Non-Breaking Inhomogeneous Wave Fields}\label{sec:Edensity}

\begin{figure}
\centering
\includegraphics[scale=0.35]{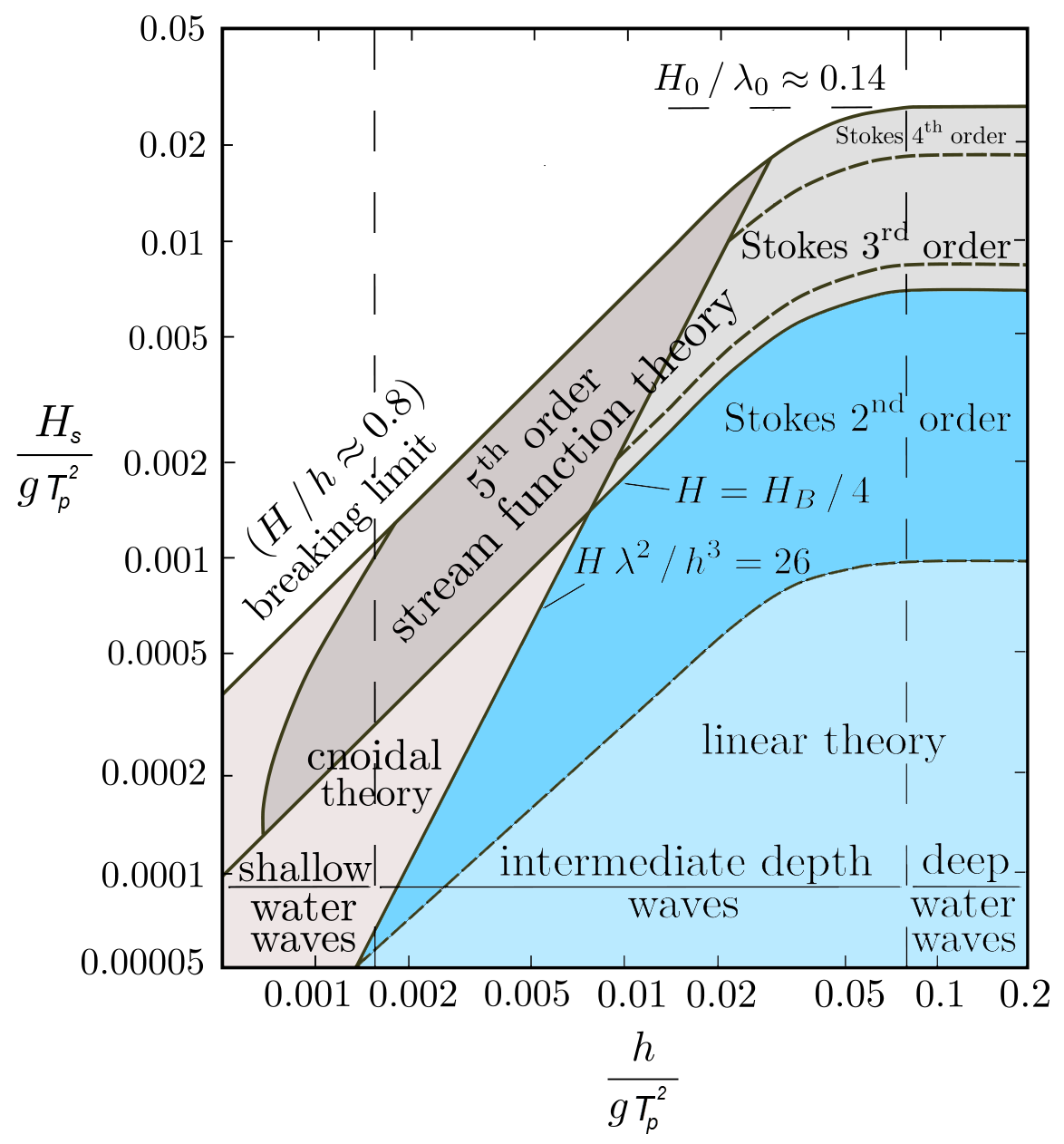}
  \caption{\justifying{\citeauthor{LeMehaute1976}'s diagram describing several water wave theories for different water depth and wave steepness \ac{values}}.}
  \label{fig:LMdiag}
\end{figure} 
Following \sm{\citet{Dalrymple984},} we start by considering the total energy \jk{density, i.e., the energy} per \jk{unit} width $\pf{\mathscr{E}}$ contained within a wavelength, \jk{given by}:
\begin{eqnarray}
 \mathscr{E} = \frac{1}{2\lambda} \int_{0}^{\lambda} \left[ \Big( \zeta + h  \Big)^{2}  + \frac{1}{g} \int_{-h (x)}^{\zeta}  \Big( u^{2} + w^{2} \Big) \, dz \right]\, dx \quad ,
\label{eq:Energy}
\end{eqnarray}
where $\rho$ is the density, $g$ is the gravitational acceleration, $\lambda$ the wavelength, $h$ the water column depth, $x$ the direction of motion and $z$ the vertical axis, $\zeta$ is the sea surface elevation and $u = \partial \Phi/\partial x$ and $w = \partial \Phi/\partial z$ are the velocity components derived from the velocity potential $\Phi$. For waves of finite amplitude, we must calculate a finite upper bound of vertical integration in the kinetic energy formulation of Eq.~(\ref{eq:Energy}). In principle, energy integrals are simple to deal with, but in practice, they are very sensitive to specific conditions of the wave process at hand. \xxx{For a uniform water depth,} Several authors have theoretically assessed how the partition of \jk{the} wave energy between \jk{the} potential and \jk{the} kinetic energy takes place \citep{Platzman1947,Starr1947,Higgins1975b}. More recently, \citet{Henry2021} has re-examined this problem by estimating energy excess (or variations) for an exact nonlinear wave model. Harmonic solutions such as $\zeta(x,t) = a \cos{(kx - \omega t)}$ are quite far from the wave breaking \jk{regime we intend to cover as we seek for a} theory applicable to all regimes of \citet{LeMehaute1976} diagram (see \jfm{Figure} \ref{fig:LMdiag})\jk{. Therefore}, we must reassess the energetics problem through an approximation. To the best \jk{of our} knowledge, \ac{such an analytical or numerical description of energy variation of waves covering a variety of conditions like} a sharp spatial inhomogeneity of the wave characteristics (wavelength, group speed, wave period, etc.), a steep bottom slope, and wave breaking\jk{, is still missing}. This section attempts to unify all \jk{conditions} that affect irregular wave energetics. Although the latter question is of the utmost importance to energy fluxes under a range of engineering problems, namely wave energy conversion \citep{Thomas2008}, wave-current interaction \citep{Constantin2020}, wave-vegetation systems \citep{Jacobsen2016}, sediment transport \citep{Wu2023}, the wave energy density also defines wave statistics \ac{following} the framework of \citet{Mendes2022} and is the main application sought in this work. 

\pf{This section is organized as follows: \jfm{subsection} \ref{sec:sA} summarizes the literature knowledge on precise formulations for energy density, while \jfm{subsections} \ref{sec:bound}-\ref{sec:bound2} provide new theoretical extensions to include wave breaking. Firstly, we address the upper bound problem of the kinetic energy integration (\jfm{subsection} \ref{sec:bound}-\ref{sec:bound1}), then the influence of wavenumber gradient and bottom slope on horizontal and vertical velocities (\jfm{subsection} \ref{sec:bound2}). This new model has a theoretical range of validity for wave steepness covering all regimes $\varepsilon \leqslant \tanh{(kh)}/7$ (linear, second- and higher-order Stokes, Cnoidal, etc.) in the range $kh < 2$ (breaking in the coastal zone only), but it is limited to slopes with small reflection rates $K_{R} \leqslant 20\%$ and thus $\nabla h \lesssim 0.25$ \citep{Battjes1974}, and strictly smooth and planar slopes. The latter limitation implies that the model can only account for plunging and spilling breakers, and assumes no air-sea interaction. Another important limitation of the model is the absence of the undertow (due to breaking, see \citet{Svendsen1984}) and tidal effects, which requires an analysis of wave-tide and wave-current interaction for the extreme wave statistics. Therefore, although the model covers the entire range of $H_s/h$, large values of the breaker index ($H_s/h$) and very intense breaking reduce the accuracy of the model due to the fact that the undertow is neglected.}

\subsection{\sm{State-of-the-art Energetics without Wave Breaking}}\label{sec:sA}

\sm{The approach for extending the energetics over flat bottoms to a sloping bottom has been obtained in \citet{Mendes2023b}\jk{. This approach has been} recently verified numerically regardless \jk{the} \ac{of} shoaling length \citep{Zhang2024}. By} defining the variables $\varphi = k (z+h)$, \pf{with wavenumber $k = 2\pi /\lambda$ and dimensionless depth} $\Lambda_{\pf{\ast}} = kh$\pf{, as well as} $\phi = kx - \omega t$ \pf{with angular frequency $\omega = 2\pi / T$ and wave period $T$}, the second-order regular wave velocity potential $\pf{\Phi (x,z,t)}$ \sm{within this approach} can be written \pf{for an amplitude $a$}:
\begin{equation}
\Phi = \frac{a\omega}{k} \frac{\cosh{\varphi} }{\sinh{\Lambda_{\pf{\ast}}}} \sin{\phi} + \left(  \frac{3ka}{8}  \right) \frac{a\omega}{k} \frac{\cosh{(2\varphi)} }{\sinh^{4}{\Lambda_{\pf{\ast}}}} \sin{(2\phi)} \, .
\end{equation}
Ergo, the surface elevation becomes:
\begin{eqnarray}
 \zeta (x,t) &=&  a \cos{\phi} + \frac{ka}{4} \sqrt{\Tilde{\chi}_{1,\pf{\ast}}} \, a \cos{(2\phi)} \quad ,
 \label{eq:zetairregular}
\end{eqnarray}
where  $\Tilde{\chi}_{1,\pf{\ast}} = (3 - \tanh^{2}{\Lambda_{\pf{\ast}}})^2/\tanh^{6}{\Lambda_{\pf{\ast}}}$ is the super-harmonic term. If we ignore the longitudinal evolution of variables, such as wavelength, the following expressions hold for the vertical velocity:
\begin{eqnarray}
\hspace{-0.3cm}
u_0 \equiv \pd{\Phi}{x} = a\omega \left\{ \frac{\cosh{\varphi} }{\sinh{\xxx{\Lambda_{\pf{\ast}}}}} \cos{\phi}  + \left(  \frac{3ka}{4}  \right) \frac{\cosh{(2\varphi)} }{\sinh^{4}{\Lambda_{\pf{\ast}}}} \cos{(2\phi)} \right\} \, ,
\\
\nonumber
\hspace{-0.3cm}
w_0 \equiv \pd{\Phi}{z} = a\omega \left\{ \frac{\sinh{\varphi} }{\sinh{\xxx{\Lambda_{\pf{\ast}}}}} \sin{\phi}  + \left(  \frac{3ka}{4}  \right) \frac{\sinh{(2\varphi)} }{\sinh^{4}{\Lambda_{\pf{\ast}}}} \sin{(2\phi)} \right\} \, ,
\end{eqnarray}
\sm{where the \ac{${(\cdot)}_{0}$} subscript denotes the solution for flat bottoms in the absence of any form of wave breaking.}
We know that the kinetic energy of the \sm{slope-independent} term leads to \citep{Mendes2021b}\footnote{\pf{The complete integration must be carried out over a broadband wave field. 
Nevertheless, \citet{Mendes2021b} demonstrated that integration of the terms 
dependent on the free-surface elevation 
$\zeta = \sum_i \zeta_i = \sum_i a_i \cos{\phi_i} + \frac{k_ia_i}{4} \sqrt{\Tilde{\chi}_{1,\pf{\ast}}} \, a_i \cos{(2\phi_i)}$ 
and on the velocity components 
$\partial\Phi/\partial x_j = u_j = \sum_i \partial\Phi_i/\partial x_j$ 
(with $u_1 = u$, $u_2 = w$) 
yields an expression that is formally identical to the monochromatic case, 
apart from the substitutions $ka \to \pi \varepsilon \mathfrak{S}$ and $kh \to k_p h$ 
that emerge as a direct consequence of the integration. Therefore, in what follows, the amplitude $a$ should be understood as 
$a = \mathfrak{S}H_s/2\sqrt{2}$. For the sake of simplicity, we henceforth display monochromatic solutions before integration, followed by the broad-banded energetic solution as the outcome.}}:
\begin{eqnarray}
\hspace{-0.5cm}
 \mathscr{E}_{k0}^{\pf{\dagger}} &\approx&   \int_{-h}^{0} \int_{0}^{\lambda} \left(    u_0^{2} + w_0^{2} \right) \,  \frac{dx \, dz}{2 \lambda g} \sm{
 \, \approx \frac{a^2}{4}  \left[ 1   +  \left(  \frac{\pi \varepsilon \mathfrak{S}}{4}  \right)^2 \chi_1   \right] \quad ,
 }
 \label{eq:energ0}
 \end{eqnarray}
 where we have denoted  $\chi_1 =  9 \cosh{(2\Lambda)} \sinh^{-6}{\Lambda}$ \pf{with} \ac{$\Lambda = k_{p}h$} \pf{stemming from the peak wavenumber (peak period), $\varepsilon = H_s / \lambda_z$ the wave steepness based on the significant wave height $H_s$ and is measured from the zero-crossing wave period $T_z$, and $\mathfrak{S}$ is the vertical asymmetry between crests and troughs (with $\mathfrak{S} = 1$ for a harmonic wave). In anticipation of the analytical work and experimental description of the incoming sections, a secondary definition of wave steepness will also be used in the context of wave breaking, namely $\epsilon = k_p H_s$}.

\subsection{\sm{The Vertical Integration Bound Problem}}\label{sec:bound}

The usual upper bound of integration \jk{of the second integral in Eq.~(\ref{eq:Energy}}) taken to be the $\zeta (x,t)$ is not strictly correct in the case of shoaling for several reasons. \xxx{First and foremost}, \ac{the surface elevation will not be a representative measure of the sea state over the region of spatial averaging of the kinetic energy,} unless we extend the interval of integration in $x$ to a very large spatial series \jk{(several wavelengths along the shoal)}, which may extend far beyond the shoaling length itself. Much like one has to use a representative measure for the wavelength, in this case the zero-crossing measure, one must then find a representative average measure for the surface elevation as an upper bound. \xx{Secondly, $\zeta (x)$ would cover several wave theories with different formulations that cannot be expressed continuously, but only a piecewise formulation, which is not of our interest, as described below,}
\begin{equation}
 \mathscr{E}_{k} =   \pf{ \sum_{m} \frac{1}{\Delta x_m} \int_{-h}^{\zeta_{m}} \int_{0}^{\Delta x_m }}   \, \frac{ \left(    u_0^{2} + w_0^{2} \right) \, dx \, dz}{2  g} \,\, , 
\end{equation}
with $\sum_{\pf{m}} \Delta x_{\pf{m}} = \lambda $. An additional issue with this formulation is the emergence of non-periodic integrals. Traditionally \citep{Dalrymple984}, energetic integrals are not carried over arbitrary lengths that exceed or are less than a wavelength $L \neq \lambda$. Evaluating arbitrary-length integrals in this context requires cumbersome calculations of multiple wave theories as a function of the normalized arbitrary distance $L/\lambda$, which has been shown to have no significant impact on random wave statistics \citep{Zhang2024,Zhang2026}.

\begin{figure*}
\minipage{0.44\textwidth}
   \includegraphics[scale=0.6]{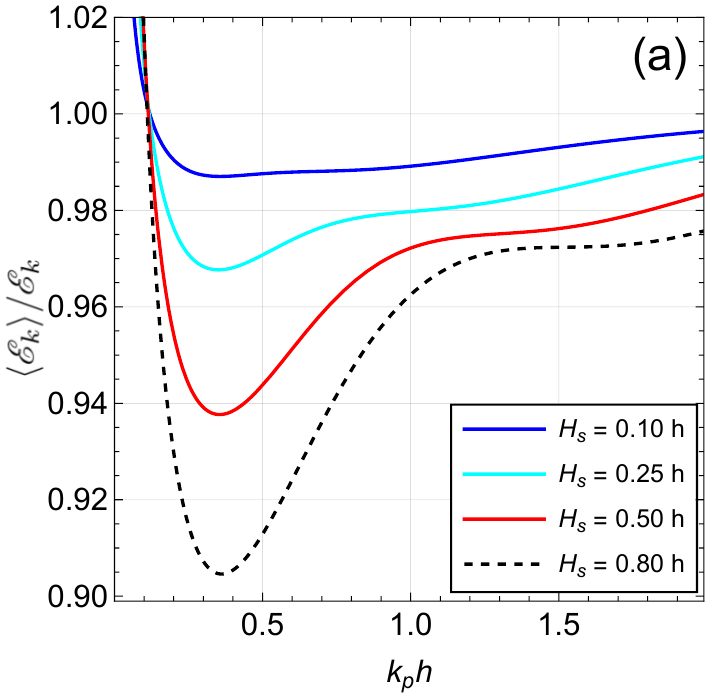}
\endminipage
\hfill
\minipage{0.55\textwidth}
   \includegraphics[scale=0.57]{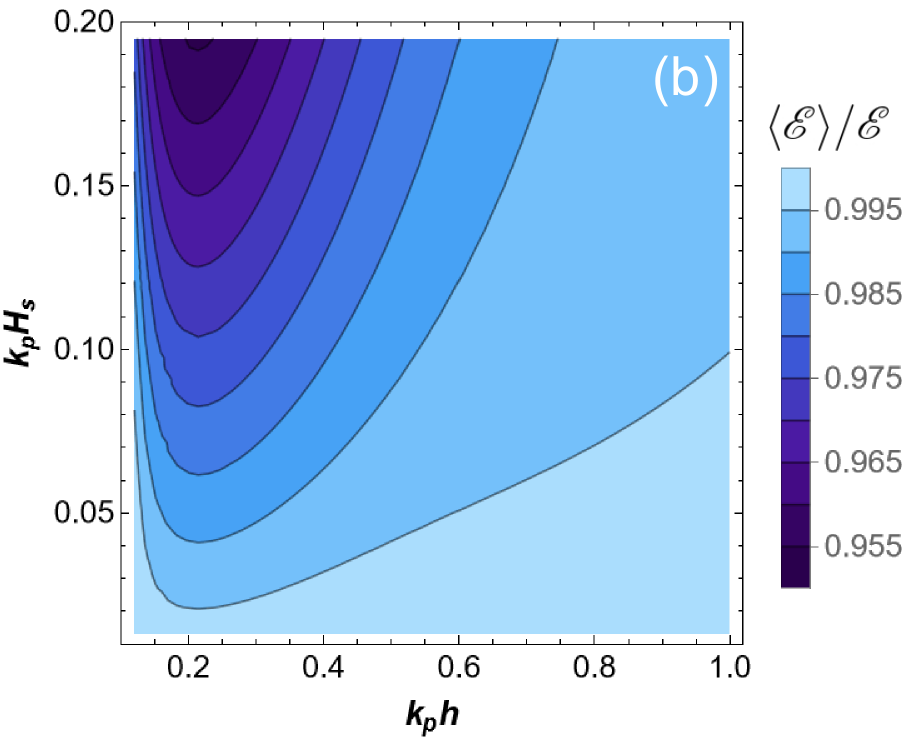}
\endminipage
    \caption{\justifying{(a) Ratio of rms kinetic energy to the \xy{mean} kinetic energy computed with the rms surface elevation ($H_s/4$), and (b) ratio for the total energy counterparts.}}
    \label{fig:RMSEnergy}
\end{figure*}
Since Gaussian seas are zero-mean and oscillations around this value are tiny due to low nonlinearity, the \ac{root-mean-square (rms)}  \jk{of surface elevation}  is the proper choice. In a typical \ac{narrowband} Gaussian sea, which implies a Rayleigh distribution of wave heights, the most likely wave height is $H=0.5H_s$ while the mean height approximately amounts to $0.6H_{s}$ \sm{\citep{BMassel2017}}. A significant increase in the ratio $H_{s}/h$ affects the kinetic term within the total energy \sm{because of the upper interval in the integral}. Hence, we have to fully evaluate the second integral in Eq.~(\ref{eq:Energy}). If we consider this increase in significant wave height associated with depth decrease, we can no longer use the assumption and delimitation $\zeta \ll H_s$ as in our previous work \citep{Mendes2022,Mendes2023}. 

That said, because the maximum and minimum of the surface elevation are of the same order of magnitude in this case, we expect the most likely range of oscillation \jk{to be} $-1/4 < \zeta / H_s < 1/4$ over \jk{a} sufficiently long shoal. Even in weakly nonlinear seas, the oscillations will not be symmetrical between the wave crest and \pf{trough}, as the ratio between them grows monotonically with normalized height, see figure 31 of \citet{Mendes2021a}. The ratio between \jk{the} nonlinear crest and its height is on average 20\% higher for rogue waves than in a symmetrical sea, so that the wave asymmetry would hardly impact the upper bound for the integration of the kinetic energy. In contrast, ordinarily tall crests \jk{do not} exceed the symmetrical sea case by more than 10\% \citep{Mendes2021a,Mendes2023}. This range is compatible with the \ac{rms of} surface elevation $\sqrt{\langle \zeta^{2} \rangle} = \sigma = H_{s}/4$\pf{, where $\sigma$ is the standard deviation of the surface elevation}.

\subsection{\sm{Direct Effect of Large Wave Heights}}\label{sec:bound1}

\sm{To take into account the effects of wave breaking, the kinetic integral bound problem introduced in \jfm{section} \ref{sec:bound} will be discussed in the present sub-section}. For brevity, we leave tedious details of the full computation to \jfm{appendix} \ref{sec:THEOR2}. We \sm{may apply the upper bound to the outcome of Eq.~\ref{eq:energ0}~\citep{Ma2020,Mendes2021b}. The integration of interest is taken} up to an upper bound representative of the bulk of waves:
\begin{eqnarray}
 \mathscr{E}_{k} \approx \frac{1}{2 \lambda g} \int_{0}^{\lambda} \int_{-h (x)}^{H_{s}/4}  \left[ u^{2} + w^{2} \right]  \, dz \, dx \quad .
\label{eq:Energyk}
\end{eqnarray}
The direct effects of finite-amplitude \sm{waves on the} kinetic energy are \sm{found to be,} 
\begin{eqnarray}
\nonumber
\mathscr{E}_{k0} &\approx&    \frac{a^2}{4}  \Big[ 1 + \frac{ \sinh{(\frac{1}{2} k_p H_s)} }{ \sinh{(2\Lambda)} }  
\\
&{}& +  \left(  \frac{\pi \varepsilon \mathfrak{S}}{4}  \right)^2  \chi_1 \left(  1 + \frac{ \sinh{( k_p H_s)} }{ \sinh{(4\Lambda)} }    \right) \Big] \, ,
\label{eq:EnergykX2}
\end{eqnarray}
\sm{and} best approximated \sm{as},
\begin{eqnarray}
\mathscr{E}_{k0} \approx    \frac{a^2}{4}  \left[ 1   +  \left(  \frac{\pi \varepsilon \mathfrak{S}}{4}  \right)^2 \chi_1   \right] \left[  1 + \frac{ \sinh{( k_p H_s)} }{ \sinh{(4\Lambda)} }    \right] \, .
\label{eq:EnergykX3}
\end{eqnarray}
\sm{The above formula contains a novel term \sm{within the second brackets} that resembles the ratio between group and phase velocities of harmonic waves.}
The corrections of the type $ \sinh{( k_p H_s)}/\sinh{(4\Lambda)}$ due to finite-amplitude waves \sm{are, to leading order,} the \sm{main} difference to the result\sm{s} of \sm{Eq.~(\ref{eq:energ0})}. \sm{The main implication of such a result is that \jk{the} kinetic energy density growth for a given increase in steepness (offshore or at the surf zone) is much faster when waves are near the breaking regime.}

\xx{The integral $\int_{-h(x)}^{\zeta}$ \xy{with integration from $-h(x)$ to $\zeta$ is not a meaningful quantity} because there is no single wave amplitude. \xy{Rather,} we should compute another average, the rms of the proposed integral over the distribution of the surface elevation $f(\zeta)$,}
\begin{eqnarray}
 \xx{\langle \mathscr{E}_{k} \rangle_{\textrm{rms}} = \sqrt{\int_{-\infty}^{\infty} \mathscr{E}^2_{k} f(\zeta)   d\zeta} \quad .}  
\end{eqnarray}
\xx{If we assume a Gaussian sea with $f(\zeta) = \, e^{-\zeta^2 / 2m_0}/\sqrt{2\pi m_0}$, with $4\sqrt{m_0} = H_s$ measuring the variance of the spectrum through the significant wave height $H_s$, one can recompute Eq.~(\ref{eq:Energyk}) with the full interval $[-h(x), \zeta]$ and obtain:}
\begin{eqnarray}
\hspace{-0.4cm}
\xx{
\left\langle \mathscr{E}_{k} \right\rangle_{\textrm{rms}}^2} &\xx{=}&\xx{  \frac{\xxx{a^4}}{\xxx{16}} \Bigg\{ 1 + \frac{ \langle \sinh^2{(2 k_p \zeta)} \rangle_{\textrm{rms}} }{ \sinh^2{(2\Lambda)} }  }
\label{eq:EnergykXx2}
\\
\nonumber
&\xx{+}& \xx{  \left(  \frac{\pi \varepsilon \mathfrak{S}}{4}  \right)^4  \chi_1^2 \left[  1 + \frac{ \langle \sinh^2{(4 k_p \zeta)} \rangle_{\textrm{rms}} }{ \sinh^2{(4\Lambda)} }    \right]   } 
\\
\nonumber
&\xx{+}& \xx{2 \left(  \frac{\pi \varepsilon \mathfrak{S}}{4}  \right)^2  \chi_1 \left[  1 + \frac{ \langle \sinh{(2 k_p \zeta)}  \sinh{(4 k_p \zeta)} \rangle_{\textrm{rms}} }{ \sinh{(2\Lambda)} \sinh{(4\Lambda)} }    \right] \Bigg\}  \, ,}
\end{eqnarray}
\xx{noting that the rms of $\sinh{(p\zeta)}$ is zero for any $p \in \mathbb{N}$.}
\pf{As observed in} \jfm{figure} \ref{fig:RMSEnergy}\pf{,} \xy{using} the rms of the surface elevation as an upper interval for the integral property \xy{does not significantly affect the calculated energy}. Even within wave breaking regimes ($\pf{\epsilon} \equiv k_pH_s \approx 0.2$\pf{, where $\epsilon$ is a secondary measure of steepness}), \jfm{figure} \ref{fig:RMSEnergy}\jfm{b} displays a small decrease in total energy density in shallow water ($k_p h \approx 0.2$), otherwise negligible in the regime \xy{of interest, namely} $0.4 \leqslant k_p h \leqslant 2$.
 
In the narrow range in relative water depth ($\Lambda  \sim 0.2$) where extreme wave statistics are strongly amplified due to shoaling \citep{Benoit2023}, the finite-amplitude correction within the last pair of brackets on the r.h.s of Eq.~(\ref{eq:EnergykX3}) reduces to $1 + H_s/(4h)$. As such, the finite-amplitude correction will only become relevant if $H_{s} \gtrsim 0.4 h$. This ratio was considerably smaller in the experiments reported in \citet{Trulsen2020}, and did not exceed this threshold in other similar studies \citep{Chabchoub2019,Adcock2021c}. However, in deep-water the term $\sinh{(4\Lambda)}$ grows rapidly, and \ac{no matter how} large $H_s/h$ is, the finite-amplitude correction to the kinetic energy is negligible. \sm{Yet, the wave steepness encoded in $\varepsilon$ still matters and differentiates the energy from the linear formula ($a^2/4$) in such limiting conditions.}

\subsection{\sm{Secondary Effects due to Wavenumber and Bottom Slope}}\label{sec:bound2}

\jk{Besides the potential energy \citep{Mendes2023b} the bottom slope $\partial h/\partial x$ is also expected to arise in the kinetic energy. The sharp modification of wavenumber and related variables} $(h,k,\lambda)$ \sm{also affects the kinetic energy. Thus, these variables} will \jk{evolve} in the direction of motion in addition to the textbook solution for $u_0$ \jk{which assumes $\pd{k}{x} = \pd{h}{x} = 0$} \citep{Dalrymple984}:
\begin{eqnarray}
\hspace{-0.2cm}
\frac{u_0}{a\omega} \equiv \frac{1}{a\omega}  \pd{\Phi}{x}  =  \frac{\cosh{\varphi}}{\sinh{\Lambda}} \cos{\phi} +   \frac{3ka}{4}  \frac{\cosh{(2\varphi)}}{\sinh^4{\Lambda}}  \cos{(2\phi)}  \, .
\label{eq:EnergykX02}
\end{eqnarray} 
\jk{Since we are considering  unidimensional inhomogeneity of the wave field,}
\begin{equation}
\sm{
\pd{\phi}{x} =k + x \pd{k}{x} \quad ; \quad   \pd{\varphi}{x} = (z+h)\pd{k}{x} + k \pd{h}{x} \quad ,
}
\end{equation}
\sm{\ac{and} for simplicity and brevity of the tedious algebra, we henceforth use the notation $\partial/\partial x \equiv \nabla$,}
\begin{equation}
\sm{
\sm{\nabla} \phi=k + x \sm{\nabla} k \quad ; \quad   \sm{\nabla} \varphi = (z+h)\sm{\nabla} k + k \sm{\nabla} h \quad .
}
\end{equation}
The \sm{change in} horizontal velocity $\Delta u = u-u_0$ is \pf{then a very long function containing the new terms $x \nabla k$, $(z+h) \nabla k$ and $k\nabla h$.} Finite-amplitude corrections to the kinetic energy $\Delta\mathscr{E}_{k} \pf{=\mathscr{E}_{k} -  \mathscr{E}_{k0}}$ due to the spatial evolution of the physical variables across the shoal become,
\begin{eqnarray}
 \pf{\Delta\mathscr{E}_{k}}  &\approx& \frac{1}{2 \lambda g} \int_{0}^{\lambda} \int_{-h}^{H_{s}/4}  \left(  2 u_0 \Delta u + (\Delta u)^{2} \right) \, dx \, dz \, .
\label{eq:EnergykX4}
\end{eqnarray}
\pf{For the sake of brevity and to focus on the resulting changes in the energy, we refer the reader to} Eq\pf{s}.~(\ref{eq:EnergykX00}-\ref{eq:EnergykX0}) \pf{for the full expressions}.
From the dispersion relation \xx{and assuming a slow bathymetry change ($\nabla h \ll kh$)}, \sm{one can approximately link changes on the  depth, wavelength and wavenumber \xxx{($\tilde{\nabla} h = \pi\nabla h / kh_{0}$)},}
\begin{equation}
\sm{\nabla} \lambda   \approx \frac{\pi }{  2  } \cdot \frac{\sm{\nabla} h}{ kh_{0} \left( 1 + \frac{x\sm{\nabla} h}{h_{0}}  \right)   } \equiv  \frac{\tilde{\nabla} h}{ 2 \left( 1 + \frac{2x}{\lambda} \tilde{\nabla} h \right)   } \equiv \frac{\tilde{\nabla} h}{2 f(x,\tilde{\nabla} h)}.
\label{eq:nablalambda}
\end{equation}
\xx{Such an approximation captures the leading order transformation of the wavelength gradient that has to be spatially integrated in \pf{eq.~(\ref{eq:EnergykX4})} while avoiding intractability of the latter. In \jfm{appendices} \ref{sec:THEOR2}-\ref{sec:refraction} we examine the relevance of computing $\nabla \lambda$ in full.}
Eq.~(\ref{eq:nablalambda}) is a linearization of Eq.~(\ref{eq:nablalambda0}) \jk{obtained} by fitting $\cosh^2{\Lambda} + \Lambda \approx 4 \Lambda$ (\jfm{Figure} \ref{fig:approxNABLA22}).
\begin{figure*}
\hspace{0.0cm}
\minipage{0.44\textwidth}
   \includegraphics[scale=0.58]{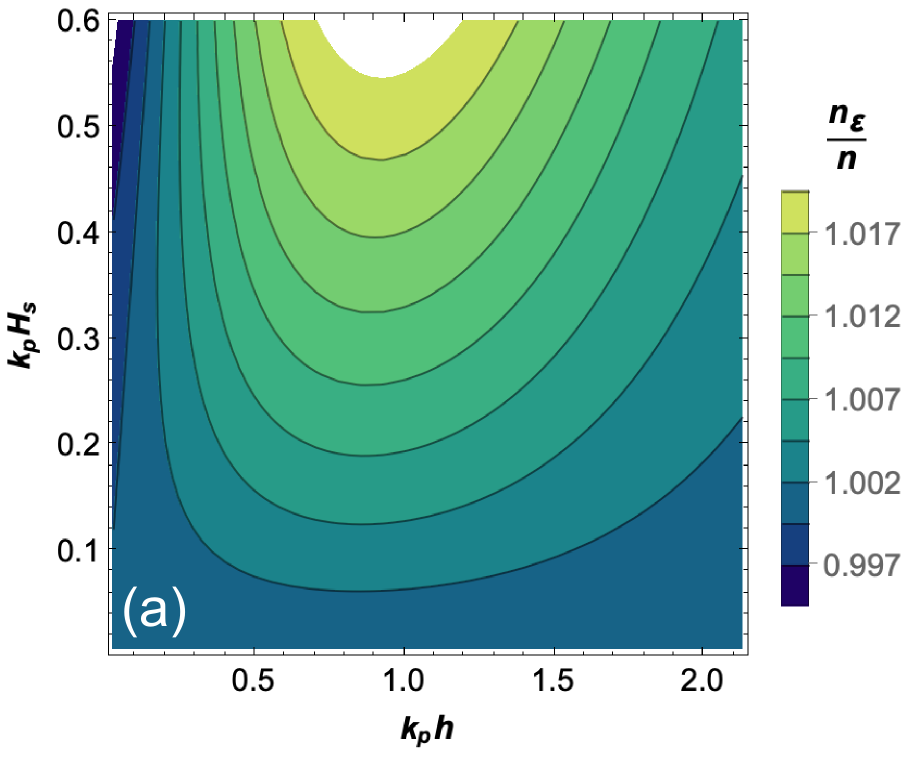}
\endminipage
\hfill
\minipage{0.52\textwidth}
   \includegraphics[scale=0.58]{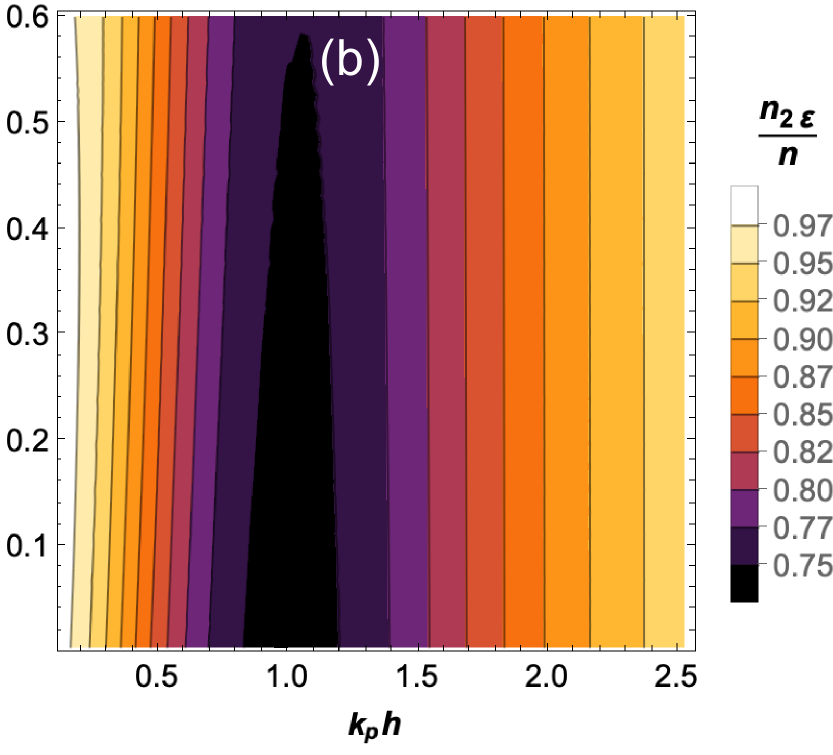}
\endminipage
\caption{\justifying{Comparison between the group-to-phase velocity ratio modified by steepness (a) $n_{\varepsilon}$, (b) $n_{2\varepsilon}$ and their steepness-independent counterpart \pf{(group velocity ratio)} $n \pf{ = c_g/c_p}$}.}
\label{fig:Groupp}
\end{figure*}

Recalling Eq.~(\ref{eq:EnergykX0}) and that cross-order trigonometric functions \sm{like $\sin{(m \varphi)} \sin{(n \varphi)}$} vanish after integration for \sm{$m \neq n$}, as well as $\sm{\nabla} k / k = - \sm{\nabla} \lambda / \lambda$, we find to leading order (see \pf{eq.~(\ref{eq:EnergykX52})} for full calculations):
\begin{widetext}
\begin{eqnarray}
\Delta\mathscr{E}_{k}  \approx  \frac{(a\omega)^2}{2 \lambda g} \int_{-h}^{H_{s}/4} \int_{0}^{\lambda}  dx \, dz \, \Bigg\{ -  \frac{2x \sm{\nabla} \lambda}{\lambda} \left[   \frac{\cosh^2{\varphi}}{\sinh^2{\Lambda}} \cos^2{\phi} +  \left(  \frac{3ka}{4}  \right)^2  \frac{\cosh^2{(2\varphi)}}{\sinh^8{\Lambda}}  \cos^2{(2\phi)} \right] \Bigg\}  \approx   \frac{ka^2 }{\sinh{(2\Lambda)}} \sm{\mathcal{I}_{1}} \quad .
\label{eq:EnergykX5}
\end{eqnarray}
\end{widetext}
According to the approximated spatial evolution of wavelength of Eq.~(\ref{eq:nablalambda}), we compute $\mathcal{I}$ whilst implementing the transformation from regular to irregular waves \pf{(see eqs.~(\ref{eq:EnergykX52}-\ref{eq:EnergykX66}) for a thorough computation)}:
\begin{widetext}
\begin{eqnarray}
\mathcal{I}_{\xxx{1}} \approx - \frac{\tilde{\nabla} h}{ \lambda^2} \int_{-h}^{H_{s}/4} \int_{0}^{\lambda}    \left[   \cosh^2{\varphi} \cos^2{\phi} +  \left(  \frac{\pi \varepsilon \mathfrak{S}}{4}  \right)^2 \chi_1  \frac{\cosh^2{(2\varphi)}}{\cosh{(2\Lambda)}}  \cos^2{(2\phi)} \right]  
\frac{x \,  dx \, dz}{\left( 1 + \frac{2x}{\lambda} \tilde{\nabla} h \right)}   \quad .
\label{eq:EnergykX6}
\end{eqnarray}
\end{widetext}
\sm{We remark that the vertical integration over the $z$-axis will result in Eq.~(\ref{eq:EnergykX3}). Thus, we now focus on the horizontal integration.} The lowest order in steepness \sm{gives rise to} a spatial function:
\begin{equation}
    \mathcal{I}_{1x} = -\frac{\tilde{\nabla} h}{ \lambda^2} \int_{0}^{\lambda} 
\frac{x       \cos^2{\phi} }{\left( 1 + \frac{2x}{\lambda} \tilde{\nabla} h \right)} \,  dx \equiv \pf{f_{\tilde{\nabla} h}}   \, ,
\end{equation}
\jk{where we approximate $f$ as a linear function $\pf{f_{\tilde{\nabla} h}} \cong - (3/8)\tilde{\nabla} h$ in the range $| \tilde{\nabla} h | < 1/3$, covering $>90\%$ of natural and artificial bottom slopes \citep{Seelig1983,Piper2005,Becker2008}. For steeper slopes,  cubic functions have to be added (see \jfm{Figure} \ref{fig:approxNABLA22} and \jfm{appendix} \ref{sec:THEOR2}).}
\sm{After integrating over the $z$-axis,} the \sm{approximated} spatial-evolution-dependent correction to the kinetic energy becomes,
\begin{eqnarray}
\hspace{-0.3cm}
\frac{\Delta\mathscr{E}_{k}}{\pf{f_{\tilde{\nabla} h}}} \approx  \frac{a^2}{4}   \left[ n_{\varepsilon}    +  n_{2\varepsilon} \left(  \frac{\pi \varepsilon \mathfrak{S}}{4}  \right)^2 \chi_1     \right]   \left[  1 + \frac{ \sinh{( k_p H_s)} }{ \sinh{(4\Lambda)} }    \right]   
\end{eqnarray}
where \sm{we have found novel analytical terms},
\begin{eqnarray}    
n_{\varepsilon}  &=& \frac{1}{2} \left[ 1 + \frac{2\Lambda + \frac{1}{2} k_p H_s}{\sinh{(2\Lambda)} + \sinh{(\frac{1}{2} k_p H_s)}}  \right] \,\, ;  \label{eq:n1} \\
n_{2\varepsilon} &=& \frac{1}{2} \left[ 1 + \frac{4\Lambda +  k_p H_s}{\sinh{(4\Lambda)} + \sinh{( k_p H_s)}}  \right] \, .
\label{eq:n2}
\end{eqnarray}
\pf{Note that these functions carry a resemblance to the ratio between group and phase velocities \citep{BMei2005,Holthuijsen2007}:}
\begin{equation}
\pf{
n = \frac{c_g}{c_p} = \frac{1}{2} \left[ 1 + \frac{2\Lambda }{\sinh{(2\Lambda)}}  \right] \quad .
}
\end{equation}
\pf{Hence, $(n_{\varepsilon},n_{2\varepsilon})$ generalize the group-to-phase velocity ratio $n$ to any wave regime in wave steepness.}
\sm{We observe the new parameters controlling the integral properties factored out in a first-of-its-kind formula. In addition \xxx{to} the wave steepness and harmonics playing an important role \citep{Ma2020}, \ac{we can now characterize} how wave breaking alters the energetics through $H_s/h$, followed by the slope magnitude $\sm{\nabla} h$ and the group-to-phase velocity ratios $(n_{\varepsilon}, n_{2\varepsilon},n)$.}
\sm{This result is the first demonstration that integral properties of waves subject to sharp changes to their characteristics will be corrected by a non-unitary physical term\jk{. In other words}, the typical formula $1+ \mathcal{O}(\varepsilon^2)
$ often seen elsewhere \citep{Ma2020,Mendes2022} is a \jk{limiting} case of an otherwise steepness-dependent group-to-phase velocity ratio $n_{\varepsilon}$.}

Naturally, in the limit of deep-water linear (Airy) waves ($\Lambda \rightarrow \infty , \varepsilon \rightarrow 0$) we find that $n_{\varepsilon} = n = n_{2\varepsilon} = 1/2$. As shown in \jfm{Figure} \ref{fig:Groupp}, the effect of steepness is \sm{small} on the group velocity ratios. \sm{In fact}, $n_\varepsilon$ \sm{typically deviates from} $n$ \sm{by a few percent and reaches a maximum of 2\% in wave breaking regime, whereas} $n_{2\varepsilon} \sm{< n}$ in all cases. \sm{The decrease of $n_{2\varepsilon}/n$ in intermediate waters $(0.8 \leqslant k_p h \leqslant 1.5)$ can impact the \xxx{the stochastics of the shoaling process}. But this effect is negligible \ac{for wave statistics} because of the very small \ac{value of} kurtosis $\mu_4 \sim (\varepsilon / k_p h)^2$ in this region~\citep{Zhang2024}. Strong rogue wave amplification, however, appears for $k_p h \leqslant 0.5$ \citep{Benoit2023}. Otherwise, $n_{2\varepsilon}$ also stays within a few percent of $n$}. \sm{Ergo, the simplest revised formulation for} the kinetic energy correction \sm{is}:
\begin{eqnarray}
\hspace{-0.3cm}
\Delta\mathscr{E}_{k}  \approx    \frac{a^2}{4}   \left[ 1   +   \left(  \frac{\pi \varepsilon \mathfrak{S}}{4}  \right)^2 \chi_1   \right] \left[  1 + \frac{ \sinh{( k_p H_s)} }{ \sinh{(4\Lambda)} }    \right] \pf{f_{\tilde{\nabla} h}} \,  n \, .
\label{eq:EnergykXX}
\end{eqnarray}
Hence, taking into account corrections due to the finite amplitude of waves and the spatial evolution of wave variables along the shoal, we recalculate the total energy density consisting of potential and kinetic terms $\mathscr{E} = \mathscr{E}_p + \mathscr{E}_{k0} + \Delta\mathscr{E}_{k}$:
\begin{eqnarray}
\hspace{-0.3cm}
&{}& \frac{2\mathscr{E}}{a^2}
\approx  1 + \frac{\pf{f_{\tilde{\nabla} h}} \,  n}{2} + \frac{ \sinh{( k_p H_s)} }{2 \sinh{(4\Lambda)} } \left[ 1 + \pf{f_{\tilde{\nabla} h}} \,  n \right]  
\label{eq:finiteampgamma0}
\\
\nonumber
\hspace{-0.5cm}
&+& \frac{1}{2} \left(  \frac{\pi \varepsilon \mathfrak{S}}{4}  \right)^2 \left\{ \Tilde{\chi}_1 + \chi_1  \left[  1 + \frac{ \sinh{( k_p H_s)} }{ \sinh{(4\Lambda)} }    \right]  \left[ 1 + \pf{f_{\tilde{\nabla} h}} \,  n \right] \right\}  \, .
\end{eqnarray}
In \pf{the} derivation \pf{of eq.~(\ref{eq:finiteampgamma0})}, we did not consider \pf{all the terms for the kinetic energy described in eq.~(\ref{eq:EnergykX52}) nor} the influence of the wavenumber evolution within functions of the type $\cos{(kx)}$ as a quickly varying \pf{function} within the WKB \pf{(\textit{Wentzel–Kramers–Brillouin}) approximation}. Moreover, the refraction of the wave field could have been modified by higher-order functions of the slope $\nabla h$ due to $k(x)$\footnote{\pf{The WKB approximation and modified mild-slope equation are occasionally employed to compute the full effects of bottom slopes on the scattering and refraction of linear water waves — namely, abrupt spatial changes in the wavenumber \citep{Meyer1979,Chamberlain1995,White1999,BMei2005,Bautista2011}}}. In \jfm{appendix} \ref{sec:refraction} we explore these supplementary corrections \pf{due to wavenumber refraction}, and show that they are not of leading order. \pf{However, it is important to highlight the regime in which all full computations of \jfm{appendices} \ref{sec:THEOR2}-\ref{sec:refraction} become relevant: these supplementary corrections are omitted in the present section because we are dealing with slopes of $\nabla h \lesssim 1/3$. However, if slopes are very steep ($\nabla h \gtrsim 1$), we would have to consider all the terms in the appendices in addition to new ones arising due to strong reflection. These terms would be relevant for surging and collapsing types of breakers, whereas the main result of eq.~(\ref{eq:finiteampgamma0}) directly applies to plunging and spilling breakers.}
\begin{figure*}
\centering
  \includegraphics[scale=0.67]{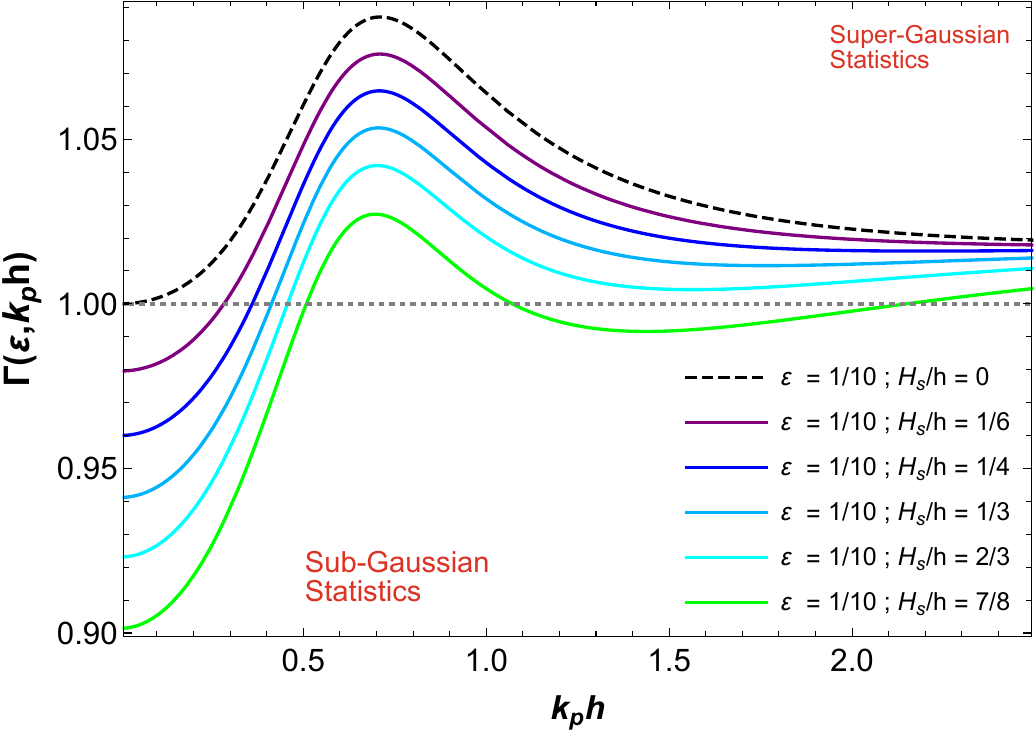}
\caption{\justifying{Effect of breaking proxy ratio $H_s/h$ on the non-homogeneous parameter $\Gamma$ defining anomalous wave statistics according to the approximation of Eq.~(\ref{eq:finiteampgamma2}). The dashed line depicts the theoretical curve for small-amplitude waves of \citet{Mendes2021b}. \sm{Super-Gaussian refer to $\Gamma - 1 > 1/100$ and sub-Gaussian to $\Gamma - 1 < - 1/100$.}}}
\label{fig:GammaHs}
\end{figure*}

\section{Effects of Energy Evolution on Wave Statistics}\label{sec:Conseq}

\sm{In this section, we apply the method of \citet{Mendes2023} to estimate the impact of the formulation for the energetics of waves breaking over relatively steep slopes \jk{developed in the previous section} (Eq.~(\ref{eq:finiteampgamma0})) on the statistical properties of waves, namely the excess kurtosis. The latter constitute a proxy for measuring the heavy tails of the exceedance probability in ocean engineering contexts \citep{Marthinsen1992,Janssen2003,Janssen2006a,Mori2007}, including for shoaling in intermediate and shallow-water regimes \citep{Trulsen2020,Doleman2021,Adcock2021c,Mori2023}.}

\subsection{\sm{Current Stochastic Knowledge}}

We assume that the statistical properties and homogeneity of a wave train travelling over a flat bottom are well-described by the linear wave theory. We also assume stationarity in time but consider spatial non-homogeneity. The Gaussian statistics of the initial wave train will be affected as soon as the wave train is \xxx{disturbed} by the bathymetry. Consequently, its exceeding probability $\mathcal{R}_{\alpha}$ will be transformed into $\mathcal{R}_{\alpha, \Gamma}$:
\begin{equation}
\hspace{0.0cm}
\mathcal{R}_{\alpha, \Gamma} \equiv \mathcal{R}_{\Gamma}(H>\alpha H_{s}) = e^{-2\alpha^{2}/\mathfrak{S}^{2}\Gamma} \quad , 
\label{eq:Rayexc}
\end{equation}
\pf{where \pf{$\alpha$ is the normalized height and the spectral parameter} $\Gamma$ measures how the variance of the surface elevation differs from the total spectral energy:}
\begin{eqnarray}
\pf{\Gamma (x) := \frac{\langle \zeta^{2} (x,t) \rangle }{ \mathscr{E} }   = \frac{\langle \zeta^{2} (x,t) \rangle (x) }{ \mathscr{E}(x)   } \quad .} 
\label{eq:GammaEnsemble}
\end{eqnarray}
Since the surface elevation $\zeta$ has a zero mean, $\langle \zeta^2 \rangle$ is the variance of the surface elevation. Applying Eq.~(\ref{eq:Rayexc}) to estimate the excess in the kurtosis of the sea surface elevation directly from \sm{energetics (}$\Gamma$\sm{)} yields \citep{Mendes2023}:
\begin{equation}
\jk{\mu_{4} = \frac{1}{9} \left[ e^{ 8 \left( 1 - \frac{ 1 }{ \mathfrak{S}^{2}\Gamma } \right) }- 1 \right] \approx \frac{1}{9} \left[ e^{ 8 \left( 1 - \Gamma^{-(1+\kappa)} \right) }- 1 \right] \quad ,}
\label{eq:kurt}
\end{equation}
where $\kappa$ is an asymmetry \pf{correction that naturally arises to fulfil} initial conditions prior to the shoal ($\mu_4 \approx 0$\pf{,} see section 3.4 of \citet{Mendes2021b}\pf{), being computed as:}
\begin{equation}
\pf{\kappa = \frac{ \ln{\mathfrak{S}(\varepsilon , \Lambda)} }{ \ln{\Gamma (\varepsilon , \Lambda)} } \quad .
}
\label{eq:kappa}
\end{equation}
According to appendix B of \citet{Mendes2023}, we may approximate $\mathfrak{S} \approx (7/6) (1+2\varepsilon)$. Without the presence of wave breaking, vertical asymmetry grows monotonically but very slowly and the $1+\kappa$ will be mostly unchanged unless $H_s$ reaches the surf zone.

\subsection{Solution \ac{Framework} for Finite Amplitude Waves}\label{sec:THEOR}

We carry the calculation of the non-homogeneous $\Gamma$ to leading order in $\tilde{\nabla} h$ as depicted in \jfm{Figure} \ref{fig:GammaHs}, correcting \citet{Mendes2022} to account for finite amplitude waves. Note that at the peak amplification $\Lambda \sim 1/2$ we have $n \sim 3/4$, thus $\pf{f_{\tilde{\nabla} h}} \,  n \sim -\tilde{\nabla} h/4$:
\begin{widetext}
\begin{equation}
 \Gamma \approx \frac{ 1 + \frac{\pi^{2} \varepsilon^{2} \mathfrak{S}^{2}}{16}     \, \Tilde{\chi}_{1} }{  1 - \frac{ \tilde{\nabla} h }{  8 } +  \frac{ \sinh{(k_{p}H_{s})}  }{2 \sinh{(4k_{p}h)} }  \left[ 1 - \frac{ \tilde{\nabla} h }{  4}  \right]  +   \frac{\pi^{2} \varepsilon^{2} \mathfrak{S}^{2}}{32}   \, \left( \Tilde{\chi}_{1}  + \left[ 1 + \frac{ \sinh{(k_{p}H_{s})}  }{ \sinh{(4k_{p}h)} } \right] \left[ 1 - \frac{ \tilde{\nabla} h }{  4 }  \right]  \chi_{1} \right) +  \check{\mathscr{E}}_{p2} } \quad ,
\label{eq:finiteampgamma1}
\end{equation}
\end{widetext}
where $\check{\mathscr{E}}_{p2}$ is the \pf{dimensionless (or spectral)} net potential energy due to the change in mean water level along the shoal \sm{\citep{Mendes2022}}:
\begin{equation}
\jk{    
\check{\mathscr{E}}_{p2} = \left( \frac{5\varepsilon^2}{\Lambda^2} \right) \tilde{\nabla} h  (1+\tilde{\nabla} h )\lesssim \frac{ \tilde{\nabla} h}{ 5} \quad .
}
\label{eq:ep2}
\end{equation}
At the maximum amplification of \jk{kurtosis} ($k_p h \sim 0.7$), $\check{\mathscr{E}}_{p2}$ tends to compensate the $- \frac{ \tilde{\nabla} h }{ 8 }$ term of Eq.~(\ref{eq:finiteampgamma1}). Furthermore, the typical value of $n_{2\varepsilon}/n_{\varepsilon}$ in the regime of interest in intermediate waters is of $7/8$. \jk{Considering this deviation from the usual approximation to unity} weakens both terms \jk{featuring} $ 1 - \frac{ \tilde{\nabla} h }{  4 }$. In fact, the effect of $ 1 -  \tilde{\nabla} h /  4  \lesssim 1.05  $ adjusted by $n_{2\varepsilon}/n_{\varepsilon}$ is even smaller for wave trains travelling from deep-water towards shallow-water. Hence, we can further approximate Eq.~(\ref{eq:finiteampgamma1}) in the steep slope regime where the slope effect saturates \ac{following}:
\begin{equation}
 \Gamma \approx \frac{ 1+   \frac{\pi^{2} \varepsilon^{2} \mathfrak{S}^{2}}{16}     \, \Tilde{\chi}_{1} }{  1 +  \frac{ \sinh{( k_{p}H_{s})}  }{2 \sinh{(4k_{p}h)}   }+   \frac{\pi^{2} \varepsilon^{2} \mathfrak{S}^{2}}{32}   \, \left( \Tilde{\chi}_{1}  + \left[ 1 + \frac{ \sinh{(k_{p}H_{s})}  }{ \sinh{(4k_{p}h)} } \right]  \chi_{1} \right) }    \,\, .
\label{eq:finiteampgamma2}
\end{equation}
Note that \ac{we do not favour a particular definition of} steepness \pf{$(\varepsilon, \epsilon)$} \ac{and keep them both in the solution as they lead to different analysis}\ac{. Thus,} the role of $H_s/h$ \jk{appears as hyperbolic sines} in the limit of $k_p h \rightarrow 0$.
The denominator has increased in magnitude, whereas the numerator has not changed in comparison with the result of section 3 of \citet{Mendes2021b}. Thus, we expect a significant decrease \jk{of} $\Gamma$ towards a weaker deviation from a Gaussian sea, \jk{therefore less pronounced extreme statistics. It can} even \jk{reach} a sub-Gaussian behaviour near wave breaking conditions as depicted in \jfm{Figure} \ref{fig:GammaHs}. If, however, waves are travelling over a slope starting already in shallow-water, then $ 1 -  \tilde{\nabla} h /  4 \lesssim 2$ and the approximation of Eq.~(\ref{eq:finiteampgamma2}) overestimates $\Gamma$ in breaking conditions. Nonetheless, this has no qualitative impact on the trend of the wave statistics. Indeed, their evolution in deep-water already shows that we achieve strong sub-Gaussian statistics \ac{in shallow-water}. Therefore, if the evolution already starts in \ac{the latter regime,} \jk{the statistics will be more} sub-Gaussian.
\begin{figure*}
\hspace{0.4cm}
\minipage{0.42\textwidth}
   \includegraphics[scale=0.68]{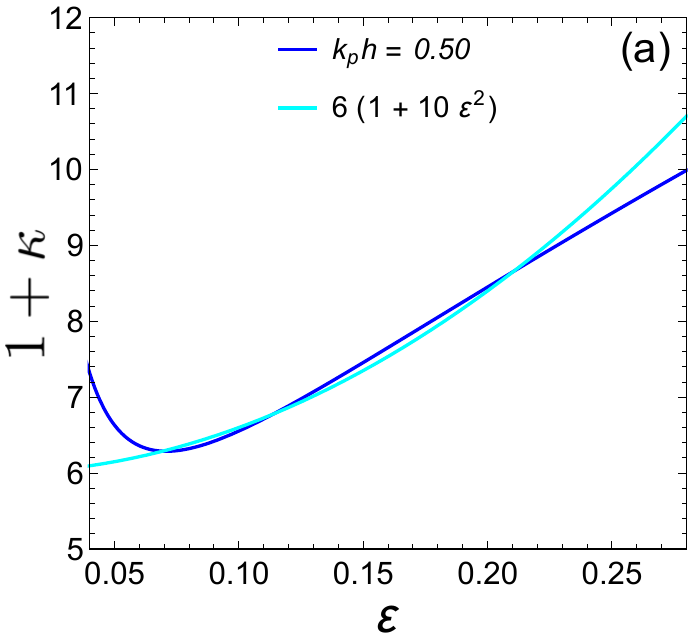}
\endminipage
\hfill
\minipage{0.55\textwidth}
   \includegraphics[scale=0.68]{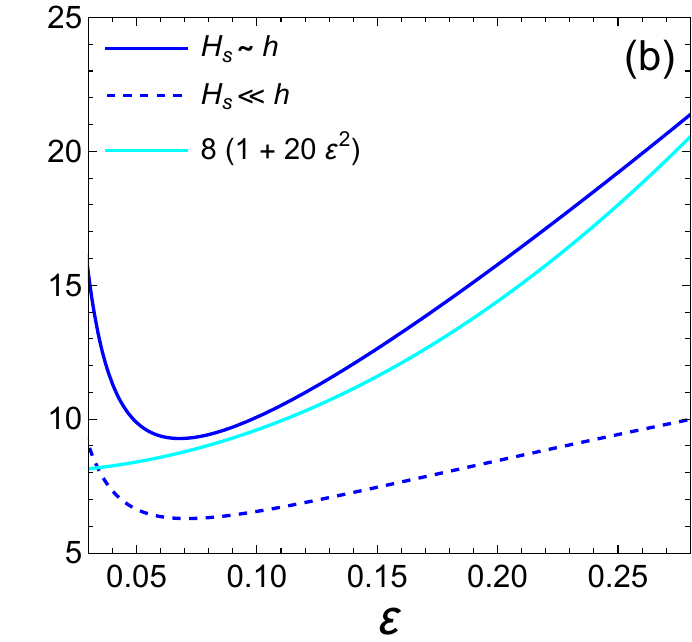}
\endminipage
   \caption{\justifying{Asymmetry parameter $\kappa$ as a function of steepness and wave breaking proxy $H_s/h$ \sm{at fixed relative water depth $k_{p}h = 0.5$}. Blue lines depict the exact calculation, whereas cyan curves show polynomial approximations. \pf{The dashed blue curve of (b) is the same as the solid blue curve in (a), to differentiate between the exact models for small amplitude (a) and finite amplitude (b).}}}
  \label{fig:Asymm}
\end{figure*}

\subsection{\sm{Generalized Formulation of} Excess Kurtosis }
\label{sec:kurtosis}

\jk{Injecting $\mathfrak{S}(\varepsilon)$ into on Eq.~(\ref{eq:kurt})}, we compute $\kappa$ through Eq.~(\ref{eq:finiteampgamma2}) and display its features \ac{and evolution as a function of $\varepsilon$} in \jfm{Figure} \ref{fig:Asymm}. For small-amplitude waves of second-order in steepness ($0.05 \lesssim \varepsilon \lesssim 0.10$), \jfm{Figure} \ref{fig:Asymm}\jfm{a} validates the approximation $1+\kappa = 6$ used by \citep{Zhang2024}. However, \jfm{Figure} \ref{fig:Asymm}\jfm{b} demonstrates that this is no longer true if the waves are of finite amplitude and near wave-breaking conditions. In this case, $1+\kappa \gtrsim 10$.  

\begin{figure*}
\centering
   \includegraphics[scale=0.83]{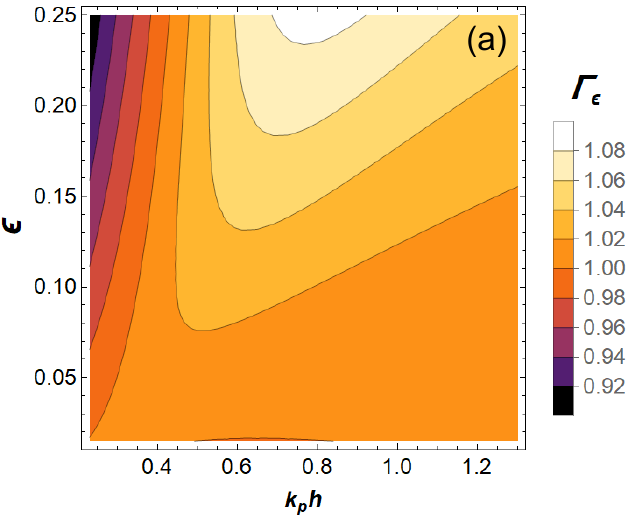}
   \includegraphics[scale=0.58]{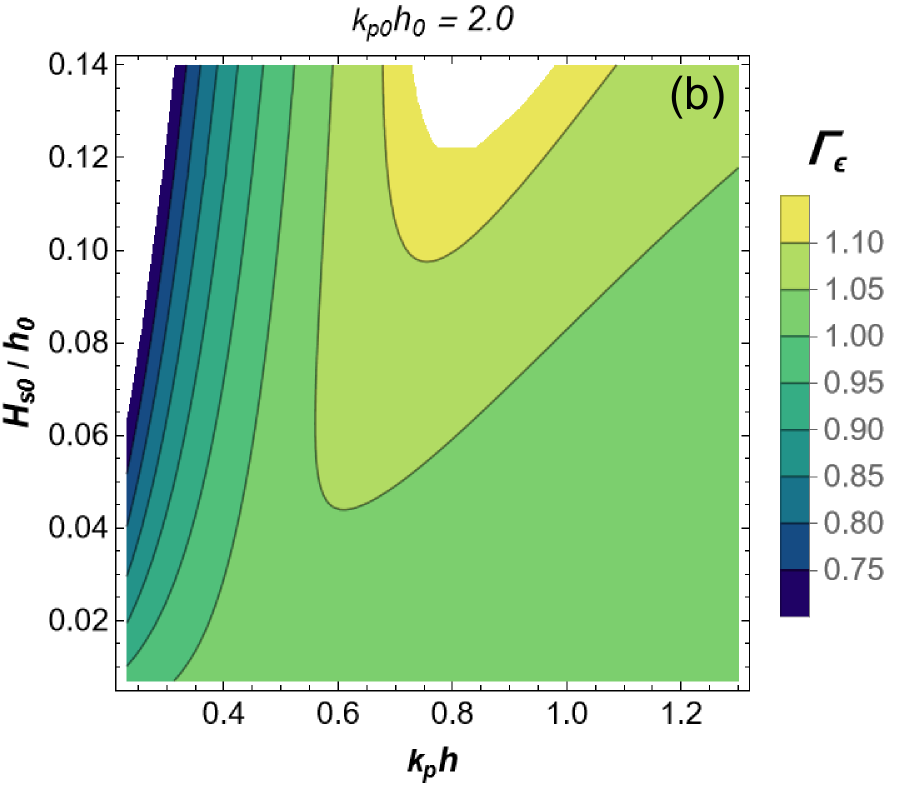}
   \caption{\justifying{(a) Evolution of the non-homogeneous parameter $\xxx{\Gamma}$ with relative water depth \pf{as a function of secondary steepness $\epsilon = k_p H_s$ and dimensionless depth $k_p h$. (b) We replace the steepness $\epsilon$ by the initial depth ratio $H_{s0}/h_0$ with offshore relative water depth $k_{p0}h_0 = 2$ while the steepness evolves with the linear shoaling coefficient.}}}
  \label{fig:GammaApproxX}
\end{figure*}

The exponential form for the excess kurtosis is evidently very cumbersome given the complexity of Eq.~(\ref{eq:finiteampgamma2}) and that \ac{of} $1+\kappa \sim 6 (1 + \mathcal{O}(\varepsilon^2))$ \pf{for small amplitude waves far from wave breaking regimes, and $1+\kappa \sim 8 (1 + \mathcal{O}(\varepsilon^2))$ for otherwise waves nearing or at breaking regimes ($H_s \rightarrow h$)}. Following the steps of Section 2.2 of \citet{Zhang2024}, we attempt to perform a Taylor expansion in $\epsilon$ around 0 on Eq.~(\ref{eq:kurt}). Denoting $\epsilon \equiv k_{p}H_{s} \approx (\pi / \sqrt{2}) \varepsilon$, a first Taylor expansion is performed on Eq.~(\ref{eq:finiteampgamma2}), leading to a finite-amplitude correction to Eq. (2.16) of \citet{Zhang2024}:
\begin{eqnarray}
\nonumber
\Gamma_\epsilon &\approx& 1 +  \frac{\epsilon^2}{\Lambda^2}   - \frac{ \sinh{\epsilon}  }{2 \sinh{(4\Lambda)}   } \left( 1 +   \frac{4 \epsilon^{2} }{3\Lambda^6}    \right) \quad , 
\\
&\approx&  1 +  \left( \frac{ \epsilon }{\Lambda} \right)^2   - \frac{ \sinh{\epsilon}  }{2 \sinh{(4\Lambda)}   } \left[ 1 +  20 \left( \frac{ \epsilon }{\Lambda} \right)^2    \right]  \,\, .
\label{eq:GammaApprox}
\end{eqnarray}
The second step requires us to expand the power $\Gamma^{1+\kappa} \approx 1 + (1+\kappa) (\Gamma - 1)$ since $\Gamma - 1 \ll 1$. Thus, the main part of the exponential calculation boils down to $1 - \Gamma^{-1-\kappa} \approx 1 + [1+(1+\kappa) (\Gamma - 1)]^{-1} \approx 1 - [1-(1+\kappa) (\Gamma - 1)] \approx (1+\kappa) (\Gamma - 1)$. Therefore, we achieve: 
\begin{eqnarray}
\nonumber
\mu_{4}  &\approx& \frac{1}{9} \left[ e^{ 8 (1+\kappa) (\Gamma - 1) }- 1 \right] \\
\nonumber
&\approx&  \frac{1}{9} \left[ 1 + 8 (1+\kappa) (\Gamma - 1)  + \frac{64}{2!} (1+\kappa)^2 (\Gamma - 1)^2  - 1 \right]
\\
&\approx& \frac{(1+\kappa)}{9} \Big[  8 (\Gamma - 1)  + 32 (1+\kappa) (\Gamma - 1)^2  \Big] \quad .
\label{eq:kurt2}
\end{eqnarray}
On the other hand, \jfm{Figure} \ref{fig:GammaApproxX} shows that in the region of highest amplification of extreme waves ($k_p h \lesssim 1/2$) \jk{$\Gamma$ does not exceed 1, so that} the sub-Gaussian behaviour will not be present. In contrast, the expansion of Eq.~(\ref{eq:kurt2}) is not heavy-tailed at $k_p H_s \gtrsim 0.3$. In other words, \pf{while obtaining a Taylor expansion that captures} the exponential behaviour of Eq.~(\ref{eq:kurt}) \pf{is straightforward in the range $0 \leqslant k_pH_s \leqslant 0.3 $, approximation for the heavy-tailed regime ($ k_pH_s > 0.3 $) by Taylor expansion (such as in Eq.~(\ref{eq:kurt2})) is quite difficult} at a reasonable order. \pf{If we keep the Taylor series up to third-order correction, a}pplying Eq.~(\ref{eq:GammaApprox}) into Eq.~(\ref{eq:kurt2}) \pf{leads to},
\begin{eqnarray}
\nonumber
\mu_{4} &\approx&  \frac{(1+\kappa)}{9} \Bigg[  8 \left(\frac{\epsilon}{\Lambda} \right)^2  + 32 (1+\kappa)\left(\frac{\epsilon}{\Lambda} \right)^4  
\\
&{}& - \frac{ \sinh{\epsilon}  }{ \sinh{(4\Lambda)}   } \left( 1 +   200 \left(\frac{\epsilon}{\Lambda} \right)^2    \right) \Bigg]  \quad .
\label{eq:kurtApp1}
\end{eqnarray}
\begin{figure*}
\hspace{0.0cm}
   \includegraphics[scale=0.7]{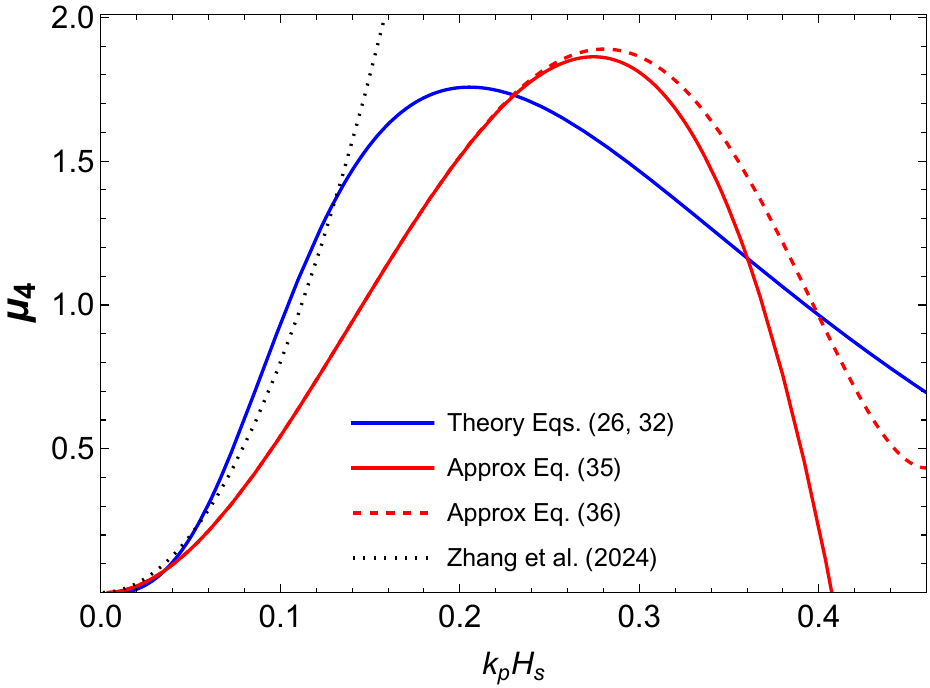}
   \caption{\justifying{Growth and saturation of \jk{maximum of} the excess kurtosis \jk{along the propagation} of irregular waves fields travelling from deep-water up to $k_p h = 1/2$ atop a breakwater. The \jk{global maximum} of kurtosis \jk{corresponds to} $H_s \approx h/2$.}.}
  \label{fig:GammaApproxX2}
\end{figure*}
We approximate this expression as a polynomial fit valid for $\Lambda > 0.5$ and $k_p H_s \le 0.5$:
\begin{eqnarray}
\nonumber
\mu_{4} &\approx&    8 \left(\frac{\epsilon}{\Lambda} \right)^2  + 32 \left(\frac{\epsilon}{\Lambda} \right)^4 \cdot \left(\frac{1+\epsilon}{12\epsilon^2}\right)  
\\
&-&    \frac{ \sinh{\epsilon}  }{ \sinh{(4\Lambda)}   } \left[ 1 +   200 \left(\frac{\epsilon}{\Lambda} \right)^2  - 40 \left(\frac{\epsilon}{\Lambda} \right)^8  \right] \,\, .
\label{eq:kurtApp2}
\end{eqnarray}
Note that \sm{$1+\kappa \sim 9$ and} the expression $(1+\epsilon)/12\epsilon^2$ arises from computing $1+\kappa$ \jk{appearing in Eq.~(\ref{eq:kurt2}) by using Eq.~(\ref{eq:GammaApprox}) for small-amplitude waves instead of the full model of Eq.~(\ref{eq:finiteampgamma2}). It has a similar magnitude as} the positive term $(40\sinh{\epsilon}/\sinh{(4\Lambda)} ) (\epsilon / \Lambda)^8$ that better captures the heavy tail of the curve. \xx{Regarding the \xy{complementary} effects arising from the difference in computing the rms energy or the energy of the rms surface elevation as an upper bound (see Eq.~(\ref{eq:EnergykXx2})), the rms energy would slightly decrease the term $\sinh{\epsilon} \rightarrow (1-\delta) \sinh{\epsilon}$ with $\delta \ll 1/2$. Consequently, the tail of the curve for kurtosis in Eq.~(\ref{eq:kurtApp2}) would be slightly heavier (blue solid line in \jfm{Figure} \ref{fig:GammaApproxX2}).}

\begin{table*}
  \centering
\begin{tabular}{rrrrrrrrrrrrrrr}    
\toprule
\emph{Run} &  \emph{$k_{p1}h_{1}$} &  \emph{$k_{p2}h_{2}$}  &  \emph{$k_{p1}H_{s1}$}  & \emph{\pf{$k_{p2}H_{s2}$}} &   \emph{$H_{s1}$}  &   \emph{\pf{$H_{s2}$}}  &  \emph{$\lambda_{p1}$}  &  \emph{$\lambda_{p2}$} &  \emph{$h_{1}$} &  \emph{$h_{2}$}  &  \emph{$  \frac{H_{s1}}{h_{1}} $}&  \emph{$ \frac{H_{s2}}{h_{2}} $}&   \emph{$ Waves$}  \\
\midrule
\bf{T} & \bf{1.80} & \bf{0.54} & \bf{0.050} & \bf{0.102} & \bf{0.014}  & \bf{\pf{0.015}} & \bf{1.75} & \bf{0.93} & \bf{0.50} & \bf{0.08} & \bf{0.028} & \bf{0.188} & \bf{--}  \\
1  &    1.40  & 0.50     &  0.032    &  \pf{0.085}   &    0.012  &  \pf{0.018}  &    2.30  &    1.30  &    0.50  &    0.10  &    0.024  &    0.176  &   25,000  \\
2  &    1.40  & 0.50     &  0.039    &  \pf{0.107}   &    0.014  &  \pf{0.022}  &    2.30  &    1.30  &    0.50  &    0.10  &    0.029  &    0.221  &   25,000  \\ 
3  &    1.40  & 0.50     &  0.044    &  \pf{0.125}   &    0.016  &  \pf{0.026}  &    2.30  &    1.30  &    0.50  &    0.10  &    0.032  &    0.258  &   30,000  \\
4  &    1.40  & 0.50     &  0.074    &  \pf{0.182}   &    0.026  &  \pf{0.038}  &    2.30  &    1.30  &    0.50  &    0.10  &    0.053  &    0.377  &   30,000  \\
5  &    1.40  & 0.50     &  0.103    &  \pf{0.246}   &    0.038  &  \pf{0.051}  &    2.30  &    1.30  &    0.50  &    0.10  &    0.076  &    0.510  &   30,000  \\
6  &    1.40  & 0.50     &  0.156    &  \pf{0.333}   &    0.057  &  \pf{0.069}  &    2.30  &    1.30  &    0.50  &    0.10  &    0.115  &    0.689  &   30,000  \\
7  &    1.40  & 0.50     &  0.182    &  \pf{0.365}   &    0.067  &  \pf{0.076}  &    2.30  &    1.30  &    0.50  &    0.10  &    0.133  &    0.756  &   25,000  \\ 
8  &    1.40  & 0.50     &  0.218    &  \pf{0.414}   &    0.080  &  \pf{0.086}  &    2.30  &    1.30  &    0.50  &    0.10  &    0.160  &    0.857  &   25,000  \\
9  &    2.82  & 0.50     &  0.346    &  \pf{0.494}   &    0.054  &  \pf{0.035}  &    1.00  &    0.45  &    0.44  &    0.04  &    0.122  &    0.884  &    2,500  \\
10 &    2.43  & 0.50     &  0.325    &  \pf{0.453}   &    0.060  &  \pf{0.047}  &    1.15  &    0.65  &    0.45  &    0.05  &    0.134  &    0.952  &    2,500  \\
\bottomrule
\end{tabular}
\caption{\justifying{Summary of approximate conditions \pf{immediately} before the shoal \pf{(gauge 1)} set for the wavemaker and atop the \pf{center of the} shoal \pf{(gauge 7)} for each run. Case T \xxx{(in bold)} depicts the equivalent values of run 1 of \citet{Trulsen2020}, which were not experimentally tested. Each experimental realization was performed with 2500 waves, and for most runs, we carried out 10-12 realizations with 2,500 waves each.
Indices 1 and 2 stand for pre-shoal and atop the breakwater, respectively.}} 
\label{tab:1}
\end{table*}
\begin{table*}
  \centering
\begin{tabular}{r|rrrrrrrrrrrrrr}    
\toprule
\emph{Measure} & \emph{G1} & \emph{G2} & \emph{G3} & \emph{G4} & \emph{G5} & \emph{G6} & \emph{G7} & \emph{G8} & \emph{G9} & \emph{G10} & \emph{G11}& \emph{G12}   \\
\midrule
$x/\lambda_{p1}$ &  0 & 0.40 & 0.60 &  0.80 &  0.95   &  1.10  &  1.25  & 1.30 &  1.45  &  1.58  &  1.65  &  2.05  \\
$x$ (m)          &  0 & 0.94 & 1.41 &  1.88 &  2.23   &  2.59  &  2.94  & 3.06 &  3.41  &  3.71  &  3.88  &  4.82  \\
\bottomrule
\end{tabular}
\caption{\justifying{\pf{Summary of wave gauge locations (G1-G12), both in absolute distance relative to the start of the breakwater (see \jfm{Figure} \ref{fig:expfig}\jfm{b}) and relative the to the mean offshore peak wavelength. In total the experiment contained eight gauges, in the first part we used gauges G1-G7 and G12, then we moved the gauges G4-G7 from the first half of the bar crest (where they were initially positioned atop the shoal) to the second half, thus becoming gauges G8-G11.}}} 
\label{tab:2}
\end{table*}
In addition to \jk{linearizing the exponential in} the kurtosis, these consecutive Taylor expansions help understanding the roots for the peak in kurtosis and its saturation and even decay. For instance, the first term \pf{of Eq.~(\ref{eq:kurtApp2})} $8 \left(\epsilon/\Lambda \right)^2$, accounts for the effect of bound harmonics on the exceedance probability and is the leading term for weakly nonlinear seas. It\jk{s order} is \jk{consistent with} computations of deep-water regimes following the Zakharov equation \citep{Janssen2006a} and other shoaling\ac{-related} frameworks \citep{Mori1997,Adcock2021c}. The second term of \pf{Eq.~(\ref{eq:kurtApp2})} is proportional to $\epsilon^4$\jk{, accounting for \pf{fourth}}-order in steepness\jk{, whereby the kurtosis depends on the second order. Interestingly, the integral properties give rise to next-order contributions, despite the necessity of reformulating the surface elevation}, thus, providing the sought\sm{-after} effective theory. Finally, the term \sm{$\sinh{\epsilon}$} \ac{in Eq.~(\ref{eq:kurtApp2})} describes how the finite amplitude waves affect not only the bound harmonics (polynomials) but also the kinetic energy. The novelty and rather unusual shape of this term in the context of statistical hydrodynamics relies on the unified description of integral properties up to wave breaking. As seen in \jfm{Figure} \ref{fig:GammaApproxX2}, \jk{Eq.~(\ref{eq:kurtApp2}) approximates well} the full model. First and foremost, the kurtosis grows quickly as a function of $\epsilon^2$ at a given relative water depth, as expected from the small-amplitude formula \citep{Zhang2024} for arbitrary slopes. The rapid growth of kurtosis along steep slopes is consistent with bound-harmonics effects as described in \citet{Adcock2021c} or \citet{Janssen2006a}. Secondly, the kurtosis reaches its peak \jk{for $ H_s / h \approx 0.5$} as wave \ac{breaking} starts to occur and the excess in kinetic energy dominates the bound harmonics. Note that in the context of shoaling, this regime \ac{had not yet been captured by existing theoretical frameworks}. Finally, the kurtosis decays because the \jk{breaking} term $200 (\epsilon / \Lambda)^2 \sinh{\epsilon}/ \sinh{(4\Lambda)}$ strongly compensates the leading-order solution of \citet{Zhang2024}, that is, the waves must break due to high values of $H_s/h$ in addition to being very steep. Nevertheless, \ac{and} as pointed out in \citet{Zhang2024}, these consecutive Taylor expansions serve to estimate the maximum kurtosis atop the shoal or at its end. \jk{Estimating the spatial resolution requires the full model.} \pf{Finally, we highlight that the combined effects of the wavenumber refraction and WKB approximation only slightly affect the results of \jfm{Figure} \ref{fig:GammaApproxX2}, see its revisited plots of \jfm{appendix} \ref{sec:refraction} in \jfm{Figure} \ref{fig:GammaWKBRefrac}.}

\section{Experimental Study}\label{sec:Exp}

\yh{\xt{A series of wave-flume experiments over a breakwater were carried out to validate the theory above, in particular, the results related to the kurtosis. In the current section, we \ac{\xt{provide details about}} the experimental setup, experimental results, and the comparison between the theoretically-predicted and experimental maximum kurtosis values for different $k_pH_s$, along with discussions on the agreement of the results.}}

\begin{figure*}
   \includegraphics[scale=0.55]{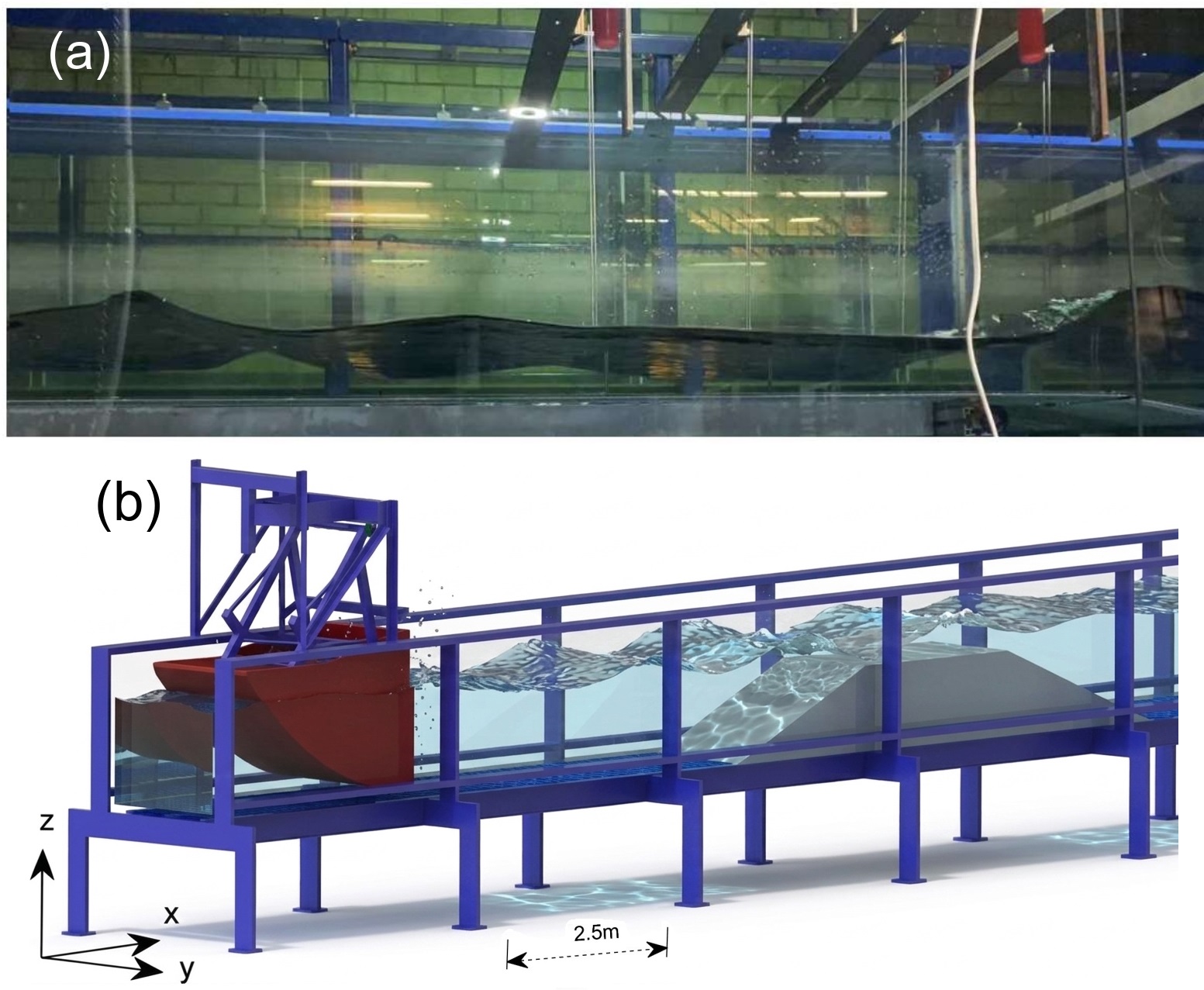}
  \caption{\justifying{(a) Photograph of the experimental setup showing the wave gauge placements over the breakwater and (b) Experimental setup limited to the wave propagation region in the flume. The piston-type wave paddle on the left generates irregular waves \jk{with JONSWAP spectrum} according to table \ref{tab:1}. A symmetric breakwater with a height of $0.4$~m and length of \pf{$6$}~m is fixed near the paddle. The length of the breakwater's upper and horizontal plane is \pf{$2$~m}. The waves are fully dissipated by an artificial grass beach at the end of the flume.}}
  \label{fig:expfig}
\end{figure*}

\begin{figure*}
    \centering
\includegraphics[width=0.37\textwidth]{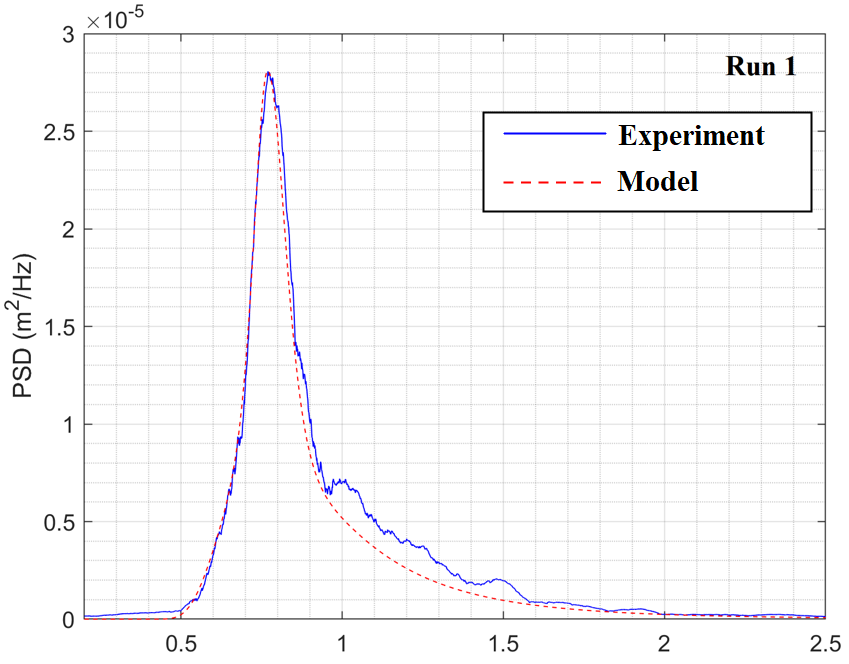}
\includegraphics[width=0.37\textwidth]{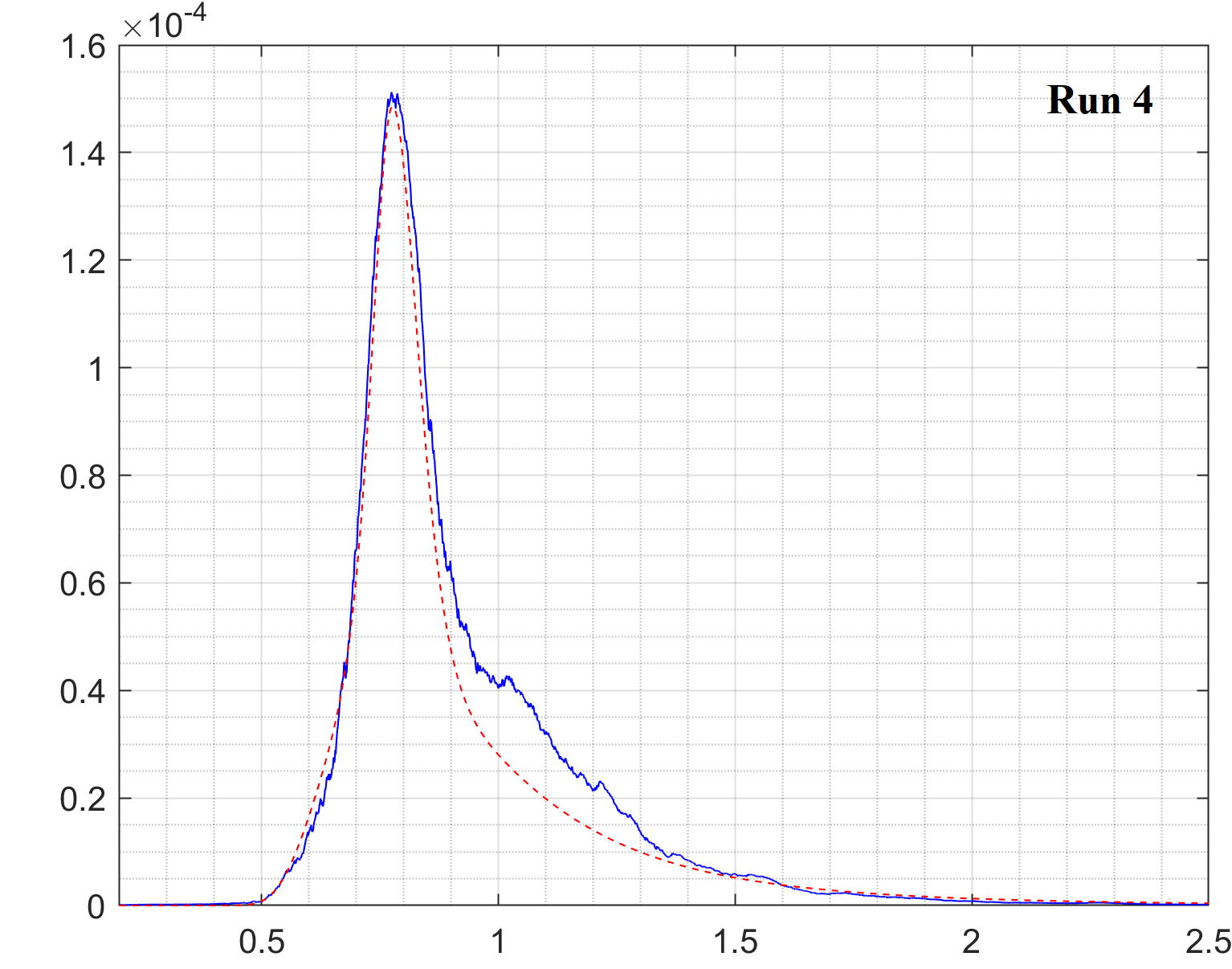}
\includegraphics[width=0.37\textwidth]{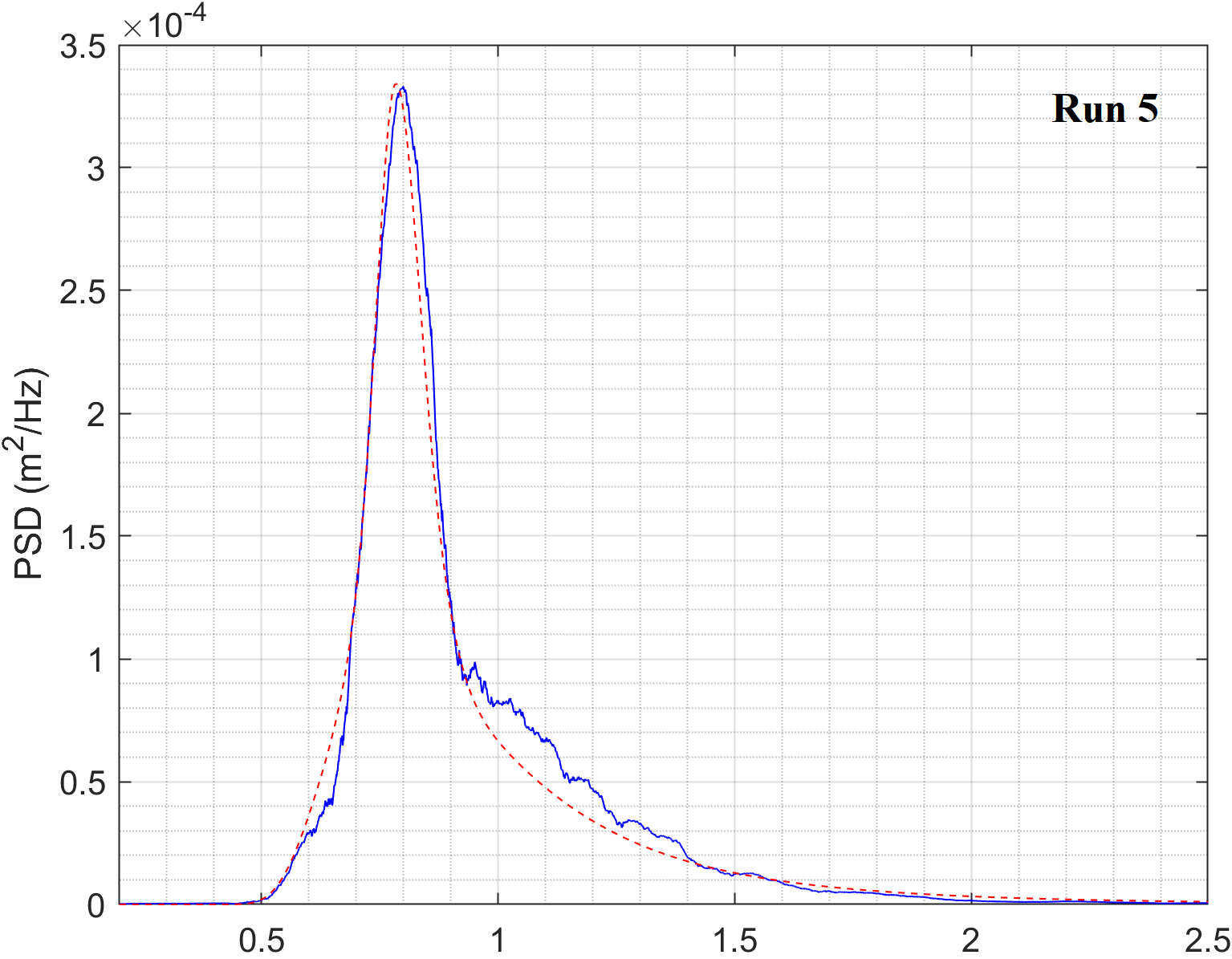}
\includegraphics[width=0.37\textwidth]{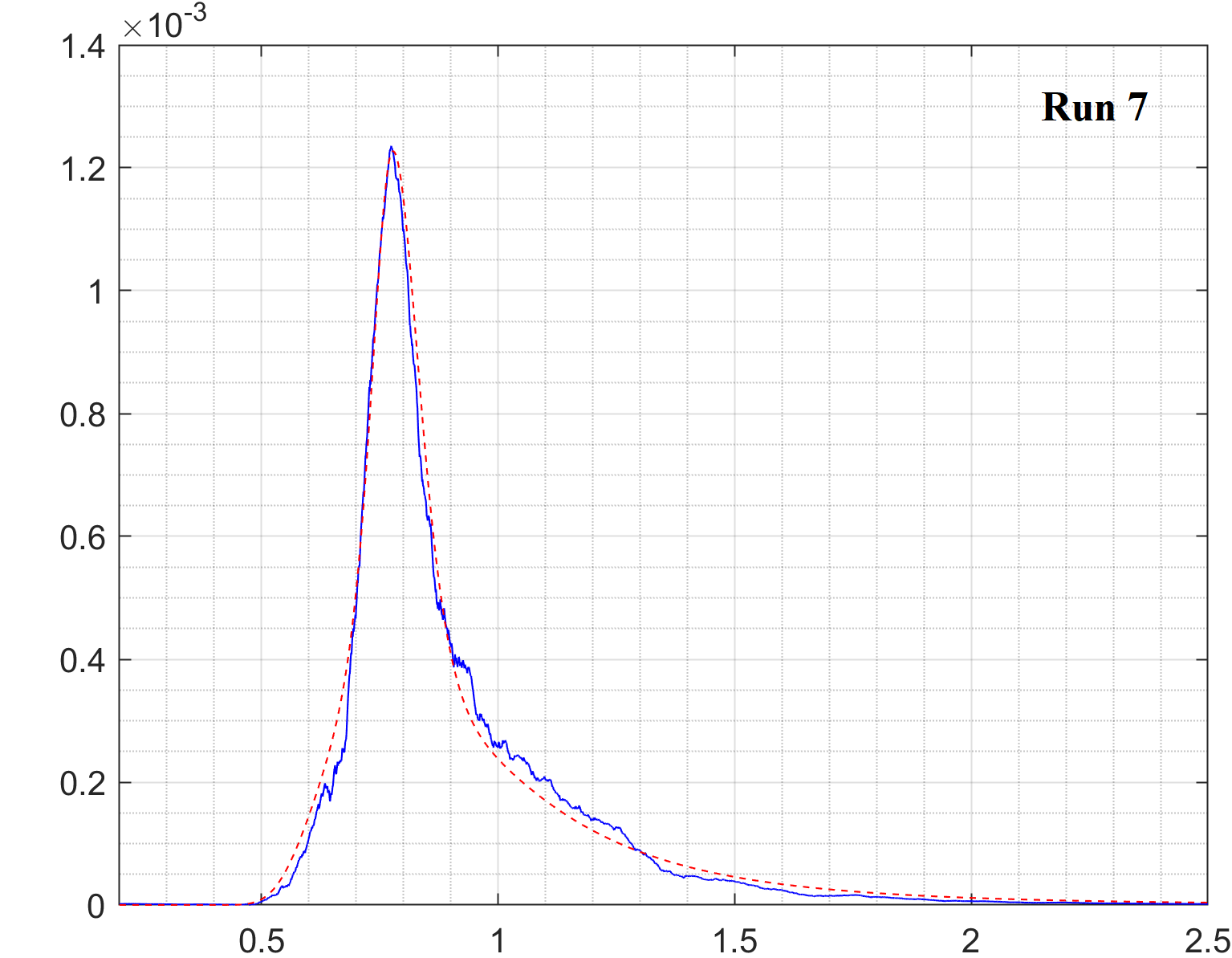}
\includegraphics[width=0.37\textwidth]{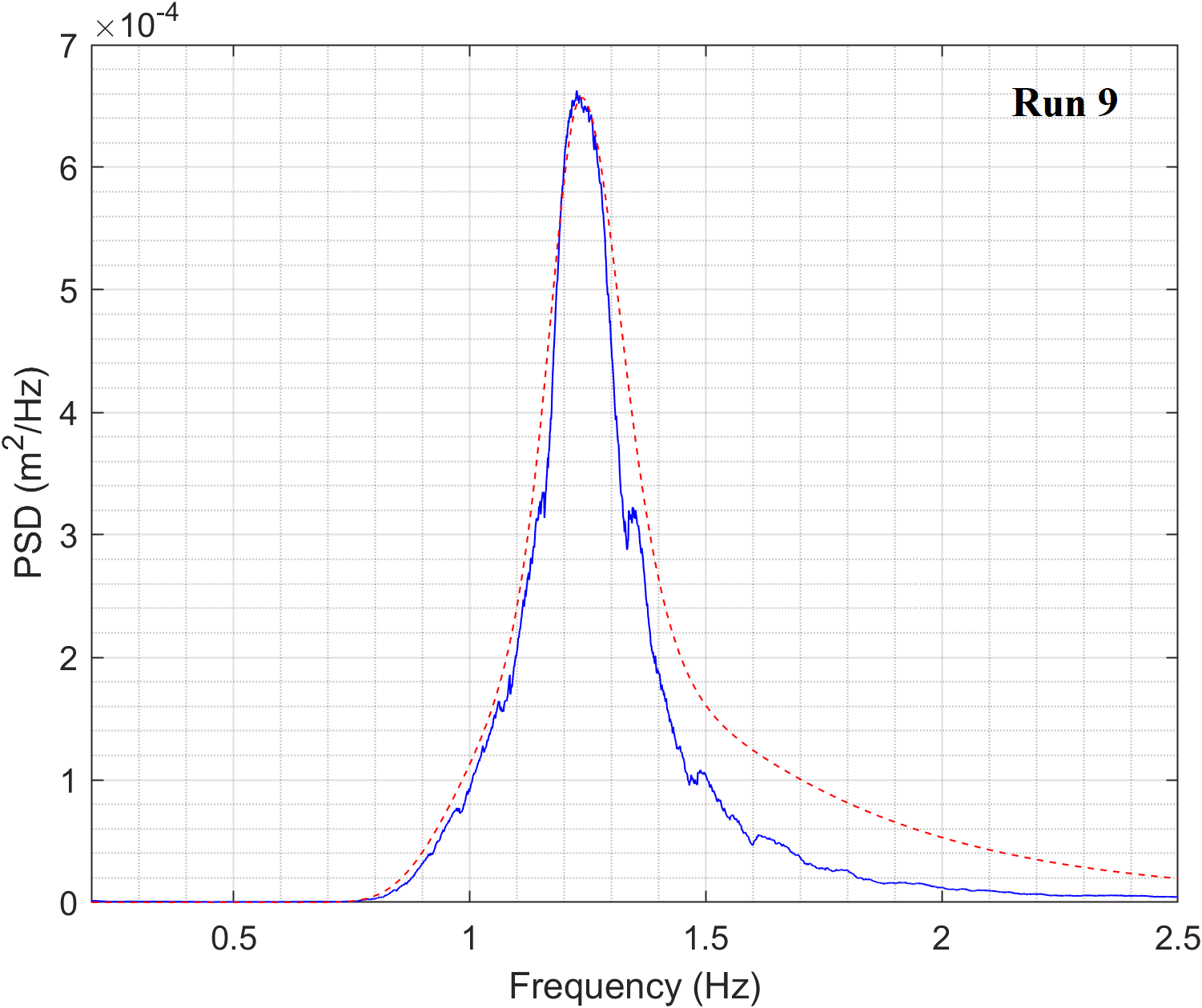}
\includegraphics[width=0.37\textwidth]{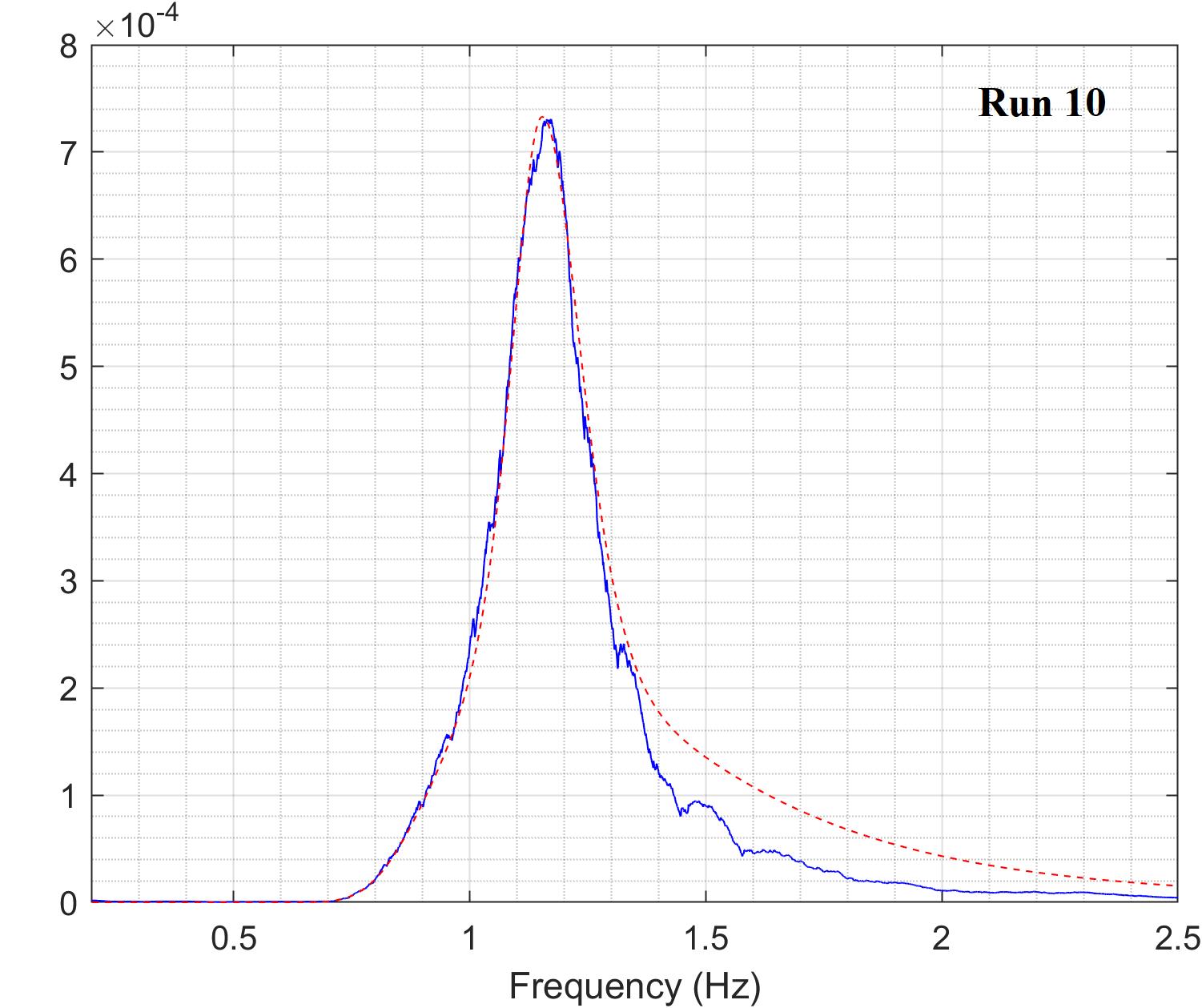}
    \caption{\justifying{Power density spectrum of the wave train just before the breakwater for runs 1, 4-5, 7, and 9-10 of \jfm{Table} \ref{tab:1} compared to the corresponding theoretical JONSWAP spectrum, see title of each plot.}} 
    \label{fig:input spectrum}
\end{figure*}
\begin{figure*}
\centering
\includegraphics[scale=0.18]{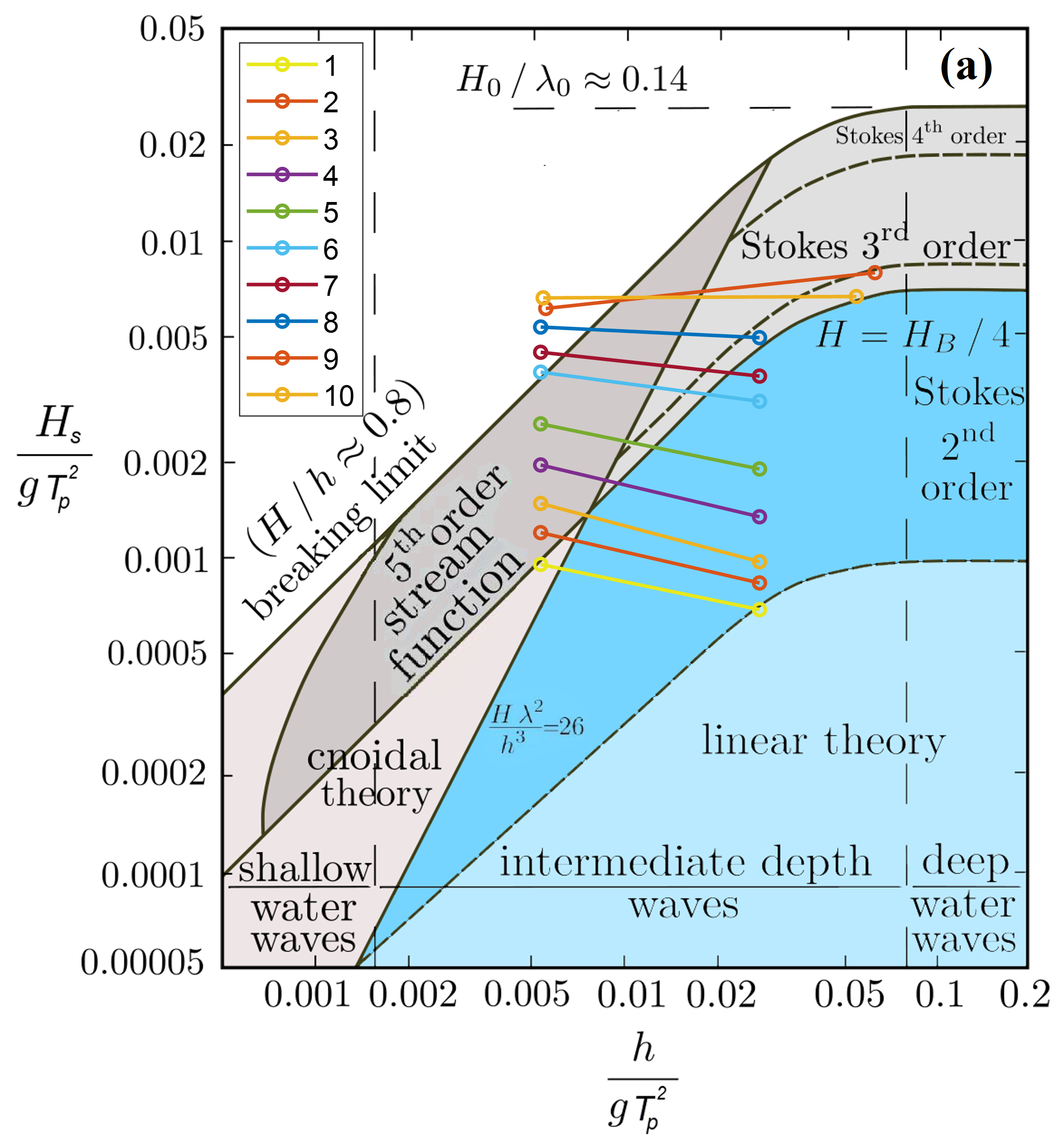}
\includegraphics[scale=0.18]{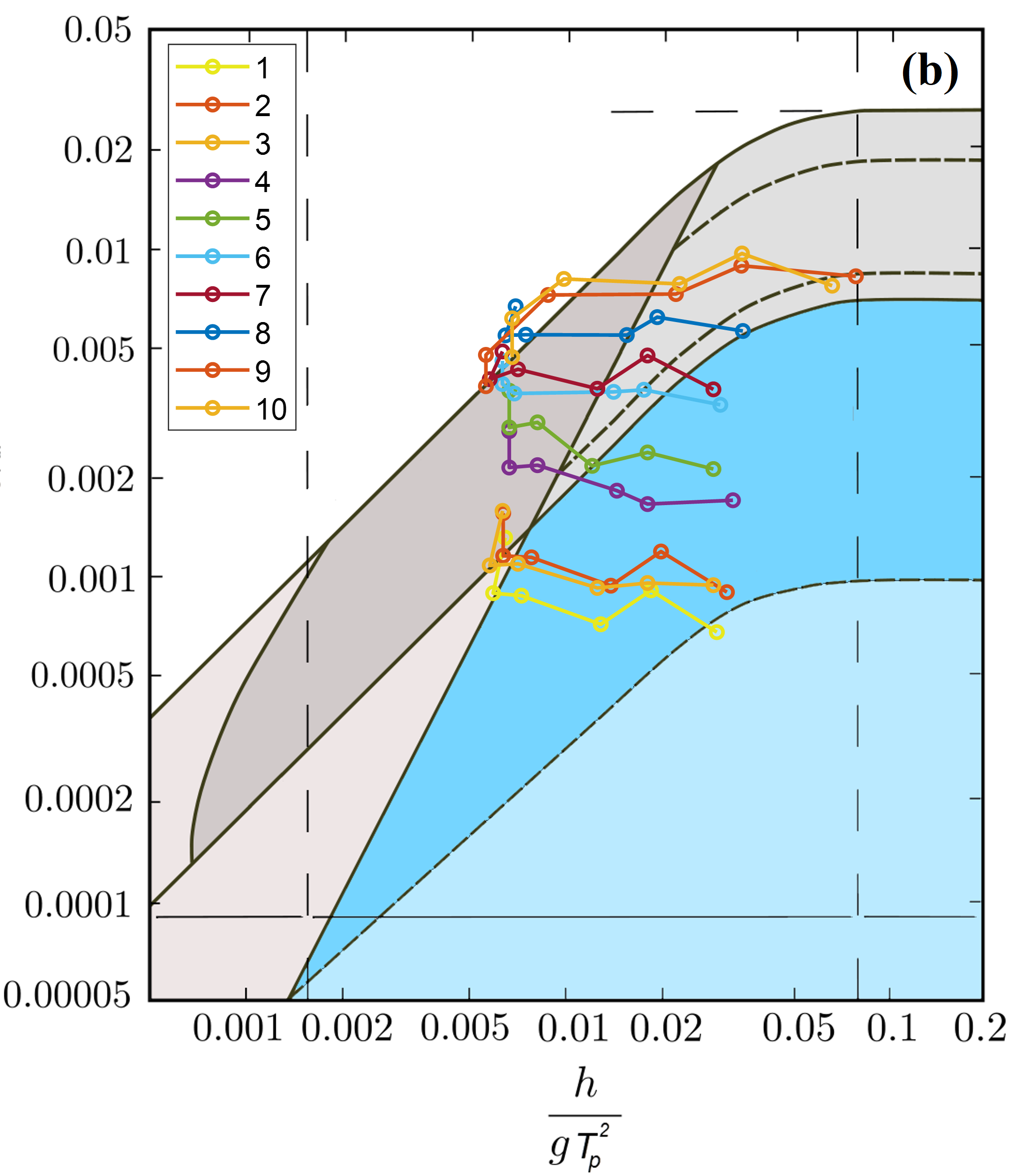}
  \caption{\justifying{Le Méhauté diagram with \pf{(a)} theoretical pre-shoal (rightmost point of the line) conditions and the wave train evolution \pf{predicted with linear shoaling} over the breakwater (left point) for the 10 cases defined in Table \ref{tab:1} and \pf{(b)} with experimental results recorded in the aforementioned 10 cases. Each curve represents the spatial evolution (from right to left) of wave height with respect to the water depth over the breakwater. Note that for cases 1--8, the curves are generated based on measurements from gauges 1--5 and 7, with the maximum wave height occurring at gauge 7, while for cases 9 and 10, the curves are limited to gauges 1--6 due to early wave breaking. Background diagram adapted from \citet{LeMehaute1976}.}}
  \label{fig:expfig2}
\end{figure*}
\begin{figure*}
    \centering
\includegraphics[width=0.31\textwidth]{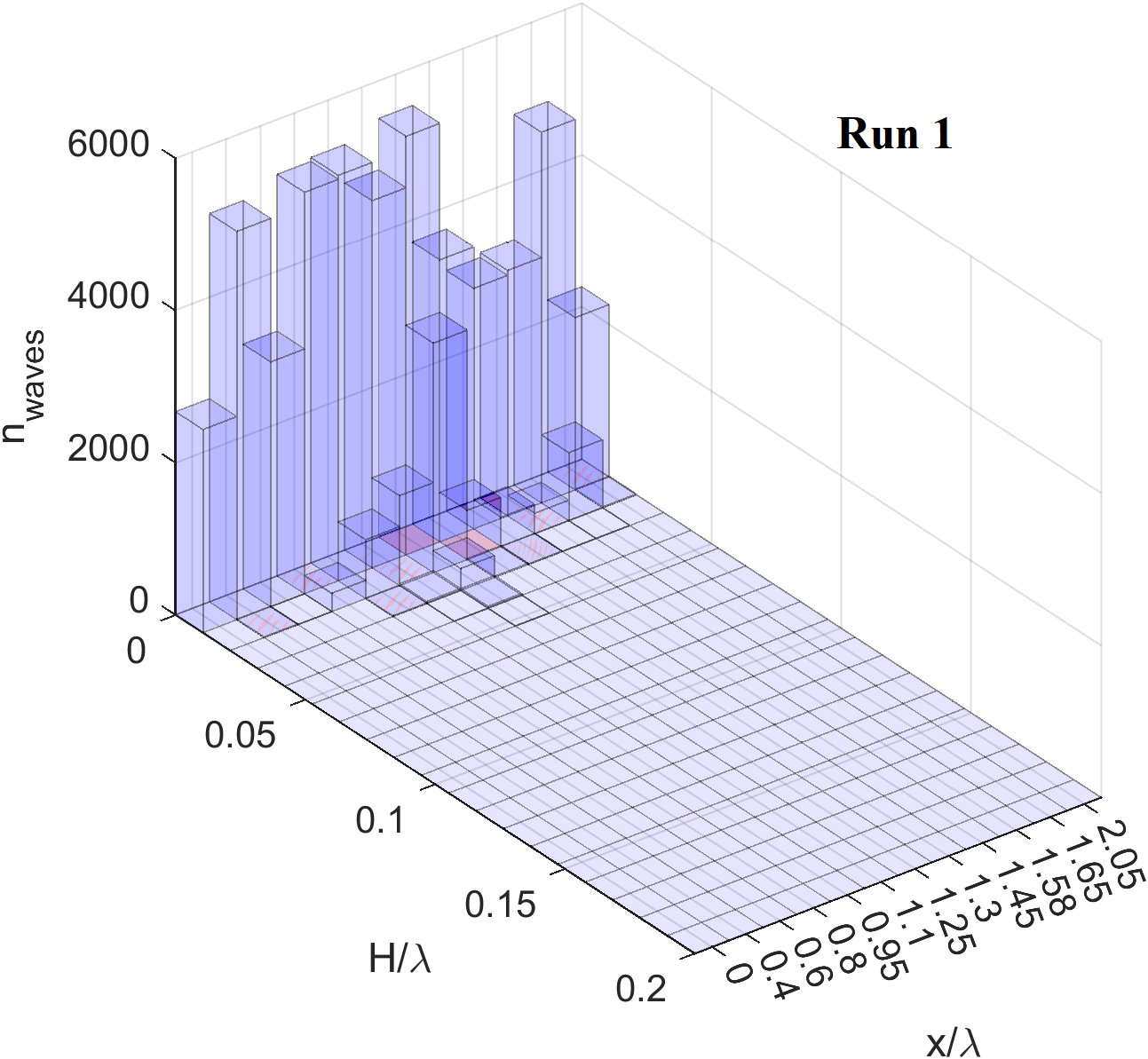}
\includegraphics[width=0.31\textwidth]{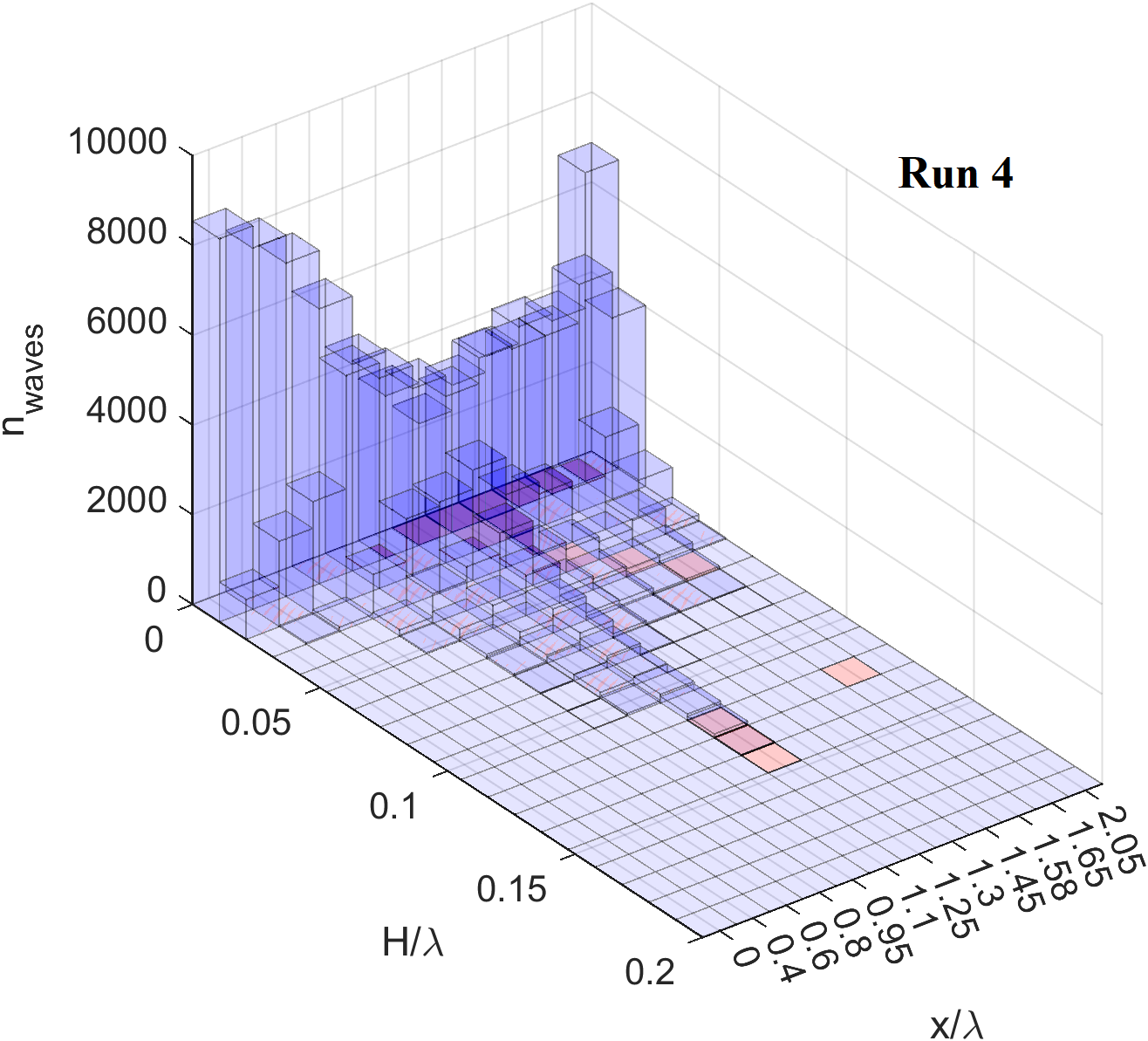}
\includegraphics[width=0.31\textwidth]{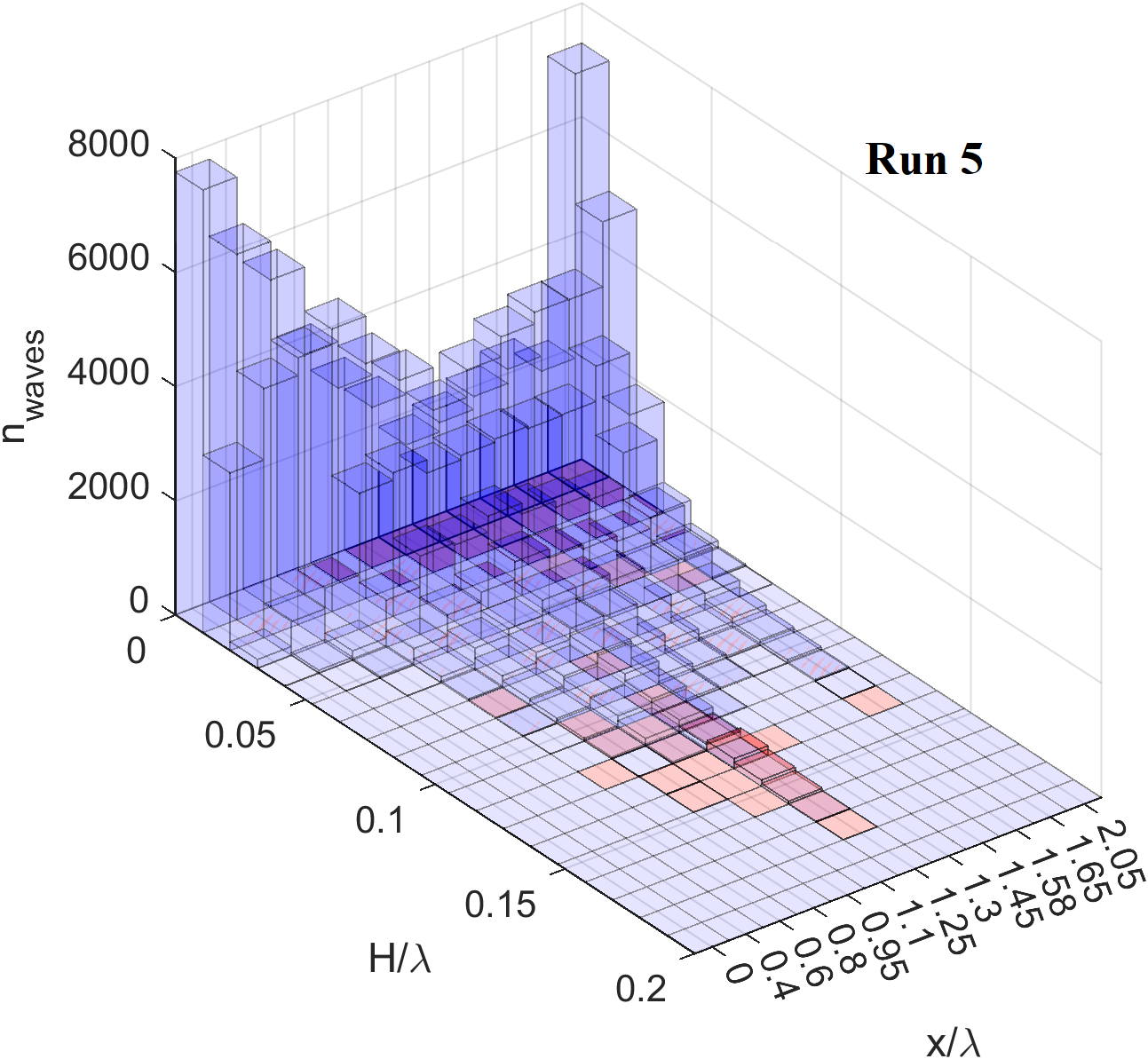}
\includegraphics[width=0.31\textwidth]{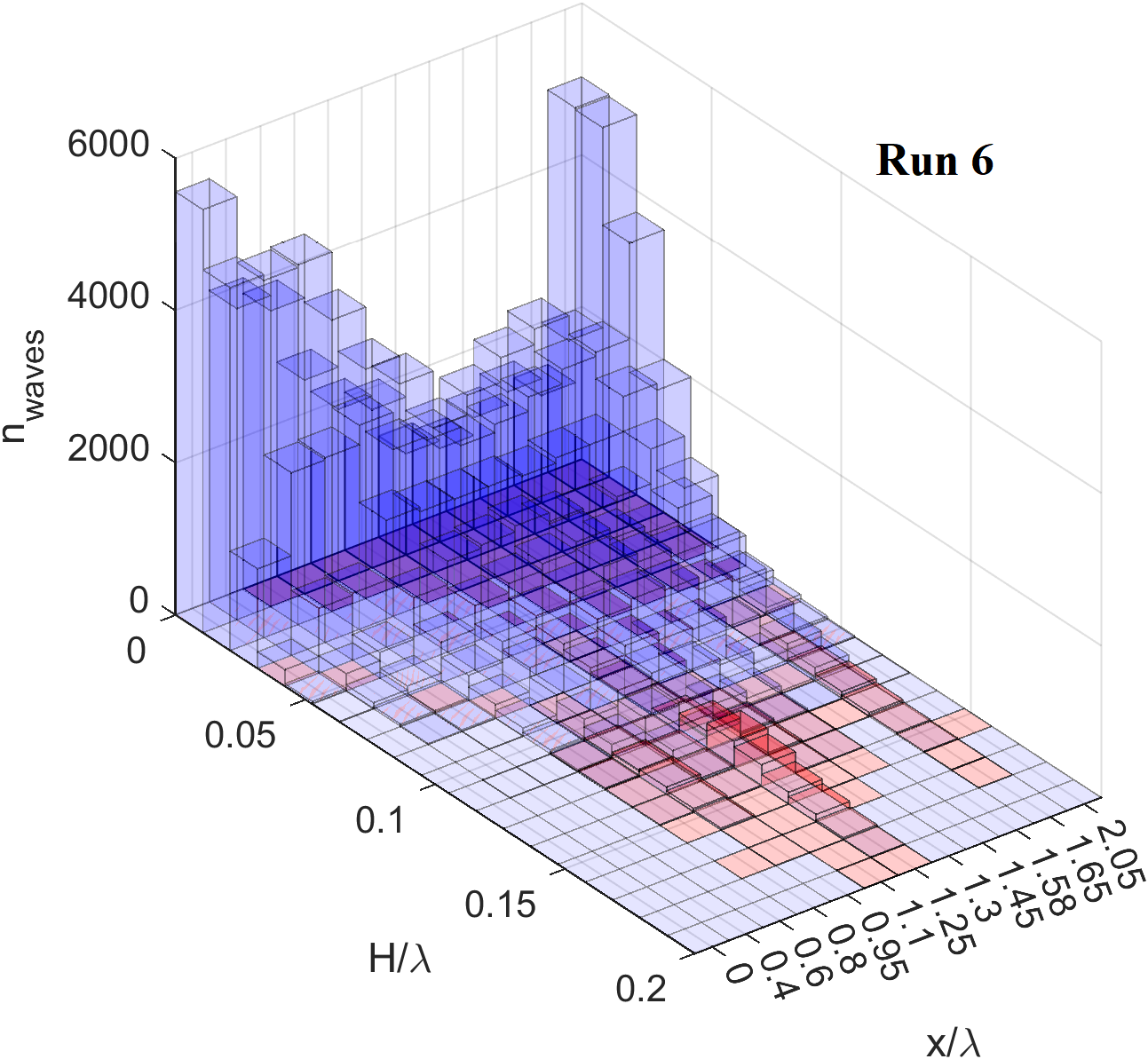}
\includegraphics[width=0.31\textwidth]{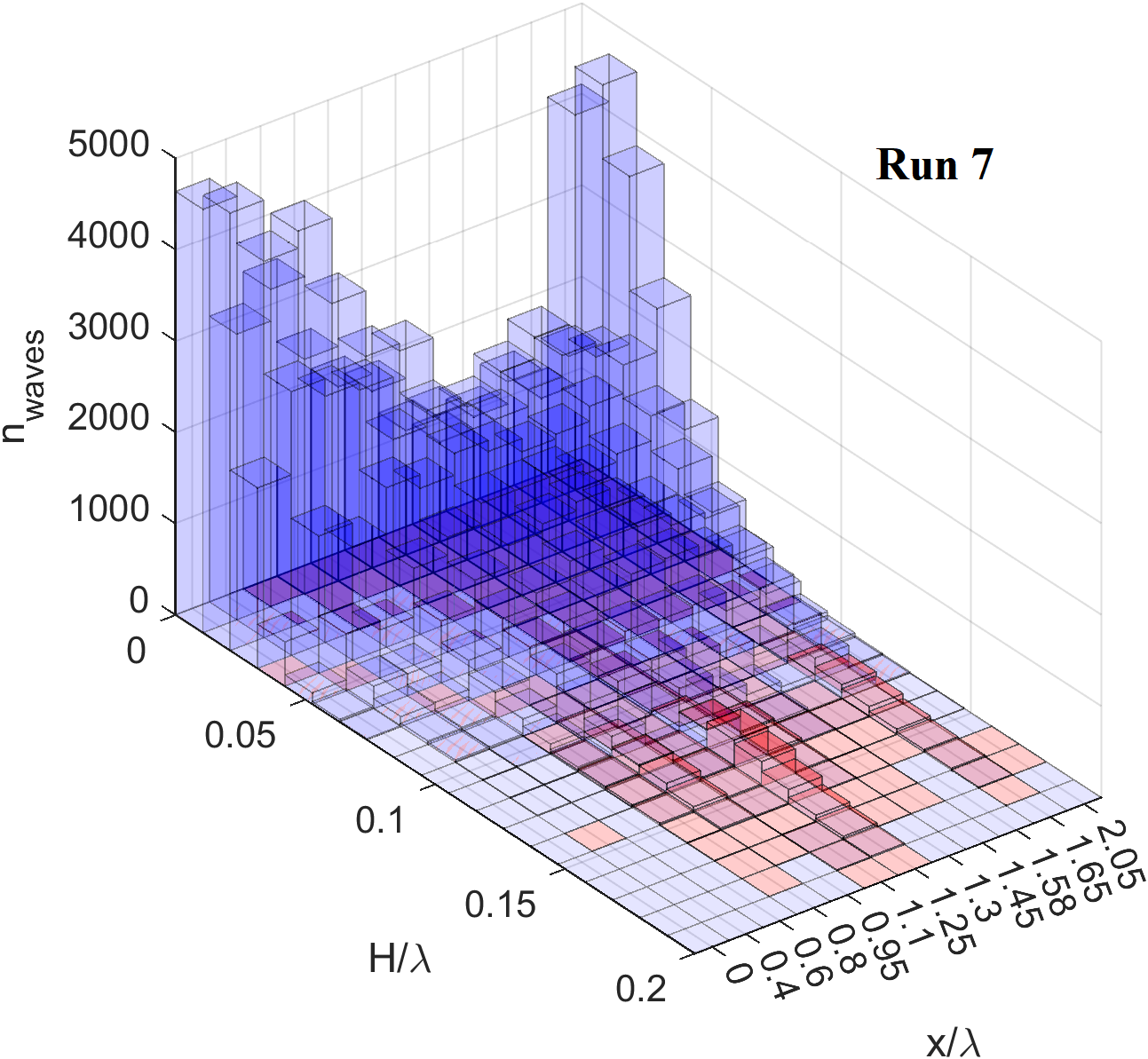}
\includegraphics[width=0.31\textwidth]{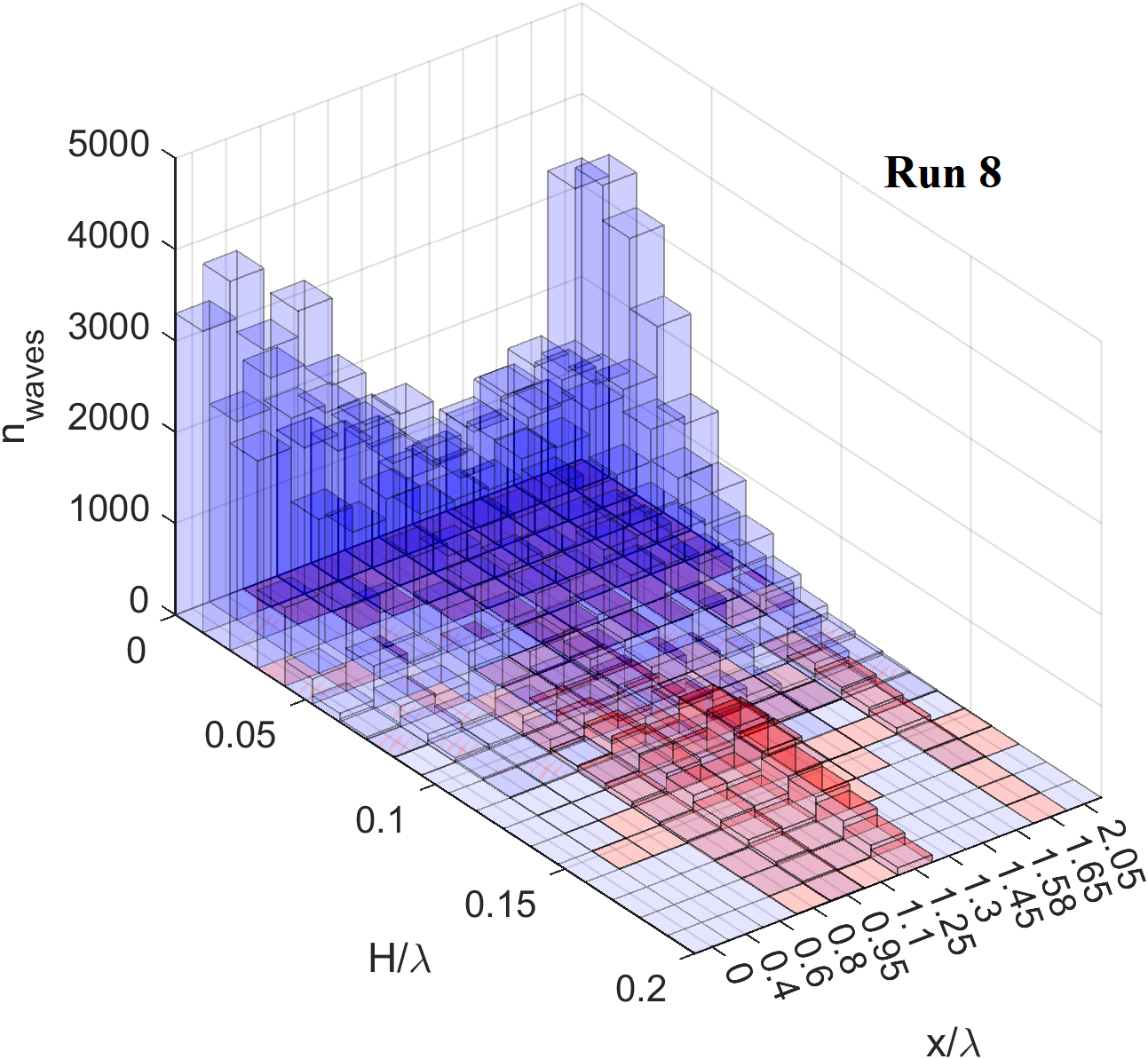}
\includegraphics[width=0.31\textwidth]{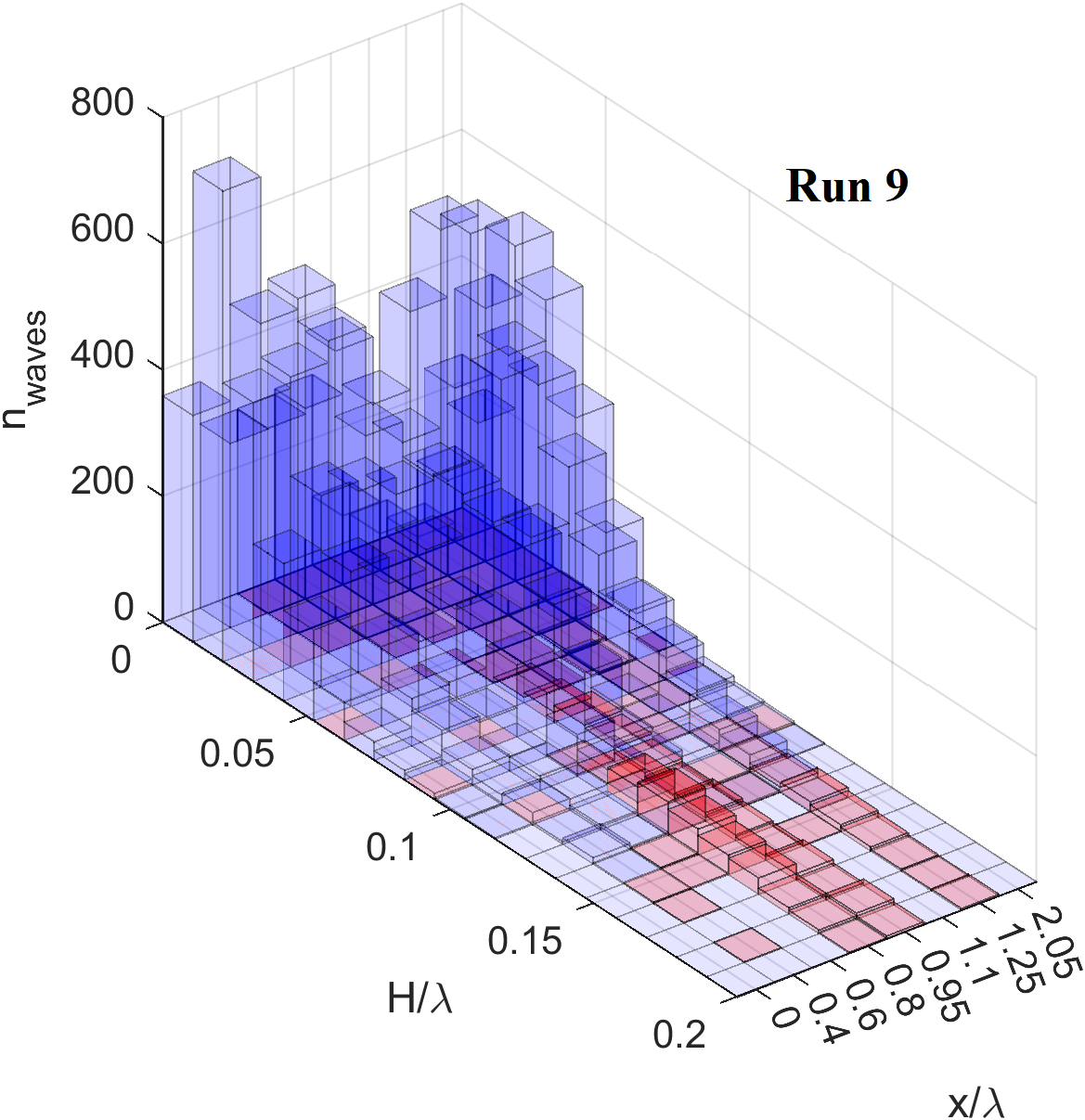}
\includegraphics[width=0.31\textwidth]{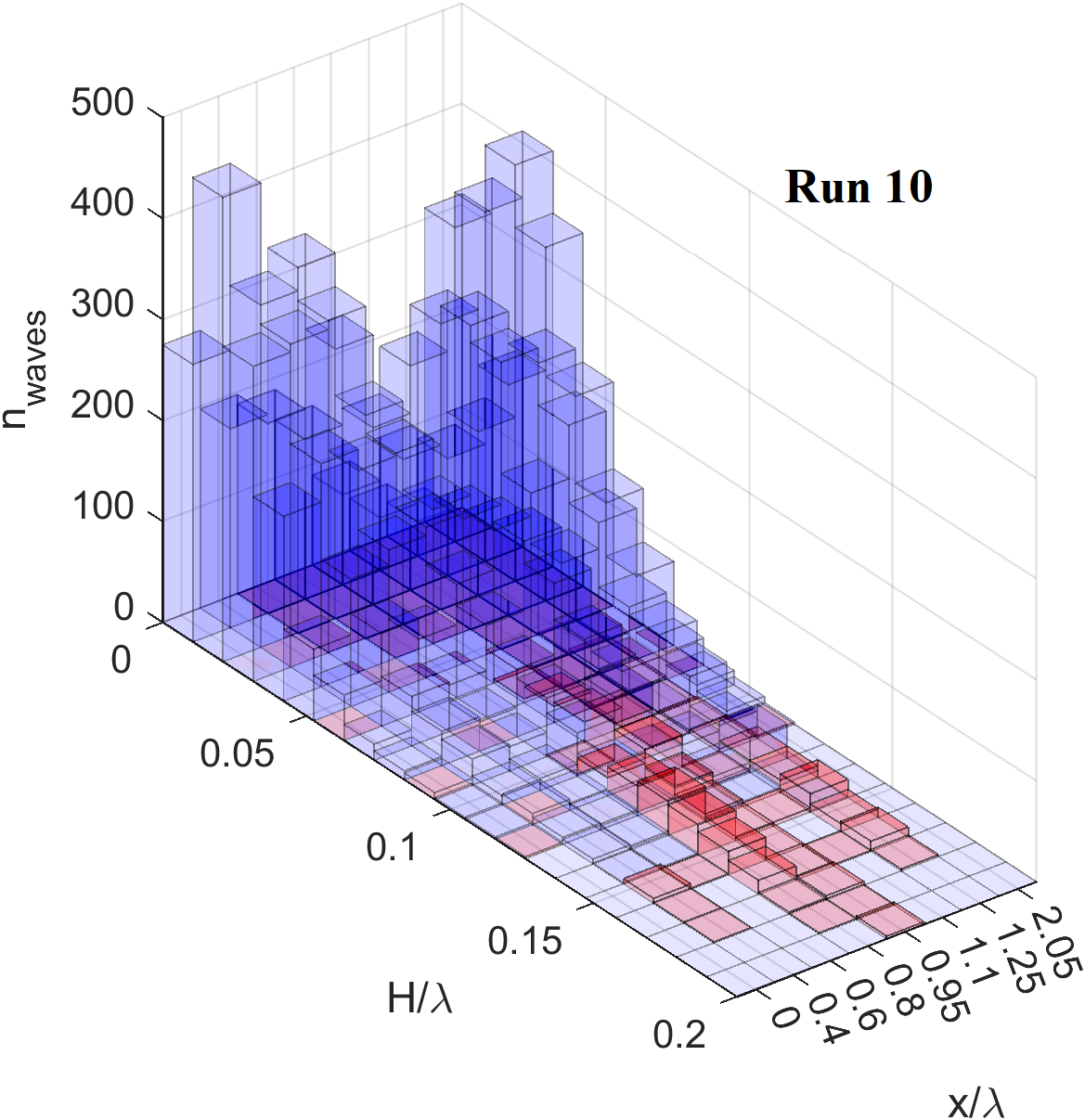}
    \caption{\pf{\justifying{Histogram of extreme waves as a function of wave steepness and location across the submerged breakwater. Blue bars depict non-breaking waves, whereas red ones highlight breaking waves according to \citeauthor{Miche1944}'s limit $H \leqslant 0.142 \lambda \tanh{kh}$.}}} 
    \label{fig:HistogramYuchen}
\end{figure*}

Laboratory experiments were carried out in a unidirectional wave flume installed at the University of Sydney (\jfm{Figure}~\ref{fig:expfig}\jfm{a,b}). 
The flume has dimensions of 30~m $\times$ 1~m $\times$ 1~m and the water depth $h$ can be varied from 0.2 to 0.8~m. A piston-type wavemaker can generate irregular waves with frequencies ranging from 0.4 to 2~Hz. \jk{We generate irregular waves satisfying a JONSWAP spectrum, parametrized by the peakedness parameter} $\gamma = 3.3$.
\yh{The power spectra of the input and generated wave train are shown in \jfm{Figure} \ref{fig:input spectrum} for four selected cases (1, 4, 7, and 9 in \jfm{Table}~\ref{tab:1}). They display a good agreement, confirming the reliability of the current wavemaker \ac{and accuracy of wave generation}}. \pf{The location of each wave gauge in its two different configurations are detailed in \jfm{Table} \ref{tab:2}.}

\begin{figure*}
    \centering
\includegraphics[width=0.45\textwidth]{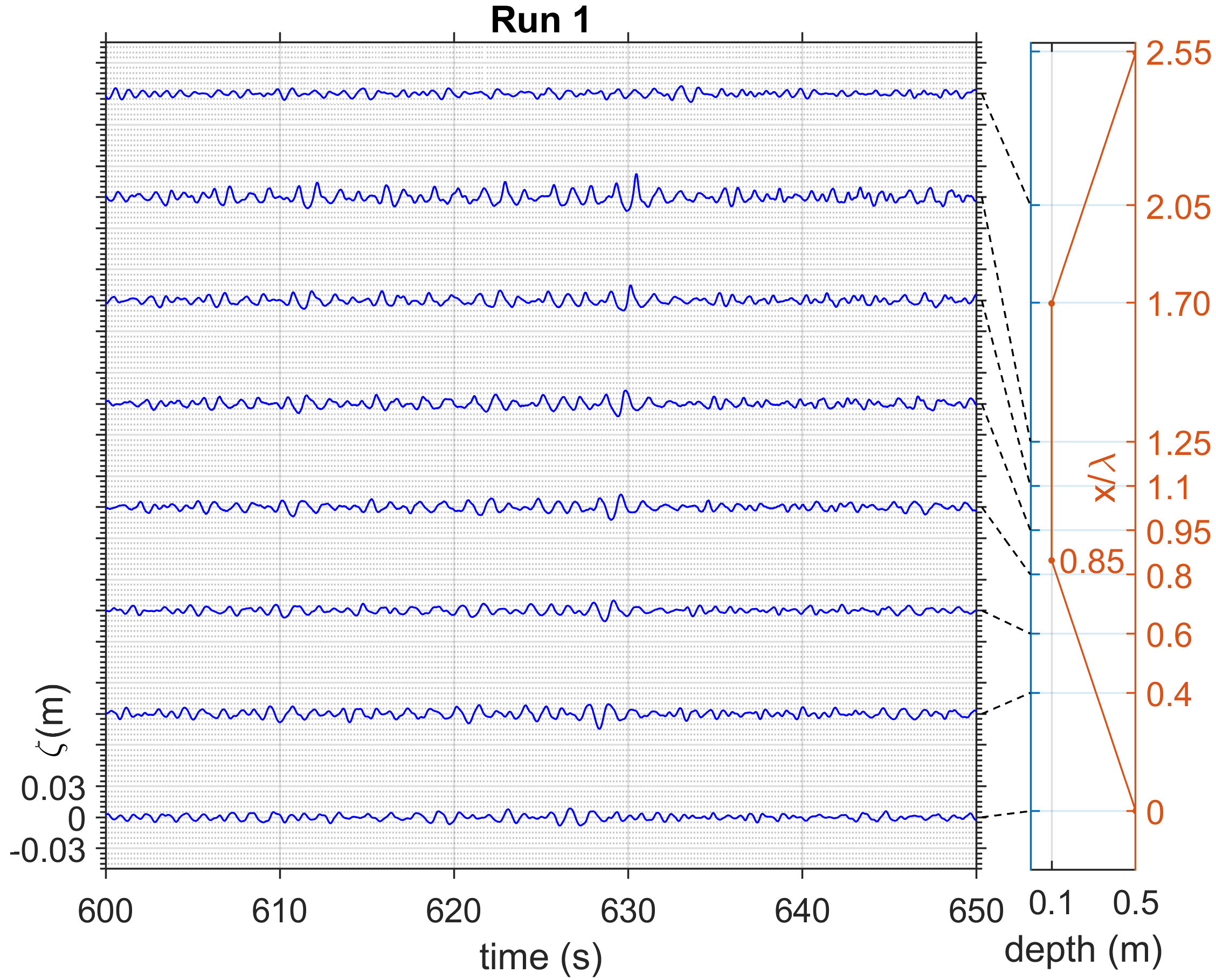}
\includegraphics[width=0.45\textwidth]{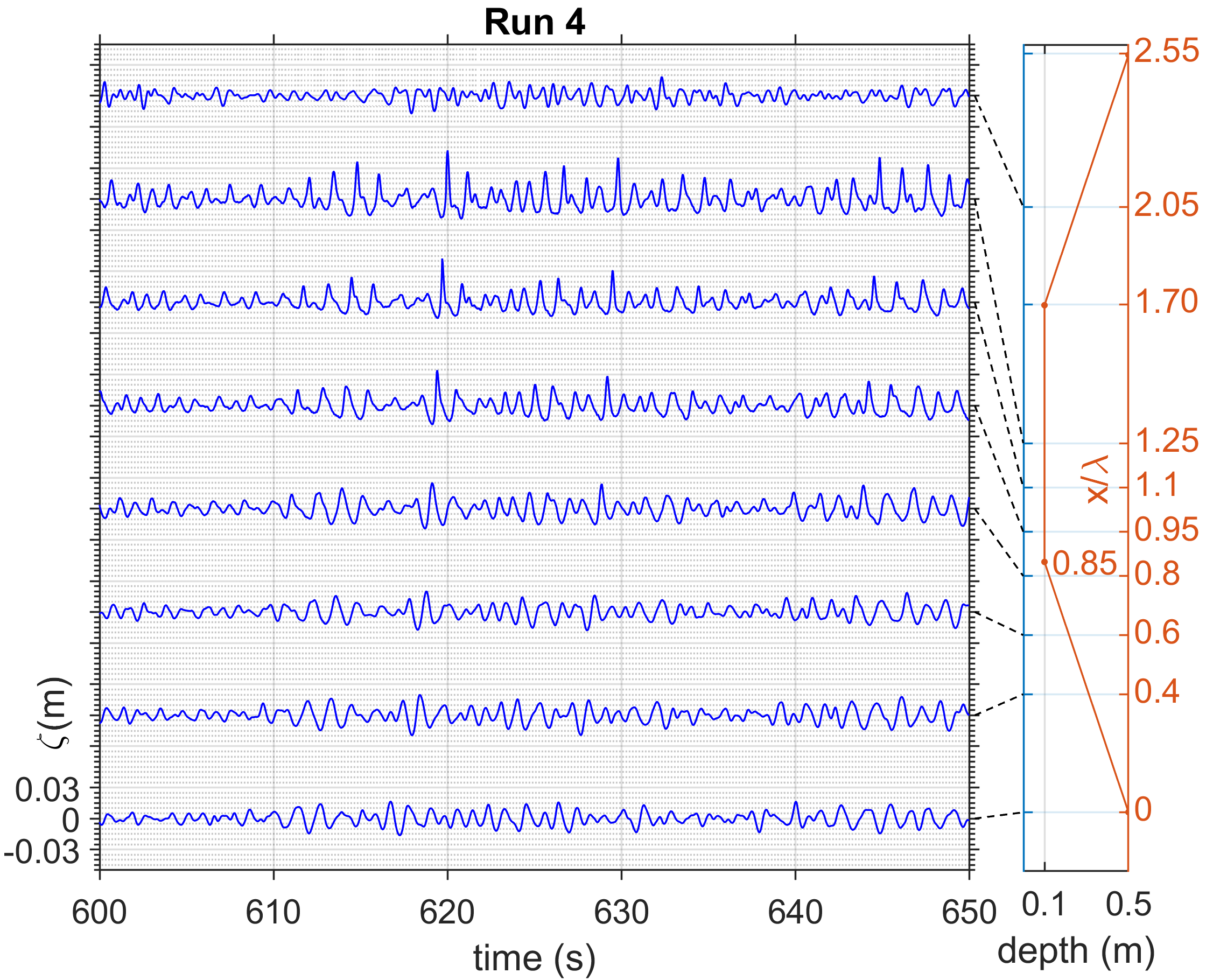}
\includegraphics[width=0.45\textwidth]{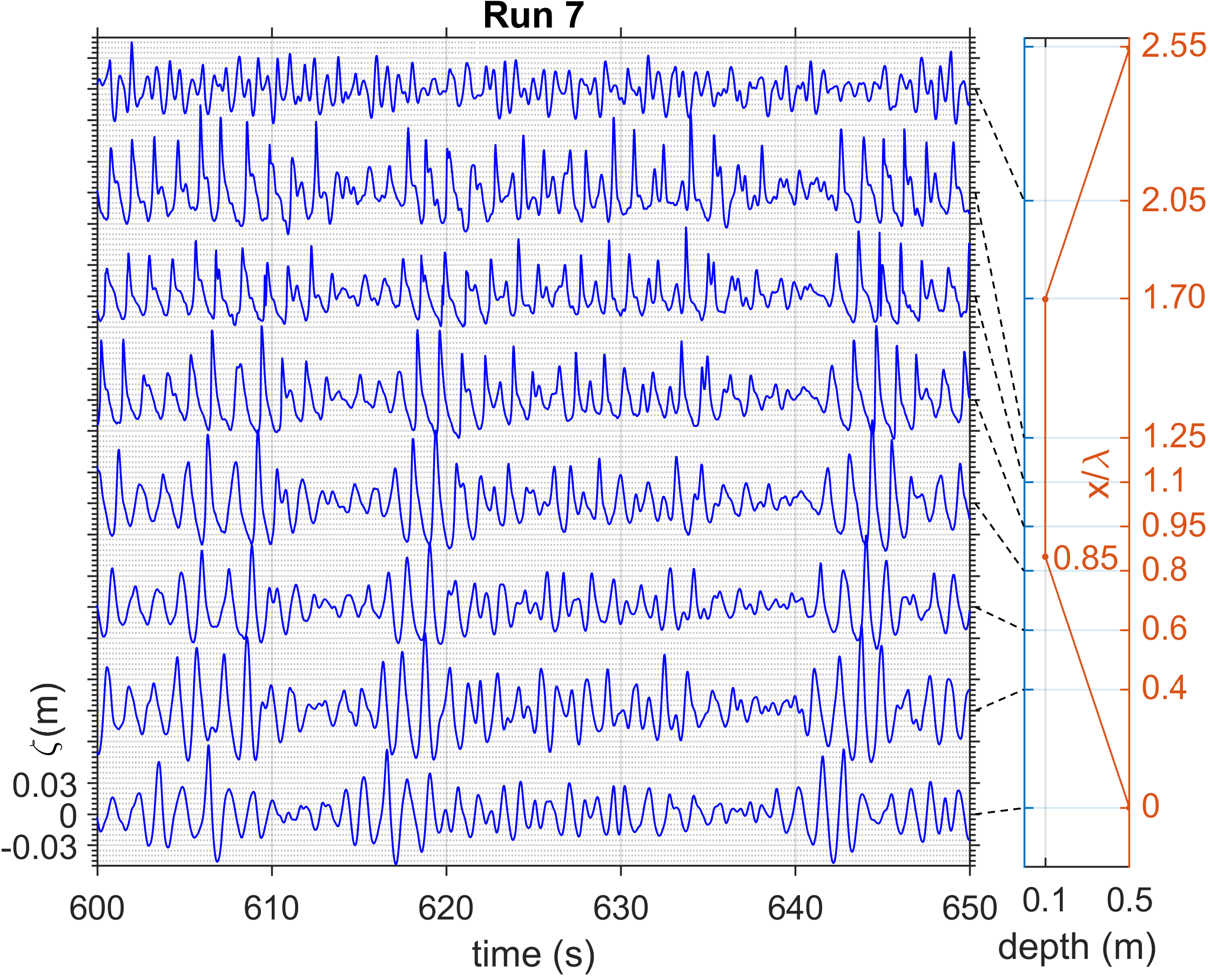}
\includegraphics[width=0.45\textwidth]{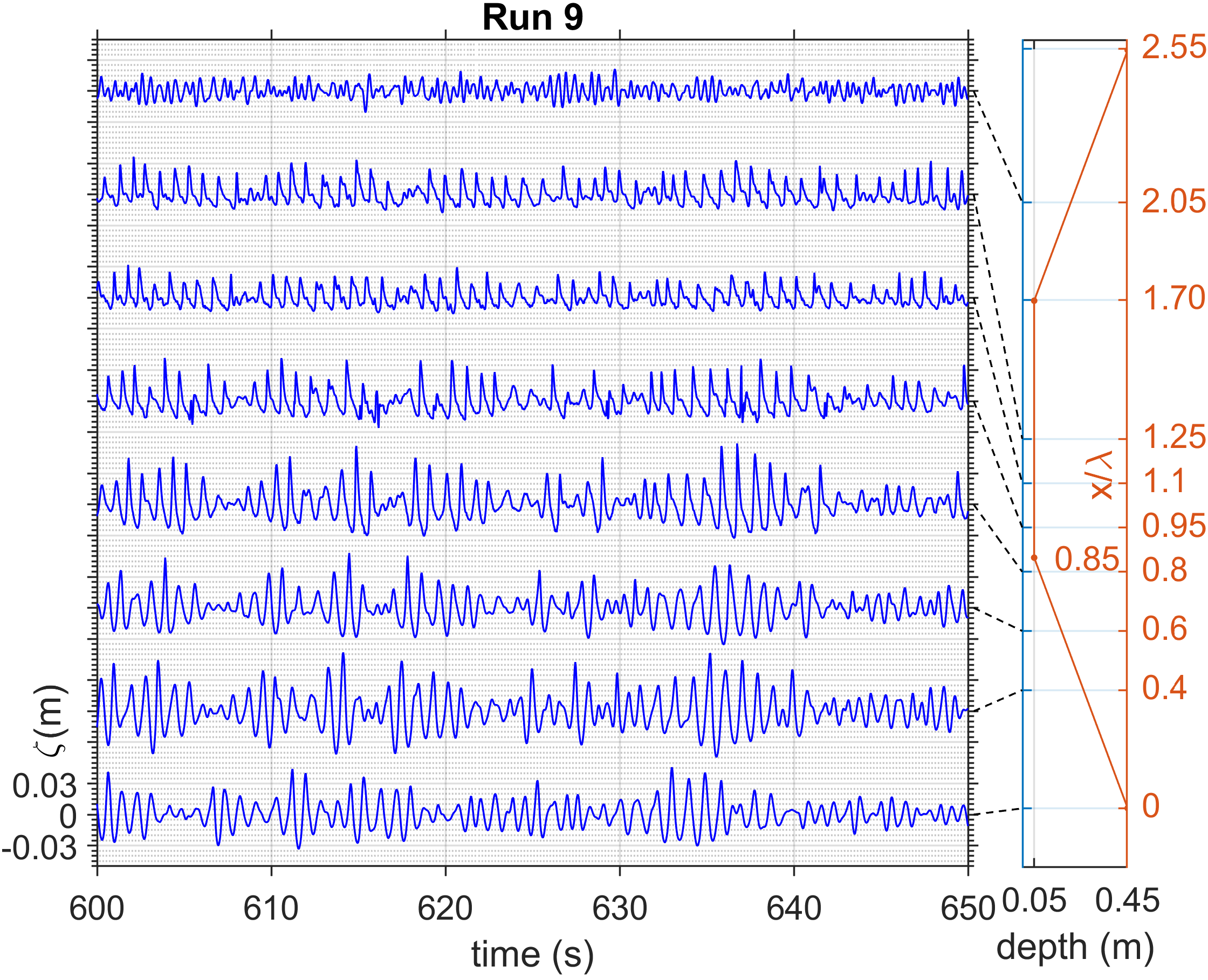}
    \caption{\justifying{Spatial evolution \pf{(normalized by the offshore peak wavelength $\lambda_{p1}$)} of the wave elevation time series \pf{$\zeta $ (m)} over the breakwater for runs 1, 4, 7, and 9 of Table \ref{tab:1}, see title of each plot. The affiliating columns on the right of each plot indicate the gauge locations relative to the breakwater and the corresponding water depths (see \jfm{Figure} \ref{fig:expfig}\jfm{b}).}} 
    \label{fig:sawtooth}
\end{figure*}
To mitigate reflection towards the region of the tank where the measurement is performed, an $8^\circ$-inclined beach covered with artificial grass is installed at the end of the flume. \pf{Calculating the wave reflection rate requires separating incident and reflected waves using methods such as linear separation with closely spaced wave gauges \citep{goda1976estimation}, which demands additional extensive measurements. The current study focuses on wave evolution over the slope, but reflection accumulates and varies spatially along the bar, making it very difficult to determine location-specific reflection coefficients. Moreover, the slope of the breakwater $1/5 \, (\approx 11.3^{\circ})$ is similar to experiments performed in the very same wave flume by \citet{he2022probability}, which showed that for $8^{\circ}$ the reflection rate was 18\%, having a small impact on extreme wave statistics. Indeed, using \citet{Battjes1974} model for reflection $K_R \approx 0.1 \xi^2 = 0.1(\nabla h)^2/\varepsilon \lesssim 0.1 \times (0.2)^2/0.03 \lesssim 13\%$.} \jk{A submerged accelerometer was installed at the surface of the breakwater, and we observed negligible vibration of the structure}. To analyse the statistical moments of out-of-equilibrium sea conditions induced by \jk{the} submerged breakwater with identical up and down slopes of $|\sm{\nabla} h| = 1/5$, we record the temporal variation of surface elevation $\zeta (t)$ prior to, over, and after the breakwater. 

The surface elevation \ac{time-series is computed with a sampling frequency of 100~Hz}, \ac{and} the corresponding control sequence of the wave paddle is generated through the Wave Synthesiser software \citep{Chabchoub2023b}. In order to cover a wide range of propagation distances with a reasonable longitudinal resolution in spite of a limited number of available wave gauges, each run is successively performed following two gauge configurations: the first has seven wave gauges evenly spaced between the start of the up-slope and the mid-point of the breakwater plateau, whereas the second configuration has the last three gauges moved to the back end of this plateau (see the locations in \jfm{Figure} \ref{fig:expfig}\jfm{b}). \sm{While several studies analyzed wave properties over mild slopes or a step under wave breaking conditions \citep{Draycott2022}, our experiments \ac{focus on extreme wave} statistics at \jk{the} relatively steep slope of a symmetrical breakwater.}

\subsection{Results of Water Surface Data Acquisition}

As shown in \jfm{Table} \ref{tab:1}, Run \pf{2} of the present experiments has similar values of both initial and final wave steepness $\epsilon = k_pH_s$ and relative water depth $k_p h$ \ac{to} run T, and as such the spatial evolution of the excess kurtosis $\mu_4$ of \jk{run} 3 is also similar 
to \jk{run} T. This strong amplification contrasts with the stability or drop in the exceedance probability, observed on flat bottoms in similar conditions \citep{Xu2021, Karmpadakis2022}. The experimental parameters, in particular the wave steepness and relative water depth, have been determined to cover regimes ranging from linear waves to breaking-dominated irregular seas. A summary of all runs (\jfm{Table}~\ref{tab:1}) can be \ac{qualitatively} assessed with regard to the physical regimes at play through the \citeauthor{LeMehaute1976}'s diagram (see \jfm{Figure} \ref{fig:expfig2}\jfm{a}), in which the water wave theories corresponding to different dimensionless water depth and significant wave height regimes are summarized. In this diagram, the initial and predicted final state of the experimental runs are shown by different line sections, evolving from right to left. Under most conditions, the significant wave height is expected to increase when the water depth decreases. \ac{The condition} $H_s/h \ge 0.8 $ delineates the wave breaking regime~\citep{Holthuijsen2007,Hallowell2015}. Visually,  a small percentage of waves started to display plunging breakers atop the breakwater as early as for the conditions of run 5 ($H_{s2} / h_{2}  \sim 0.5$), and spilling breakers were observed from run 4 onward, with $ H_{s2} / h_{2}  \sim 0.4$. \pf{This apparent initialization of wave breaking is corroborated by the statistical distribution of wave steepness varying spatially, see \jfm{Figure} \ref{fig:HistogramYuchen}}. Conditions \ac{corresponding to} run T were included to allow a direct comparison with previous experiments \citep{Trulsen2020}. Each run was repeated for up to 12 realizations featuring 2500 waves each to ensure reproducibility and to compute the standard deviation of the excess kurtosis for each case. The adopted values follow the experimental requirements for the convergence of excess kurtosis \citep{Toffoli2017}.

To verify the \ac{intensity} of wave breaking in each run\ac{, as} described in \jfm{Table} \ref{tab:1}, we have plotted the evolution of the wave train \jk{passing the shoal} overlaid on the Le Méhauté diagram (\jfm{Figure} \ref{fig:expfig2}\jfm{b}). \jk{The cases can be gathered in} four groups (1-3, 4-7, and 8-10) \jk{with} similar steepness and relative water depth\jk{, therefore, similar initial energy spectrum and evolution.}

When the initial wave steepness is relatively low, the JONSWAP spectral shape is preserved ($\gamma = 3.3$) for all gauges along the up-slope and the first \jk{one} atop the shoal (runs 1-\jk{6}). \jk{Then, higher frequencies develop and the spectrum broadens}. \pf{This frequency shift was designed, see \jfm{Table} \ref{tab:1}.} For larger initial wave steepness (run \jk{7}), the spectrum broadens already at the second wave gauge in the up-slope ($x = 0.6 \lambda $). When most waves are breaking atop the breakwater (runs 8-10, $x \gtrsim \lambda$), the spectrum is essentially flat except for a minor peak about $f_p \sim 1$~Hz, having shifted in about $\sim 20\%$ in comparison with runs of mild steepness such as runs 2 and 3. In addition, the spatio-temporal evolution of run 9 (\jfm{Figure}~\ref{fig:sawtooth}, bottom right panel) displays the vertical asymmetry and sawtooth surface elevation characteristic of the surf zone \citep{Bonneton2023}, confirming the occurrence of wave breaking. This wave shape after gauge 4, i.e. $x / \lambda = 0.8$, is observed in all cases with steepness higher than or equal to run 5. 

\begin{figure*}
\centering
   \includegraphics[scale=0.67]{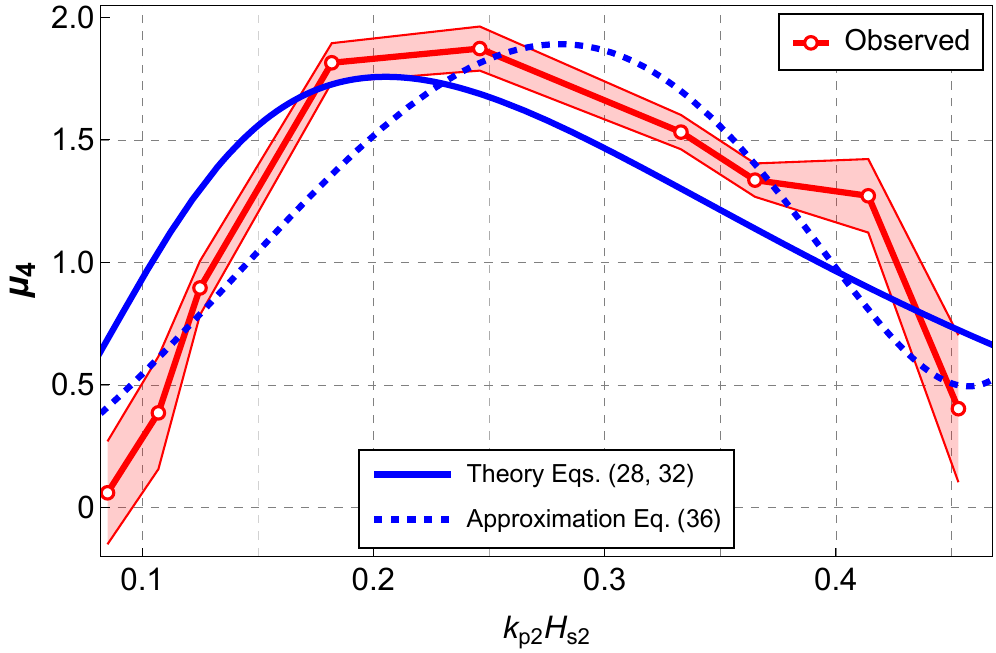}
       \caption{\justifying{Comparison between experimental \jk{(red, \pf{see runs in \jfm{Table} \ref{tab:1}})}, (derived, \jk{blue}) theoretical \pf{(eqs.~(\ref{eq:kurt},\ref{eq:finiteampgamma2}))}, and Taylor-expanded approximated \pf{(eqs.~(\ref{eq:kurtApp2})} maximum excess kurtosis atop the shoal. 
       The shaded areas correspond to the 95\% confidence
intervals.}}
  \label{fig:FINAL}
\end{figure*}
Note that we also observed a broadening of the wave spectrum \jk{before the end of} the shoal due to the strong steepening of the waves. The delay in the narrowing of the spectrum can be explained by the decomposition of the wave field into cnoidal waves in the shallow area \citep{zabusky1965interaction,trillo2016experimental}. \sm{Also, \ac{since wave} breaking is \ac{likely to introduce a subtle directionality to the waves} as it mimics the surf zone \citep{Szczyrba2023}, we verified that transverse effects were not significant\jk{. For that purpose, we compared} measurements at two gauges with the same $x$ as well as different, non-symmetrical, \ac{and} transverse positions.}

\subsection{Extreme Wave Statistics}

\ac{The Rayleigh distribution serves as a statistical model for water wave envelopes and, to a good approximation, for wave heights as well} \citep{Higgins1952}. The rationale involved in this early work, which \ac{relates to} the electromagnetic theory of signals \citep{Rice1944}, is most effective \ac{to describe narrowband and Gaussian} ocean \ac{wave} spectra \citep{Boccotti1983}. \sm{Although \citet{Higgins1963} has associated Gram-Charlier/Edgeworth series with nonlinearities in the wave field, the latter formulae are rather an approximation for linear superposition in \jk{the case of a number of waves insufficient to reach} the Gaussian limit \citep{Rice1944,BMendes2020}}. Even in this case of linear superposition, asymptotic expansions of the normal law are needed \citep{Chebyshev1887,Cramer1972}. Indeed, \citet{Higgins1963} has shown that \citet{Edgeworth1906} series can be applied, and the analytical expression for the surface elevation probability density can be described as a function of the excess in kurtosis \jk{of the surface elevation}. Partly due to the complexity of these types of distributions, the excess kurtosis \jk{of the surface elevation} is often used as a proxy to the evaluation of statistical distributions \citep{Marthinsen1992,Mori2002b,Janssen2006a}. \jk{F}or a \jk{zero-mean} random variable $X$\jk{, the excess kurtosis is defined as:}
\begin{equation}
\label{eq:skew_kurt}
    \mu_4(X) = \frac{\mathbb{E}\left[ X^4 \right]}{\left(\mathbb{E}\left[ X^2 \right]\right)^{2}} - 3 \quad,
\end{equation}
where $\mathbb{E}[X]$ denotes the expectancy of $X$. In the context of shoaling small-amplitude waves, due to bound harmonics, the spatial evolution of kurtosis and \jk{its} growth is well understood and documented \citep{Benoit2021,Mendes2021b,Adcock2021c,Benoit2023}.

In \jfm{Figure}~\ref{fig:FINAL} we observe the maximum excess kurtosis atop the breakwater as a function of wave steepness. We find a steep rise of the peak in excess kurtosis \ac{with the increase} of wave steepness up to the point \ac{of wave breaking initialization} $(k_{p2} H_{s2} \sim 0.25)$, followed by a decay of \sm{15--50\%}. Our results contrast with the aforementioned flat bottom and mild slope results \citep{Xu2021,Karmpadakis2022}, which observe a five to six times smaller peak in excess kurtosis. \jk{They even observed} negative excess kurtosis when wave breaking dominates. The likely source for this different behaviour is the milder slope in \citet{Xu2021} and \citet{Karmpadakis2022} than in our experiments. Moreover, the water depth in our experiment was not sufficiently shallow to drop the excess kurtosis below zero\jk{, as expected by the summary of theory in \jfm{Figure} \ref{fig:GammaHs}}. 

Indeed, the breaking threshold of steep Stokes waves dominated by modulation instability may be larger than predicted by empirical models for unidirectional \ac{in} deep or intermediate water conditions, see for instance Section 5.2 of \citet{Babanin2011}. This breaking threshold may increase even further when considering shallower depths \citep{mostert2020inertial} and wave directionality in finite water regimes \citep{karmpadakis2020average}. We also would like to highlight that \jk{estimations of wave breaking dissipation rates can be obtained through} large eddy simulations or direct numerical simulations \citep{kim2021large,liu2023wave}. However, in view of the number of waves and duration of the processes, this is very challenging to undertake \jk{in our conditions}.

\begin{figure}
\includegraphics[scale=0.35]{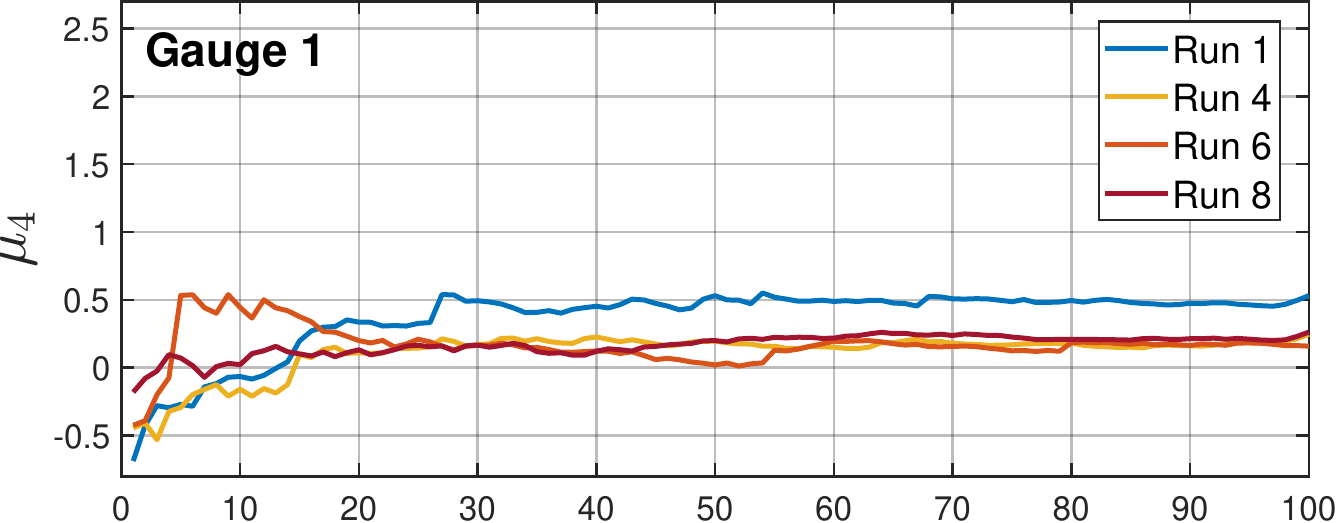}
\includegraphics[scale=0.36]{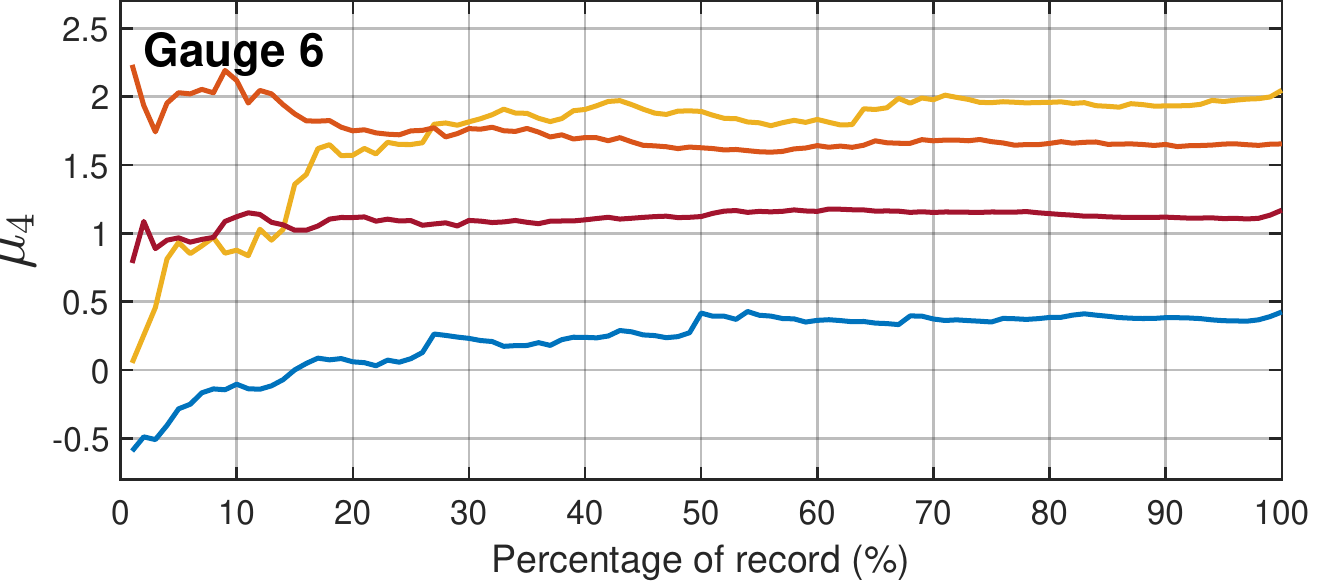}
    \caption{\justifying{\pf{Convergence of excess kurtosis for different wave gauges on configuration 1 and runs of \jfm{Table} \ref{tab:1} over one realization (2500 waves).}}} 
    \label{fig:waveheights2}
\end{figure}
\jk{The wide range of steepness and aeration of wave breaking in the devised experiments fulfils the purpose of seeking
 regimes where the assumptions discussed in \jfm{sections} \ref{sec:bound}-\ref{sec:bound2} break down}. In \jfm{Figure} \ref{fig:FINAL} we compare the experimental results for the excess kurtosis with the theory, that is, \yh{substituting} Eq.~(\ref{eq:finiteampgamma2}) into Eq.~(\ref{eq:kurt}). We also compare them with the semi-analytical approximation of Eq.~(\ref{eq:kurtApp2}). \jk{ The theory seems to break down in the range $k_{p}H_{s} \geqslant 0.43$ when wave breaking becomes dominant, featuring a faster kurtosis decay than expected. Moreover,} both theoretical formulations over-predict the kurtosis for steepness $k_pH_s \lesssim 0.1$, which can be related to the omission of the effect of reflection in the theory \citep{he2022probability,stole2018extreme}.  This is consistent with the empirical formulation for reflection $K_{R} \approx 0.1 \upxi^{2}$ over sloping beaches \citep{Battjes1974}, in which $\upxi$ denotes the surf similarity and is computed in terms of the bottom slope and mean steepness $\upxi \sim \sm{\nabla} h /  \sqrt{\varepsilon}$ \citep{Holthuijsen2007}. Wave regimes with the smallest wave steepness will therefore have the most underestimated reflection rates and thus, overestimate the kurtosis. For \ac{second-order Stokes waves in the Le Méhauté diagram}, both models \ac{agree very well with} the laboratory results \ac{of excess kurtosis} until its peak, whereas the heavy tail for conditions nearing wave breaking is also well captured. \ac{This validates} the theoretical expectations of the kurtosis growth, saturation, and decay. \jk{Although experimentation with steeper slopes could have been carried out, slope effects have been
 verified to be negligible (or saturated) beyond $\nabla h \geqslant 1/5$, both numerically \citep{Adcock2020,Trulsen2021} and
experimentally \citep{Adcock2021c}. \xx{Remarkably, contrary to the expectation of a \xy{symmetric} bell-shaped curve for the excess kurtosis, both experiment and theory demonstrate a skewed concave function that quickly rises and thereafter slowly decays.}}
\begin{figure*}
    \centering
\includegraphics[width=0.4\textwidth]{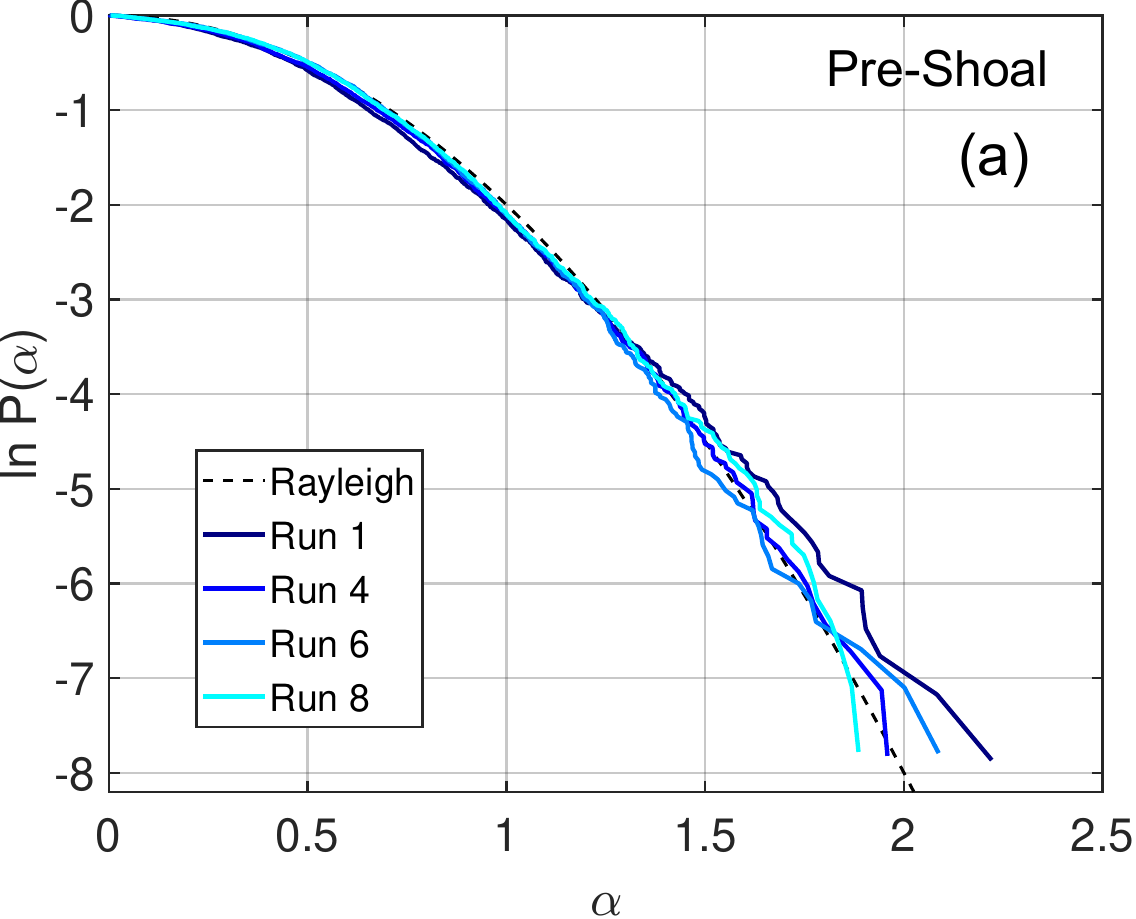}
\includegraphics[width=0.4\textwidth]{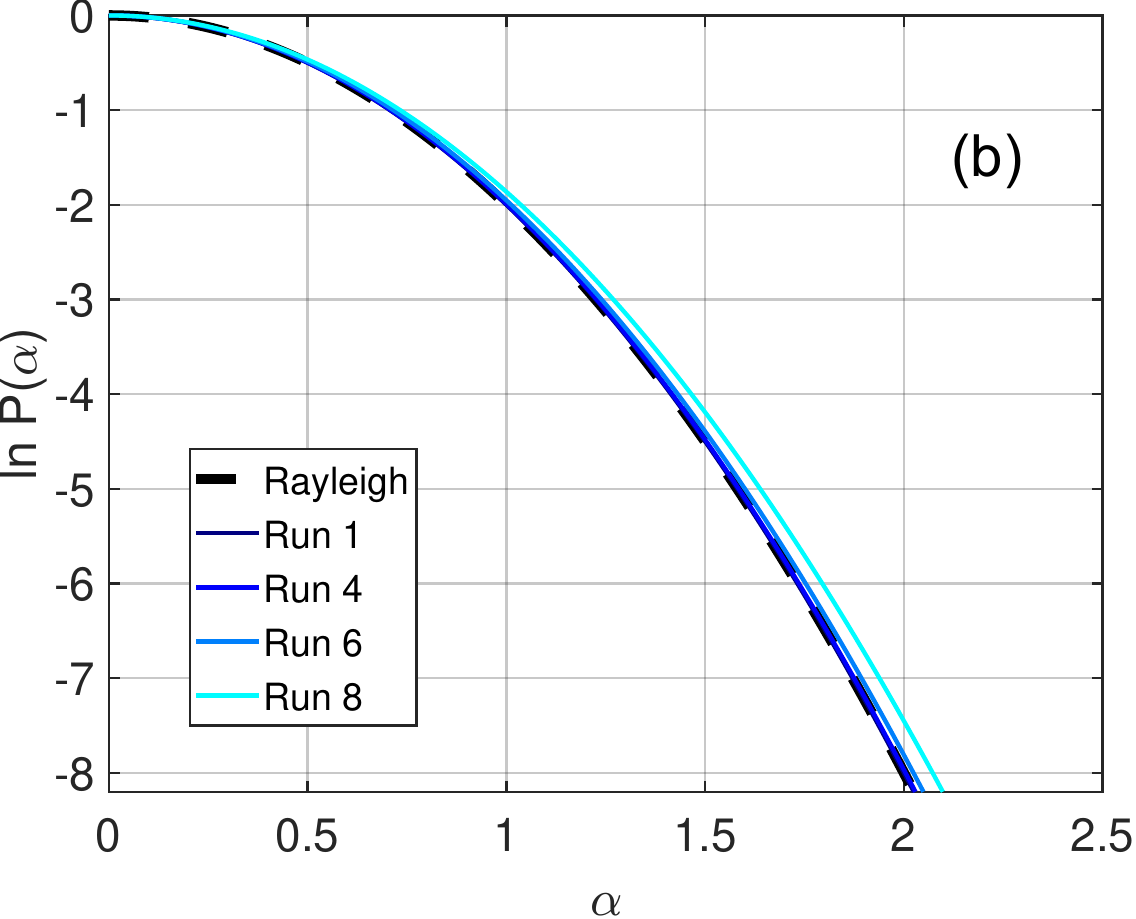}
\includegraphics[width=0.4\textwidth]{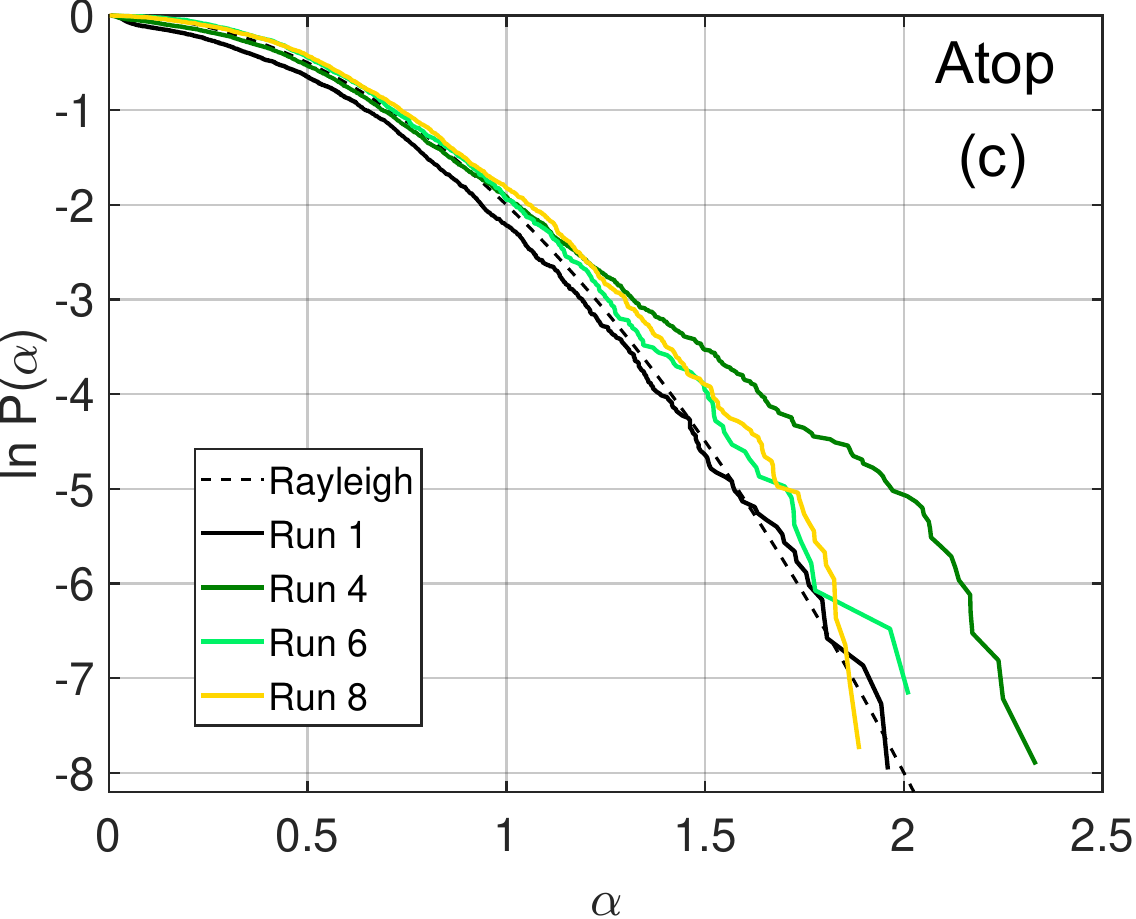}
\includegraphics[width=0.4\textwidth]{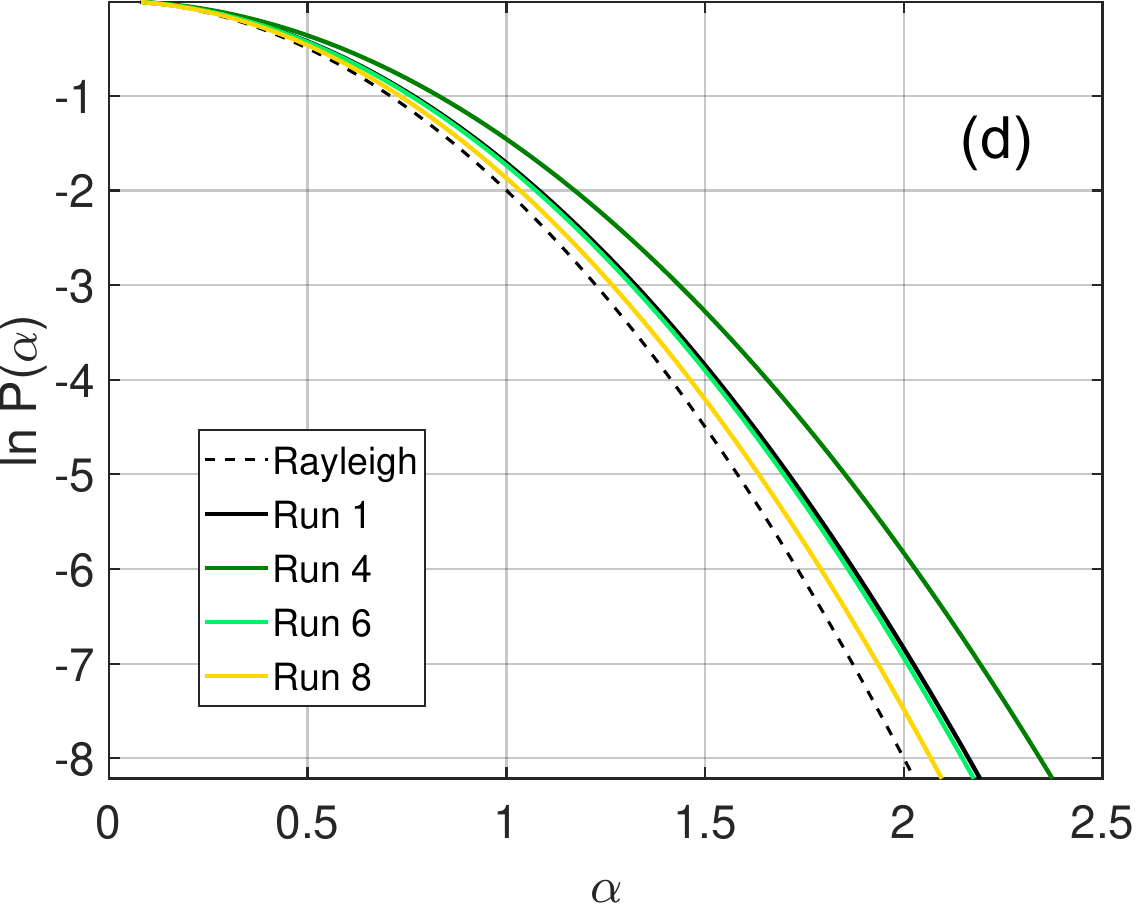}
\includegraphics[width=0.4\textwidth]{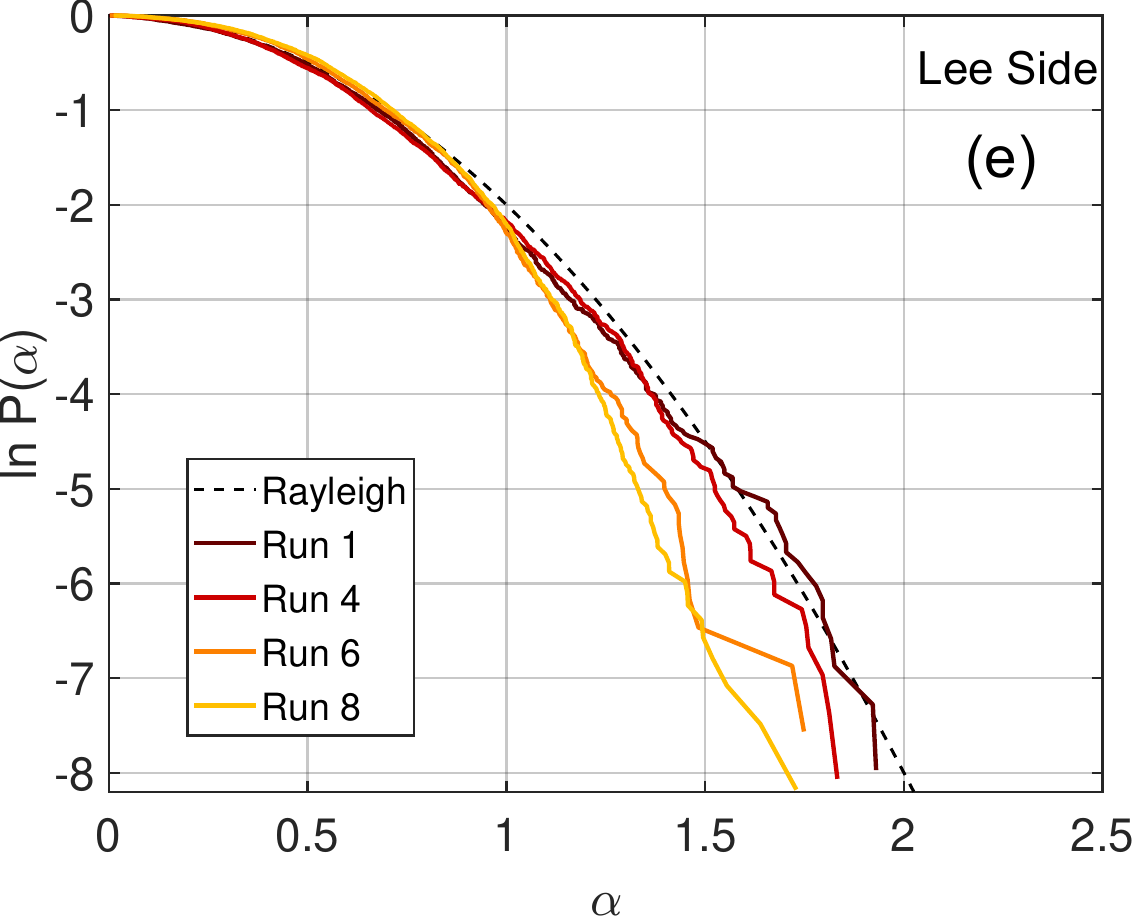}
\includegraphics[width=0.4\textwidth]{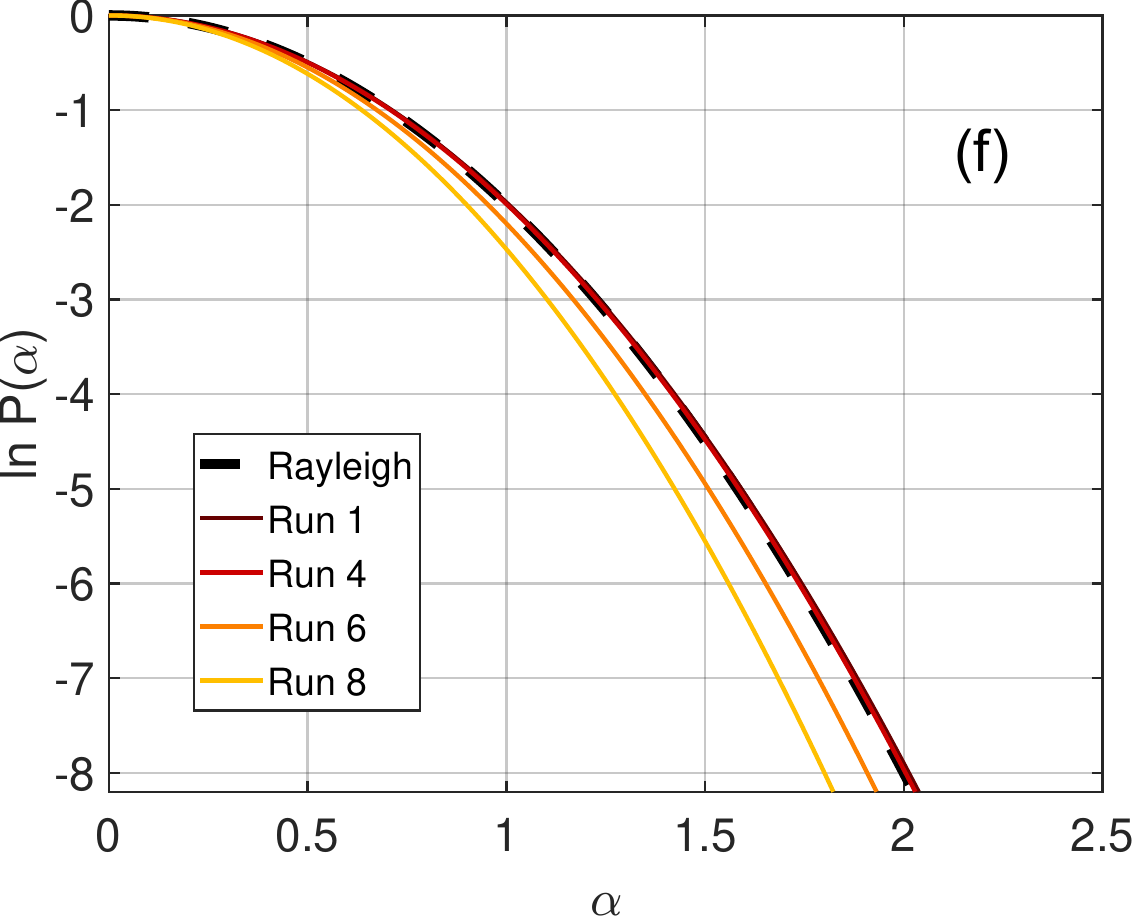}
    \caption{\justifying{\pf{Comparison of wave height exceedance probabilities $P (H > \alpha H_s)$ between (left) experimental observation and (right) theoretical computations from eqs.~(\ref{eq:finiteampgamma0}-\ref{eq:Rayexc},\ref{eq:kurt}).}}} 
    \label{fig:waveheights0}
\end{figure*}

\subsection{\pf{Data Processing \&  Statistics}}

\pf{Confidence intervals applied to \jfm{Figure} \ref{fig:FINAL} were computed with two standard deviations per realization or across subsections of a realization when applicable. Stati onarity is expected to be reached after a few thousand waves \citep{Toffoli2017,Trulsen2022},  and indeed, \jfm{Figure} \ref{fig:waveheights2} showcases the convergence for different gauges (pre-shoal and atop the shoal) is reached earlier than expected. Note that in Gauge 6, wave breaking (runs 4-8) does not seem to affect kurtosis convergence. Moreover, the use of $N=2500$ indeed ensures statistical reliability. This sample size yields a theoretical standard error for kurtosis of approximately $0.1$ based on the asymptotic variance $\text{Var}(\mu_4) \approx 24/N$ \citep{kendall1968advanced}. This is also consistent with established sampling limits for ocean data \citep{Mori2003,Bitner1990}.}

\subsection{\pf{Probability of Surface Elevation \& Wave Heights}}

\begin{figure*}
    \centering
\includegraphics[width=0.46\textwidth]{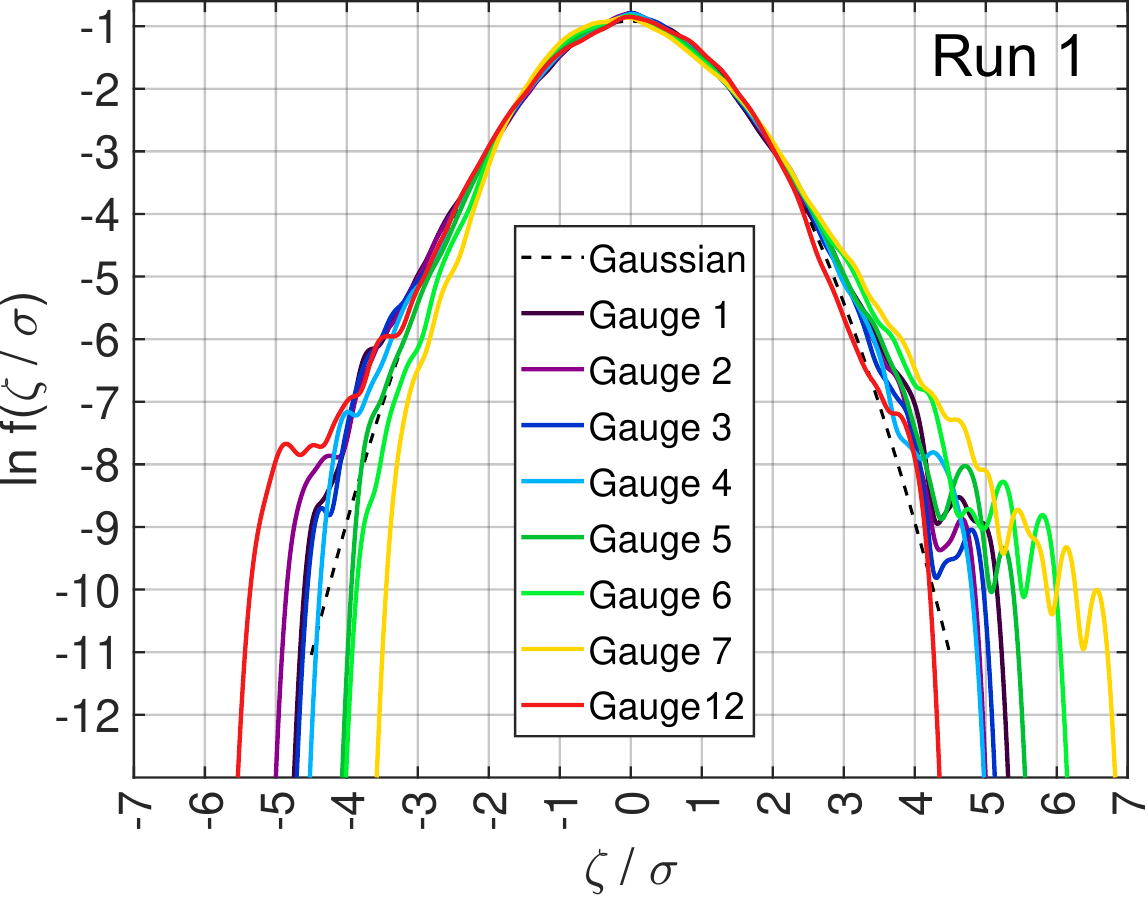}
\includegraphics[width=0.46\textwidth]{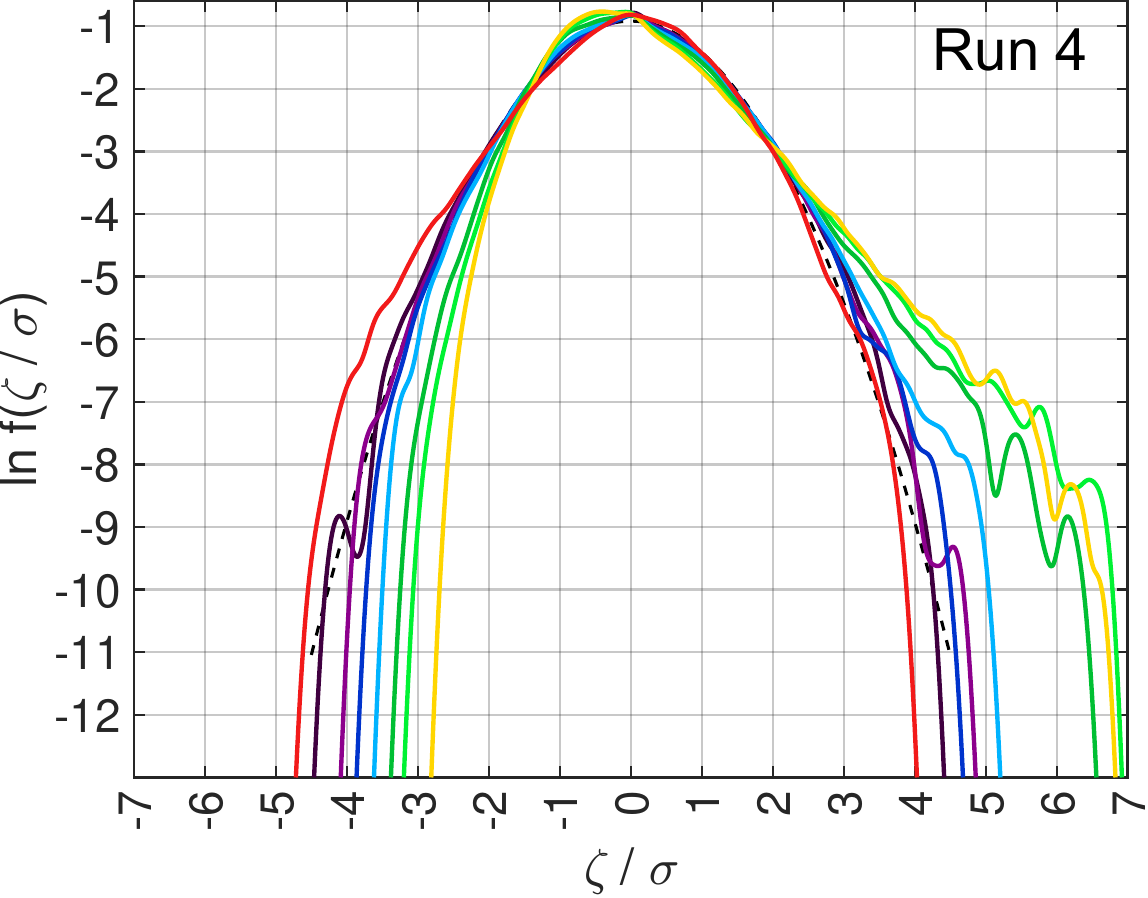}
\includegraphics[width=0.46\textwidth]{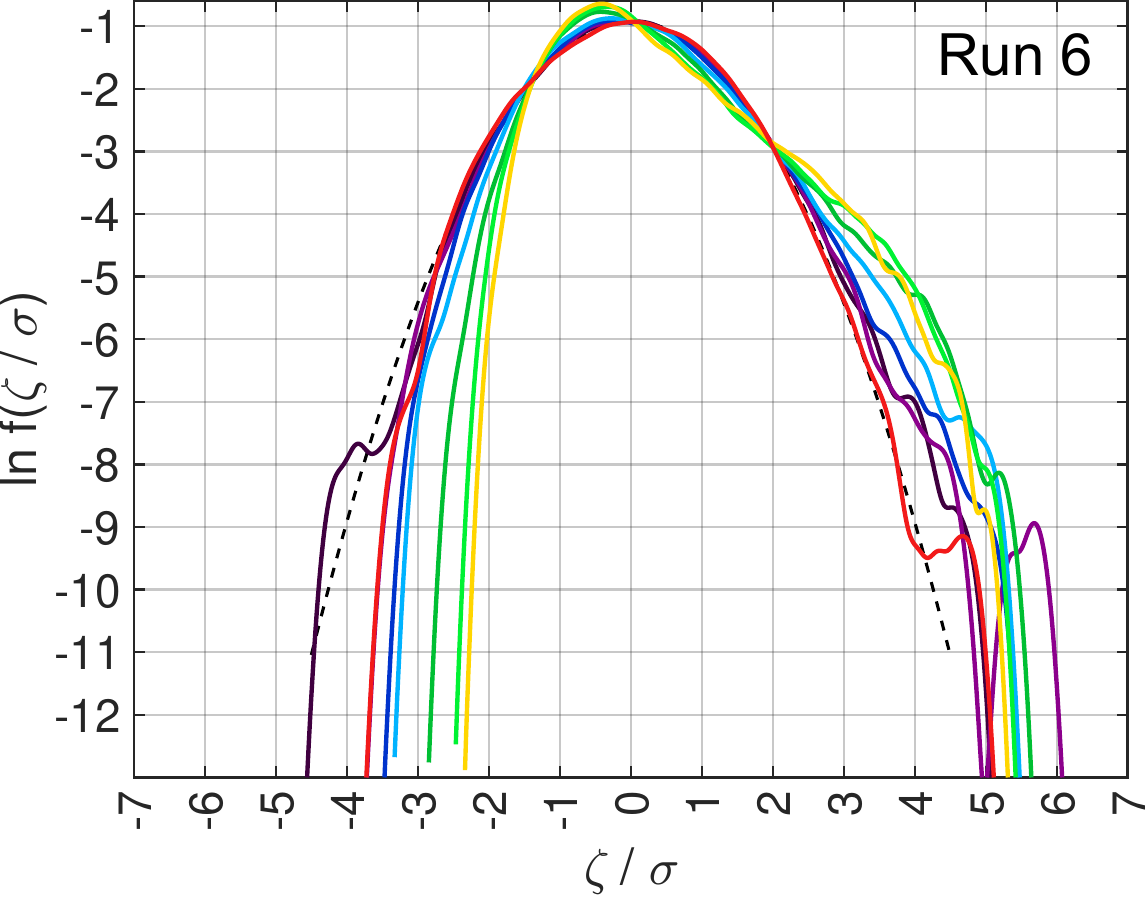}
\includegraphics[width=0.46\textwidth]{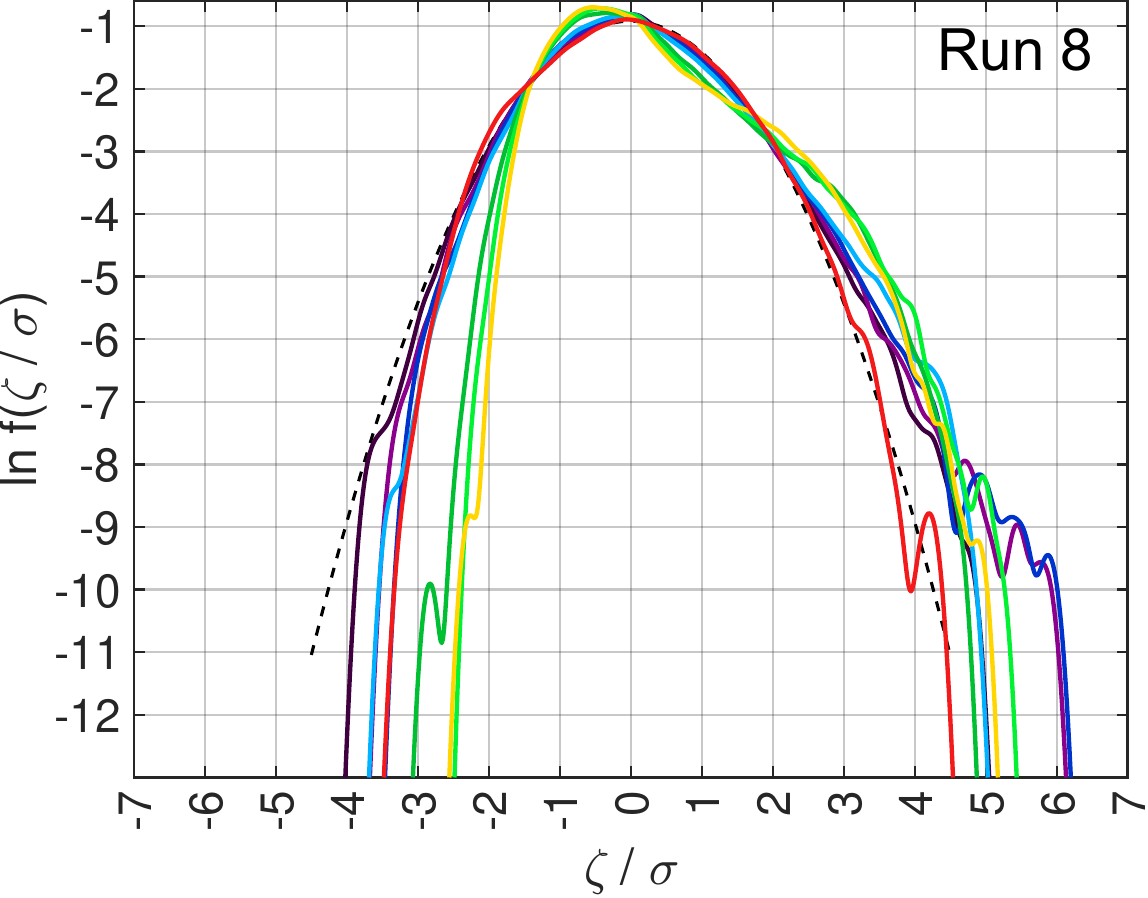}
    \caption{\justifying{\pf{Surface elevation probability densities across different runs of \jfm{Table} \ref{tab:1} and all gauges of configuration one (see \jfm{Figure} \ref{fig:expfig}\jfm{b}).}}} 
    \label{fig:waveheights}
\end{figure*}
\pf{\jfm{Figure} \ref{fig:waveheights0} depicts the exceedance probability, plotted in natural log, of normalized wave heights as observed in the aforementioned experiments. The exceedance probabilities are compared at different locations and for different runs, with \jfm{figure} \ref{fig:waveheights0}\jfm{a} highlighting pre-shoal statistics, whereas \jfm{figure} \ref{fig:waveheights0}\jfm{c} features the data at the maximum nonlinearity (equivalent to the location of \jfm{figure} \ref{fig:FINAL}) atop the shoal and \jfm{figure} \ref{fig:waveheights0}\jfm{e} displays the lee side counterpart.} 

\pf{To compare the data with the theoretical model of eq.~(\ref{eq:Rayexc}), we need to revisit the algebra for the lee side, since eq.~(\ref{eq:finiteampgamma2}) assumes a decreasing water depth $\nabla h < 0$. In eq.~(\ref{eq:ep2}) we argued that the net potential energy balances the slope effect from the wavenumber refraction in the shoaling regime. Yet, in the lee side the de-shoaling has a different effect because of the change in the sign of $\nabla h$, changing from $\nabla h < 0$ (shoaling) to $\nabla h > 0$ (de-shoaling). Moreover, the wave set-up due to both wave breaking \citep{Higgins1964,Guza1981} and breakwater geometry \citep{Higgins1967,Diskin1970,Mendes2024} will alter this interaction. Therefore, the effect of slope on the wavenumber refraction decreases the influence of the net potential energy, but for the case of the lee side the latter is much stronger than the former \footnote{\pf{We can avoid tedious algebra by using the proportionality $\check{\mathscr{E}}_{p2} = -B \nabla h (1-\nabla h)$, and as it balances a term $-A (-\nabla h)$ in the denominator of $\Gamma$ in eq.~(\ref{eq:finiteampgamma1}), then $\check{\mathscr{E}}_{p2} - A \nabla h = 0$ leads to $A = B (1-\nabla h)$. For the lee side the net energy is now $\check{\mathscr{E}}_{p2} = B \nabla h (1+\nabla h)$ because of the wave set-up, and so the balance becomes $-A\nabla h + B \nabla h (1+\nabla h)$ which reduces to $ \check{\mathscr{E}}_{p2}/(1 + \nabla h)$.}}.} 

\pf{Accordingly, eq.~(\ref{eq:finiteampgamma1}) has to be corrected by an additional term $\check{\mathscr{E}}_{p2}/(1 + \nabla h)$ in its denominator in the lee side. In this region, two factors have an influence to decrease extreme wave probabilities (and the kurtosis): wave set-up and wave breaking. Although the two effects are often connected in the surf zone, it is possible to have the former without the latter \citep{Higgins1967}.}

\pf{In the region preceding the shoal, there are small departures from the Rayleigh distribution despite a large increase in wave steepness. This effect is associated with the narrow range of relative water depth $(0.5 \lesssim k_p h \lesssim 1.0)$ required for large amplifications, see \citet{Chabchoub2019,Trulsen2020,Mendes2022}. The similarity between blue curves in \jfm{Figure} \ref{fig:waveheights0}\jfm{a} is well-explained by the theory (shown in \jfm{Figure} \ref{fig:waveheights0}\jfm{b}) considering the error on kurtosis is about 0.1. Atop the shoal, we can observe a similar good agreement between experimental (\jfm{Figure} \ref{fig:waveheights0}\jfm{c}) and proposed theoretical values (\jfm{Figure} \ref{fig:waveheights0}\jfm{d}) of height exceedance probabilities. We recall that Run 1 (\xxx{black} curves) is the reference case with very weak nonlinearity, while in Run 4 (dark green curves) we reach peak strongly nonlinear field characteristics without breaking indications. On the other hand and in Run 8 (yellow curves), very strong breaking occurs. Although the yellow curves for troughs are similar to those of Run 1, the crest curves (\jfm{Figure} \ref{fig:waveheights}, G5-G7) are still above Gaussian expectations, which demonstrates that extreme waves are still more likely even after strong breaking because of the slope presence. Without the slope we would expect the yellow curve to be below the Gaussian expectation.}

\pf{Indeed, the wave breaking lowers the probability of exceedance in the three regions, however, it only becomes sub-Gaussian on the lee side because of the set-up, as described in \cite{Mendes2022}. The theory is therefore supported by observation, and demonstrates that the probability lowers due to breaking, but not necessarily below Gaussian in all zones. However, after the peak in Run 4, the model does not capture this decrease in the region prior to the shoal. This shortcoming is likely due to the absence of reflection in the theory \citep{he2022probability}, since wave breaking was absent in this zone. Nevertheless, the changes in pdf pre-shoal are very small compared to regions 2 and 3. The main understanding drawn from \jfm{Figure} \ref{fig:FINAL}, namely that kurtosis peaks earlier than previously thought but it does not become sub-Gaussian if the bottom slope is steep, can be observed in the transition from Run 1 to Run 8 in \jfm{Figure} \ref{fig:waveheights}: green and yellow curves represent the gauges atop the shoal, and they feature the highest degree of nonlinearity (deviation from the normal distribution, dashed) which increases with steepness until breaking becomes dominant (Runs 6,8) and they become weakly nonlinear, though not sub-Gaussian.}

\section{Discussion}

\ac{Our theoretical results highlight} the following: second-order models for rogue wave statistics expect \ac{in a way or another the} relentless growth of kurtosis with an ever-increasing steepness \citep{Tayfun1980}. As pointed out in \citet{Mendes2021a}, this creates a boundless exceedance probability and consequently a boundless kurtosis. However, wave steepness \ac{cannot} grow boundlessly \citep{Stokes1847,Miche1944}. Our formulae describe \ac{the evolution of kurtosis growth, saturation, and its slow decay up to the wave breaking phase featuring sub-Gaussian statistics \citep{Trulsen2020,Benoit2021,Adcock2021c,Glukhovskii1966,Xu2021,Karmpadakis2022}}.

\sm{On the other hand, the increased exceedance probability due to rogue wave breakers can be mitigated by wave directionality\jk{, that} decreases rogue wave amplification \citep{Karmpadakis2022}. Nonetheless, typical changes of directional spread in the surf zone due to wave breaking amount to a dozen degrees \citep{Szczyrba2023}. The results of \citet{Karmpadakis2022} (see figure 8d) and \citet{Mei2023} (see figure 19) demonstrate that the decrease in kurtosis or exceedance probability due to directional spread is the smallest in the surf zone. This implies \ac{a decrease of excess kurtosis by at most} 10-20\% in the surf zone. From this, we estimate that directionality could at most reduce the maximum of excess kurtosis atop the breakwater ($\mu_4 \approx 1.9$) to  $\mu_4 \approx 1.5-1.7$.}

However, a rather strict application of the $\Gamma$-correction to the wave theory delimitation diagram introduced by \citet{LeMehaute1976} (see \jfm{Figure}~\ref{fig:LMdiag}) would suggest the co-existence of several piecewise formulations for $\Gamma$ delimited together by criteria related to the Ursell number, mean steepness and wave breaking. Unfortunately, a continuous formulation connecting all versions of $\Gamma$ for each wave theory of \citeauthor{LeMehaute1976}'s diagram is \jk{out of reach and} the task of unifying all water wave solutions is yet to be accomplished \citep{Antuono2022,zhao2024guide}. Practical applications of the cnoidal
wave theory and the $\Gamma$ parameter thereof are also very difficult to \ac{address} from an engineering perspective, because it
contains Jacobian elliptic functions and
complete elliptic integrals of the first and second kinds \citep{Iwagaki1968,Iwagaki1982,Isobe1985}. Therefore, it would not serve us to compute piecewise formulations of $\Gamma$, particularly if our goal is to provide a unified picture of rogue wave occurrence from deep or finite depth waters to shallow-waters from low to very high steepness, and from low to high height-to-depth ratios. Inevitably, we shall hold to the unification of the stochastic behaviour of irregular waves grounded on foundational integral properties of water waves, namely, their normalized variance and total spectral energy.

\section{Conclusion}

In this work, we have experimentally investigated the statistical properties of irregular long-crested waves traveling over a symmetrical and steep breakwater in intermediate water. We have characterized and quantified \ac{the amplification of excess kurtosis of surface elevation over the breakwater and thus, the increase of extreme wave events}. Our results cover up to the regime where incident waves are so steep that they resort to the finite wave amplitude theory.  When the significant wave height is small compared to the water depth atop the shoal, the excess kurtosis of the surface elevation \ac{rises quickly}. However, as waves become steeper, two different types of wave breaking emerge, growing slowly from just a few waves to full dominance. A continuous variation between the two cases shows that beyond a threshold of $k_{p2} H_{s\pf{2}} \sim 0.25$, wave breaking balances the effect of \ac{wave} steepness, thus decreasing the excess kurtosis and therefore also the rogue wave occurrence. We \ac{characterize} this significant wave height threshold from the developed theory and find good agreement with experiments. We therefore unify the picture for rogue wave amplification due to shoaling over wide ranges of relative water depth, steepness, and height-to-depth ratio, based on experimental evidence and theoretical developments.
\jk{This combined theoretical and experimental study confirms the possibility of treating irregular waves as second-order, even if they belong to a regime of high-orders in steepness, or to an even more complex solution such as cnoidal waves. }
Our results contrast to previous \ac{studies considering} flat bottoms up to the breaking limit or based on the Glukhovskiy distribution, which in the surf zone ($k_p h \lesssim 0.5$) describes the decrease of excess kurtosis to negative (sub-Gaussian) values in the wave breaking regime.

Moreover, we demonstrate that \jk{the} stochastic analysis of irregular water waves does not consist of a simple and yet cumbersome study of collections of regular waves. Rather, we hint at the existence of fundamental principles of stochastic wave analysis that cannot be transferred to the deterministic counterpart.

From a statistical point of view, early wave breaking is generally expected to decrease the probability of rogue wave breakers. However, we have shown that this probability increases if the beach slope is sufficiently steep. \sm{The latter implication of our model can explain why coastal rogues are more dangerous and numerous when they appear over steep beach slopes or seawalls, or interact with cliffs, as reported in \citet{Didenkulova2023}.} 
Since rogue wave breakers have the highest loads on ships or marine structures\jk{, they} are the most hazardous\jk{, and t}his risk is highest over steep slopes \pf{\citep{Adcock2020,Zhang2026}}. \pf{Moreover, it has been recently shown that extreme wave load kurtosis of waves travelling over a steep submerged breakwater cannot be explained by kinematics, but is rather strongly correlated with the excess kurtosis of the surface elevation \citep{Karmpadakis2026}.} Therefore, in zones prone to very steep sea conditions, breakwaters should be conceived with slopes below $\sim 1/5$ \pf{in order to avoid a large excess kurtosis, and in return higher maxima of wave loads}.

\sm{Last but not least, our results put forward a novel and broader formula for the mechanical energy of waves under all possible regimes of wave steepness, bottom slope magnitude, water depth and breaking condition. In doing so, we have found new \ac{formulae} that resemble the ratio between group and phase velocity\ac{, and a form of} generalization of the \sm{$c_g/c_p$}. It is quite possible \jk{that} the quantities $(n_{\varepsilon}, n_{2\varepsilon})$ have wider implications for wave mechanics (e.g. radiation stress and energy flux of finite amplitude waves following \citet{Klopman1990}), that may have been elusive \jk{so far}. }

\pf{Potential future refinements are necessary for tackling the limitations of our present experiments and theoretical work. Necessary investigation for the interplay between wave breaking and reflection effects on extreme wave statistics is one such example. In our present study, we limited the theory and experiment to relatively steep slopes that still had small reflection rates. In particular, it would be valuable to assess how breaking alters the stabilizing influence of high reflection rates on excess kurtosis (as found in this study for steep smooth beaches), in contrast to the behaviors reported in \citet{Zhang2026} for different slope and breaking conditions. Since wave breaking is more characteristic of coastal zones, the interaction between the reflection and the undertow caused by breaking \citep{Svendsen1984,Hoque2002,Martins2017} as well as its interplay with strong opposing or forward tides, or sheared currents on wave fields \citep{Li2023,Li2024,Li2025,Mendes2025,Li2026,Teutsch2026} have been shown to be significant at the surf zone. By omitting these processes, we have found small shortcomings in our model, particularly at low steepness (higher reflection) and very high steepness (undertow being neglected). Furthermore, although we used $H_s/h$ (a proxy similar to the breaker index) to parametrize wave breaking, a potential extension to our work would seek a joint probability distribution of wave heights (to gauge excess kurtosis and extreme wave likelihood) and wave breaking, to rigorously assess how the direct share of breaking waves modulate extreme events, thus generalizing both our present study and that of breaking wave probabilities \citep{Babanin2001,Babanin2010}. Finally, since rapid variations in meteorological conditions, such as wind speed changes over time or space, could disrupt wave equilibrium and potentially amplify kurtosis and rogue wave probability \citep{Zou2017,Trulsen2018,Toffoli2024}, it would be of paramount importance to study the effects of wave breaking and wind speed gradient concurrently.}

\section{Acknowledgements}

S.M. and J.K. were supported by the Swiss National Science Foundation under grant 200020-175697. S.M. also acknowledges the support of Nanyang Technological University Start-up Funds. Y.H. acknowledges the support from the Distinguished Postdoctoral Fellowship Scheme of the Hong Kong Polytechnic University. A.C. acknowledges support from the Okinawa Institute of Science and Technology (OIST) with subsidy funding from the Government of Japan's Cabinet Office. Authors are also grateful to Dr. Kapil Chauhan and Theo Gresley-Daines from the University of Sydney for their support and advice on the experimental setup and valuable help with the installation. 

\textbf{Declaration of Interests}. The authors report no conflict of interest.

\appendix

\section{Energetics of Irregular and Inhomogeneous Finite Amplitude Waves}\label{sec:THEOR2}

\pf{This appendix contains the complete step-by-step integration procedures and intermediate computations that support and justify the principal analytical results derived in Section \ref{sec:Edensity}, notably the expressions given in eqs.~(\ref{eq:energ0},\ref{eq:EnergykX2},\ref{eq:EnergykXX}). W}e explicitly detail some of the tedious algebra involved in the major integrals in the bulk of the text. Since the integrals of cross-order terms $\sin{\phi}\sin{(2\phi)}$ and $\cos{\phi}\cos{(2\phi)}$ of Eq.~(\ref{eq:energ0}) vanish, we have:
\begin{widetext}
\begin{eqnarray}
\nonumber
\hspace{-0.5cm}
 \mathscr{E}_{k0} &=&  \frac{(a\omega)^2}{2 \lambda g} \int_{0}^{\lambda} \int_{-h}^{H_{s}/4} \Bigg[   \frac{\cosh^2{\varphi}}{\sinh^2{\Lambda}} \cos^2{\phi} +  \left(  \frac{3ka}{4}  \right)^2  \frac{\cosh^2{(2\varphi)}}{\sinh^8{\Lambda}}  \cos^2{(2\phi)} + \frac{\sinh^2{\varphi} }{\sinh^2{\Lambda}} \sin^2{\phi}  + \left(  \frac{3ka}{4}  \right)^2 \frac{\sinh^2{(2\varphi)} }{\sinh^{8}{\Lambda}} \sin^2{(2\phi)} \Bigg] \, dx \, dz 
\\
&=&  \frac{a^2 \cdot gk \tanh{\Lambda}}{4 g \sinh^{2}{\Lambda}}   \int_{-h}^{H_{s}/4} \left[  \cosh^2{\varphi} + \sinh^2{\varphi} +  \left(  \frac{3ka}{4}  \right)^2  \frac{\cosh^2{(2\varphi)} + \sinh^2{(2\varphi)}}{\sinh^6{\Lambda}}    \right]  \, dz \, ,
\label{eq:EnergykX1}
\end{eqnarray}
\end{widetext}
Noting that a transformation of steepness between regular and irregular wave fields obeys $ka \rightarrow \pi \varepsilon \mathfrak{S} \approx \sqrt{2} k_p H_s \mathfrak{S} $ \citep{Mendes2021b}, the above integral is approximated for the regime of interest ($\Lambda \leq 2$, $k_pH_s \leq 1/2$), \jk{ covering both regular and irregular waves}:
\begin{widetext}
\begin{eqnarray}
\mathscr{E}_{k0} \approx   \frac{a^2 }{4 } \frac{2k }{\sinh{(2\Lambda)}}  \Bigg[  \frac{\sinh{(2\Lambda)} + \sinh{(\frac{1}{2} k_p H_s)}}{2k}  +  \left(  \frac{3 \pi \varepsilon \mathfrak{S}}{4}  \right)^2  \frac{\sinh{(4\Lambda)} + \sinh{( k_p H_s)}}{4k \sinh^6{\Lambda}}  \Bigg] .
\label{eq:Ekx}
\end{eqnarray}
\end{widetext}
Due to the fact that wave breaking starts to dominate for $ k_p H_s \approx 1/4$, we can approximate $\sinh{(\frac{1}{2} k_p H_s)}$ as $\frac{1}{2}\sinh{( k_p H_s)}$ with an error of less than \jk{2.5}\%.

\begin{figure}
   \includegraphics[scale=0.7]{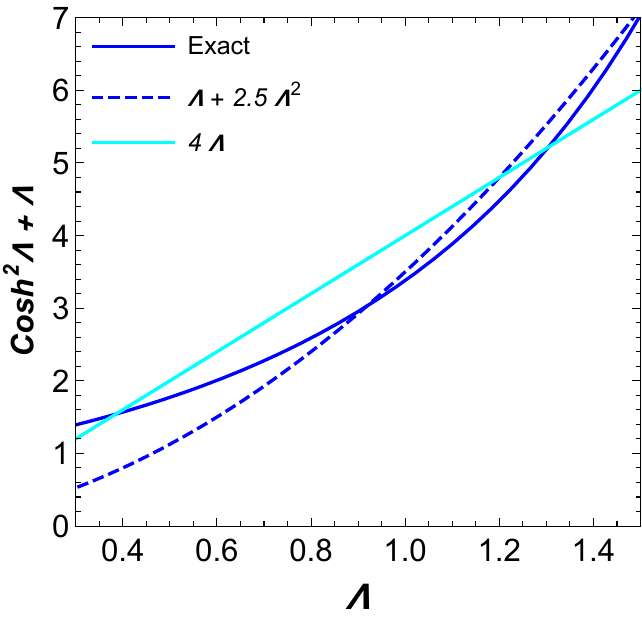}
   \caption{Polynomial approximations for the denominator of Eq.~(\ref{eq:nablalambda0}).}
  \label{fig:approxNABLA2}
\end{figure} 
\sm{Furthermore, to explore the effect of additional terms on the above kinetic energy, one must investigate how quickly the wavenumber changes. W}e know that for a pre-shoal wave period $T_{0}$, the shoaling causes a spatial variation in the wavelength, namely:
\begin{equation}
\lambda (x) = \frac{gT_{0}^{2}}{2\pi} \, \tanh{\big( k(x)h(x) \big)} \quad .
\end{equation}
Then, the derivative along the direction of motion is applied to the wavelength, reading:
\begin{eqnarray}
\sm{\nabla} \lambda = \lambda_{0} \sm{\nabla} \Big[ \tanh{\big( k(x)h(x) \big)} \Big] 
=  \frac{\lambda_{0}}{\cosh^{2}{ \Lambda(x) }} \Big[ k \sm{\nabla} h + h \sm{\nabla} k \Big]  \, .
\end{eqnarray}
Because of the identity $\sm{\nabla} k / k = - (\sm{\nabla} \lambda / \lambda)$, it now reads:
\begin{equation}
\nabla \lambda = \frac{2\pi \sm{\nabla} h}{  \cosh^{2}{ \left[ kh_{0} \left( 1 + \frac{x\sm{\nabla} h}{h_{0}}  \right)   \right] } + kh_{0} \left( 1 + \frac{x\sm{\nabla} h}{h_{0}}  \right)   }    \quad .
\label{eq:nablalambda0}
\end{equation}
\sm{We linearize the denominator of this equation with the fit $\cosh^2{\Lambda} + \Lambda \approx 4 \Lambda$ as displayed on \jfm{Figure}~\ref{fig:approxNABLA2}.}
\begin{eqnarray}
 \mathscr{E}_{k} - \mathscr{E}_{k0}  \sim \frac{1}{2 \lambda g} \int_{0}^{\lambda} \int_{0}^{H_{s}/4}  \left[ \left(    \pd{\Phi}{x}  \right)^{2} + \left(    \pd{\Phi}{z}  \right)^{2} \right]  \, dz \, dx \quad ,
\label{eq:EnergykA}
\end{eqnarray}
Noting that $ \sm{\nabla} \phi=k + x \sm{\nabla} k$ and  $\sm{\nabla} \varphi = (z+h)\sm{\nabla} k + k \sm{\nabla} h$, the horizontal velocity becomes:
\begin{widetext}
\begin{eqnarray}
\hspace{-0.2cm}
\nonumber
\frac{k  u}{a\omega} &=&    - \frac{\sm{\nabla} k}{k}     \frac{\cosh{\varphi} }{\sinh{\Lambda}} \sin{ \phi} +  \sm{\nabla} \varphi \cdot    \frac{\sinh{\varphi}}{\sinh{\Lambda}}  \sin{\phi}  - \sm{\nabla} \Lambda \,  \frac{\cosh{\Lambda}}{\sinh{\Lambda}} \cdot \frac{\cosh{\varphi} }{\sinh{\Lambda}} \sin{ \phi} +  \sm{\nabla} \phi \cdot     \frac{\cosh{\varphi}}{\sinh{\Lambda}}  \cos{\phi} +   \frac{3ka}{8}   \Bigg[  - \frac{\sm{\nabla} k}{k}   \frac{\cosh{(2\varphi)} }{\sinh^4{\Lambda}} \sin{(2\phi)} 
\\
&+&  2\sm{\nabla} \varphi \cdot  \frac{\sinh{(2\varphi)}}{\sinh^4{\Lambda}}  \sin{(2\phi)} - 4\sm{\nabla} \Lambda \, \frac{\cosh{\Lambda}}{\sinh{\Lambda}}  \cdot \frac{\cosh{(2\varphi)} }{\sinh^4{\Lambda}}\sin{(2\phi)} + 2 \sm{\nabla} \phi \cdot     \frac{\cosh{(2\varphi)}}{\sinh^4{\Lambda}}  \cos{(2\phi)} \Bigg] \quad .
\end{eqnarray}
This horizontal velocity \sm{is} expanded and compared to the otherwise slope-independent velocity $u_0$ \pf{within brackets "$\{ \}$"}:
\begin{eqnarray}
\nonumber
 \frac{  u}{a\omega} &=& \left\{  \frac{\cosh{\varphi}}{\sinh{\Lambda}} \cos{\phi} +  \left(  \frac{3ka}{4}  \right)  \frac{\cosh{(2\varphi)}}{\sinh^4{\Lambda}}  \cos{(2\phi)} \right\} + \frac{x \sm{\nabla} k}{k} \frac{\cosh{\varphi}}{\sinh{\Lambda}} \cos{\phi}   - \frac{\sm{\nabla} k}{k^2}   \frac{\cosh{\varphi} }{\sinh{\Lambda}} \sin{ \phi} + \left( (z + h) \frac{\sm{\nabla} k}{k}  + \sm{\nabla} h \right)  \frac{\sinh{\varphi}}{\sinh{\Lambda}}  \sin{\phi} 
\\
\nonumber
&-&     \left( h \frac{\sm{\nabla} k}{k}  + \sm{\nabla} h \right) \frac{\cosh{\Lambda}}{\sinh{\Lambda}} \cdot \frac{\cosh{\varphi} }{\sinh{\Lambda}} \sin{ \phi}  + \left(  \frac{3ka}{4}  \right)  \Bigg[    \frac{x \sm{\nabla} k}{k} \frac{\cosh{(2\varphi)}}{\sinh^4{\Lambda}} \cos{(2\phi)}   -  \frac{1}{2} \frac{\sm{\nabla} k}{k^2}   \frac{\cosh{(2\varphi)} }{\sinh^4{\Lambda}} \sin{(2\phi)}  
\\
&+&  \left( (z + h) \frac{\sm{\nabla} k}{k}  + \sm{\nabla} h \right)  \frac{\sinh{(2\varphi)}}{\sinh^4{\Lambda}}  \sin{(2\phi)} -  \left( h \frac{\sm{\nabla} k}{k}  + \sm{\nabla} h \right) \frac{2\cosh{\Lambda}}{\sinh{\Lambda}} \cdot \frac{\cosh{(2\varphi)} }{\sinh^4{\Lambda}}\sin{(2\phi)} \Bigg] \equiv \frac{u_0 + \Delta u}{a \omega} \quad ,
\label{eq:EnergykX00}
\end{eqnarray} 
\end{widetext}
\pf{such that the difference reads}
\begin{widetext}
\begin{eqnarray}
\nonumber
 \frac{ \sm{\Delta} u}{a\omega} &=& \frac{x \sm{\nabla} k}{k} \frac{\cosh{\varphi}}{\sinh{\Lambda}} \cos{\phi}   - \frac{\sm{\nabla} k}{k^2}   \frac{\cosh{\varphi} }{\sinh{\Lambda}} \sin{ \phi} +  \left( (z + h) \frac{\sm{\nabla} k}{k}  + \sm{\nabla} h \right)  \frac{\sinh{\varphi}}{\sinh{\Lambda}}  \sin{\phi} - \left( h \frac{\sm{\nabla} k}{k}  + \sm{\nabla} h \right) \frac{\cosh{\Lambda}}{\sinh{\Lambda}} \cdot \frac{\cosh{\varphi} }{\sinh{\Lambda}} \sin{ \phi}  
\\
\nonumber
&+& \left(  \frac{3ka}{4}  \right)  \Bigg[    \frac{x \sm{\nabla} k}{k} \frac{\cosh{(2\varphi)}}{\sinh^4{\Lambda}} \cos{(2\phi)}    - \left( h \frac{\sm{\nabla} k}{k}  + \sm{\nabla} h \right) \frac{2\cosh{\Lambda}}{\sinh{\Lambda}} \cdot \frac{\cosh{(2\varphi)} }{\sinh^4{\Lambda}}\sin{(2\phi)} 
\\
&+&  \left( (z + h) \frac{\sm{\nabla} k}{k}  + \sm{\nabla} h \right)  \frac{\sinh{(2\varphi)}}{\sinh^4{\Lambda}}  \sin{(2\phi)} -  \frac{1}{2} \frac{\sm{\nabla} k}{k^2}   \frac{\cosh{(2\varphi)} }{\sinh^4{\Lambda}} \sin{(2\phi)}   \Bigg]  \quad .
\label{eq:EnergykX0}
\end{eqnarray} 
\end{widetext}
\jk{According to Eq.~(\ref{eq:EnergykX0}), $\Delta u$ scales with $\sm{\nabla} h$}
because $\sm{\nabla} \lambda$ \sm{should} \ac{be dependent on} the slope \citep{Zhang1998}. 
Small finite-amplitude corrections to the kinetic energy due to the evolution of the physical variables across the shoal become
\begin{widetext}
\begin{eqnarray}
\hspace{-0.1cm}
\nonumber
 \Delta\mathscr{E}_{k} &\approx&    \frac{1}{2 \lambda g} \int_{0}^{\lambda} \int_{-h}^{H_{s}/4}  \left(  2 u_0 \Delta u + (\Delta u)^{2} \right) \, dx \, dz ,
 \\
\nonumber
&=&  \frac{(a\omega)^2}{2 \lambda g} \int_{0}^{\lambda} \int_{-h}^{H_{s}/4} dx \, dz \, \Bigg\{ - 2x \frac{ \sm{\nabla} \lambda}{\lambda} \left[   \frac{\cosh^2{\varphi}}{\sinh^2{\Lambda}} \cos^2{\phi} +  \left(  \frac{3ka}{4}  \right)^2  \frac{\cosh^2{(2\varphi)}}{\sinh^8{\Lambda}}  \cos^2{(2\phi)} \right]
\\
\nonumber
&+&  \left[   \frac{\cosh^2{\varphi}}{\sinh^2{\Lambda}} \cos^2{\phi} +  \left(  \frac{3ka}{4}  \right)^2  \frac{\cosh^2{(2\varphi)}}{\sinh^8{\Lambda}}  \cos^2{(2\phi)} \right] \left( x \frac{ \sm{\nabla} \lambda}{\lambda} \right)^2 
\\
\nonumber
&+& \left[   \frac{\cosh^2{\varphi} }{\sinh^2{\Lambda}} \sin^2{\phi} \,  +  \left(  \frac{3ka}{4}  \right)^2  \frac{\cosh^2{(2\varphi)} }{4\sinh^8{\Lambda}} \sin^2{(2\phi)} \,  \right] \,\,\, \left(  \frac{ \sm{\nabla} \lambda}{2\pi} \right)^2
\\
\nonumber
&+&  \left[   \frac{\sinh^2{\varphi} }{\sinh^2{\Lambda}} \sin^2{\phi} \,  +  \left(  \frac{3ka}{4}  \right)^2  \frac{\sinh^2{(2\varphi)} }{\sinh^8{\Lambda}} \sin^2{(2\phi)} \,  \right] \left(  \sm{\nabla} h - (z + h) \frac{\sm{\nabla} \lambda}{\lambda}   \right)^2
\\
\nonumber
&+&  \left[   \frac{\cosh^2{\varphi}}{\sinh^2{\Lambda}} \sin^2{\phi} +  \left(  \frac{3ka}{4}  \right)^2  \frac{4\cosh^2{(2\varphi)}}{\sinh^8{\Lambda}}  \sin^2{(2\phi)}  \right] \left( h \frac{ \sm{\nabla} \lambda}{\lambda} - \sm{\nabla} h \right)^2 \frac{\cosh^2{\Lambda} }{\sinh^2{\Lambda}}
\\
\nonumber
&+&  \left[   \frac{\sinh{(2\varphi)}}{\sinh^2{\Lambda}} \sin^2{\phi} +  \left(  \frac{3ka}{4}  \right)^2  \frac{\sinh{(4\varphi)}}{2\sinh^8{\Lambda}}  \sin^2{(2\phi)}  \right] \frac{ \sm{\nabla} \lambda}{2\pi}  \left(  \sm{\nabla} h - (z + h) \frac{\sm{\nabla} \lambda}{\lambda}   \right)
\\
\nonumber
&+&  \left[   \frac{\sinh{(2\varphi)}}{\sinh^2{\Lambda}} \sin^2{\phi} +  \left(  \frac{3ka}{4}  \right)^2  \frac{2\sinh{(4\varphi)}}{\sinh^8{\Lambda}}  \sin^2{(2\phi)}   \right]  \frac{\cosh{\Lambda} }{\sinh{\Lambda}}  \sm{\cdot} \left(  \sm{\nabla} h - (z + h) \frac{\sm{\nabla} \lambda}{\lambda}   \right) \left( h \frac{ \sm{\nabla} \lambda}{\lambda} - \sm{\nabla} h \right) \Bigg\}
\\
&+&  \left[   \frac{\cosh^2{\varphi} }{\sinh^2{\Lambda}} \sin^2{\phi} +  \left(  \frac{3ka}{4}  \right)^2  \frac{\cosh^2{(2\varphi)} }{\sinh^8{\Lambda}} \sin^2{(2\phi)}  \right] \frac{ \sm{\nabla} \lambda}{2\pi} \left( h \frac{ \sm{\nabla} \lambda}{\lambda} - \sm{\nabla} h \right) \frac{\cosh{\Lambda} }{\sinh{\Lambda}}  \equiv  \frac{ka^2 }{\sinh{(2\Lambda)}} \sum_{i=1}^{8} \mathcal{I}_{i} \,\, . 
\label{eq:EnergykX52}
\end{eqnarray} 
\end{widetext}
\pf{Because the terms $(\nabla \lambda /\lambda )^2$ and $(\nabla h)^2$ are small, and a}ccording to the approximated wavelength longitudinal evolution above, we compute $\mathcal{I}_1$ \pf{(leading term)} whilst implementing the transformation from regular to irregular waves:
\begin{widetext}
\begin{eqnarray}
\nonumber
\mathcal{I}_1 &\approx& - \frac{2}{ \lambda^2} \int_{0}^{\lambda} \int_{-h}^{H_{s}/4} dx \, dz \, x \sm{\nabla} \lambda  \left[   \cosh^2{\varphi} \cos^2{\phi} +  \left(  \frac{\pi \varepsilon \mathfrak{S}}{4}  \right)^2 \chi_1  \frac{\cosh^2{(2\varphi)}}{\cosh{(2\Lambda)}}  \cos^2{(2\phi)} \right] \quad ,
\\
\nonumber
&\approx& - \frac{\tilde{\nabla} h}{ \lambda^2} \int_{0}^{\lambda} \int_{-h}^{H_{s}/4}   \left[   \cosh^2{\varphi} \cos^2{\phi} +  \left(  \frac{\pi \varepsilon \mathfrak{S}}{4}  \right)^2 \chi_1  \frac{\cosh^2{(2\varphi)}}{\cosh{(2\Lambda)}}  \cos^2{(2\phi)} \right]  
\frac{x \,  dx \, dz}{\left( 1 + \frac{2x}{\lambda} \tilde{\nabla} h \right)}   \, ,
\\
&\sm{\approx}& \sm{- \frac{\tilde{\nabla} h}{ \lambda}  \int_{-h}^{H_{s}/4}   \left[ \mathcal{I}_{11x}  \cosh^2{\varphi}  + \mathcal{I}_{12x}  \left(  \frac{\pi \varepsilon \mathfrak{S}}{4}  \right)^2 \chi_1  \frac{\cosh^2{(2\varphi)}}{\cosh{(2\Lambda)}}   \right]  
 dz   \quad ,}
\label{eq:EnergykX66}
\end{eqnarray}
\end{widetext}
\sm{where t}he lowest order in steepness has a spatial function \jk{dependence}:
\begin{equation}
    \mathcal{I}_{11x} = -\frac{\tilde{\nabla} h}{ \lambda^2} \int_{0}^{\lambda} 
\frac{x       \cos^2{\phi} }{\left( 1 + \frac{2x}{\lambda} \tilde{\nabla} h \right)} \,  dx \quad ,
\label{eq:I1x}
\end{equation}
which can be approximated as a polynomial:
\begin{figure}
   \includegraphics[scale=0.65]{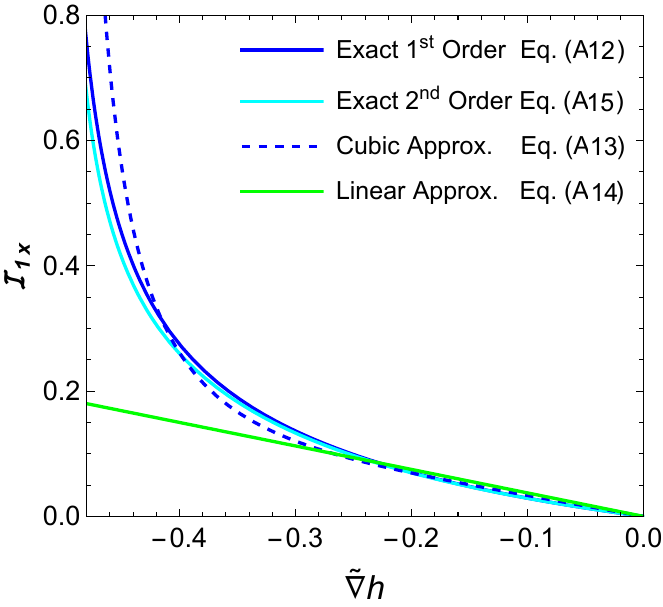}
   \caption{\justifying{Resulting \jk{Slope dependence} of different levels of approximations for the integral in Eq.~(\ref{eq:I1x}) and their comparison with the second order integral of Eq.~(\ref{eq:I2x}).}}
  \label{fig:approxNABLA22}
\end{figure} 
\begin{eqnarray}
\mathcal{I}_{11x} &\approx&  - \frac{\tilde{\nabla} h}{3}  \left[  1 - \frac{ 3 (\tilde{\nabla} h)^3  }{ (1+2\tilde{\nabla} h)   }  \right] 
\\
&\cong&  - \frac{3\tilde{\nabla} h}{8} \equiv \pf{f_{\tilde{\nabla} h}}  \quad .
\label{eq:approx}
\end{eqnarray}
The second-order term  \sm{of Eq.~(\ref{eq:EnergykX66})} reads:
\begin{equation}
\mathcal{I}_{12x} = -\frac{\tilde{\nabla} h}{ \lambda^2} \int_{0}^{\lambda} 
  \frac{x       \cos^2{(2\phi)} }{\left( 1 + \frac{2x}{\lambda} \tilde{\nabla} h \right)} \,  dx \quad .
\label{eq:I2x}
\end{equation}
Since  $\cos^2 {\phi}$ and $\cos^2 {2\phi}$ oscillate much faster than the evolution of $x$, \jk{the second-order term provides a good approximation of Eq.~(\ref{eq:approx}), that is} $\mathcal{I}_{12x} \approx \mathcal{I}_{11x}$. I\jk{ndeed, $0.95 \leq \mathcal{I}_{12x}/\mathcal{I}_{11x} \leq 1$ in the interval $\jk{0 > } \tilde{\nabla} h > -0.4$ (\jfm{Figure} \ref{fig:approxNABLA22}}), the latter being the upper limit for the theory of arbitrary slopes \citep{Mendes2022}. As such, the leading term in the \sm{slope}-dependent correction to the kinetic energy becomes
\begin{eqnarray}
\Delta\mathscr{E}_{k}  &\approx&  \frac{\pf{f_{\tilde{\nabla} h}}(a\omega)^2}{2 g \sinh^2{\Lambda}}
\\
\nonumber
&\times&\int_{-h}^{\frac{H_{s}}{4}}   \left[   \cosh^2{\varphi}  +  \left(  \frac{\pi \varepsilon \mathfrak{S}}{4}  \right)^2   \frac{\chi_1 \cosh^2{(2\varphi)}}{\cosh{(2\Lambda)}}   \right] dz \quad ,
\end{eqnarray} 
whose application of the algebraic manipulation in Eqs.~(\ref{eq:EnergykX1}-\ref{eq:Ekx})
leads to:
\begin{widetext}
\begin{eqnarray}
\hspace{-1.3cm}
\Delta\mathscr{E}_{k}  &\approx& \pf{f_{\tilde{\nabla} h}} \frac{a^2 }{4 } \left[ \left(1 + \frac{ \sinh{(\frac{1}{2} k_p H_s)} }{ \sinh{(2\Lambda)} }\right) n_{\varepsilon}   +  \left(  \frac{\pi \varepsilon \mathfrak{S}}{4}  \right)^2  \chi_1 \left(  1 + \frac{ \sinh{( k_p H_s)} }{ \sinh{(4\Lambda)} }    \right) n_{2\varepsilon} \right]  ,
\end{eqnarray}
\end{widetext}
where the group-to-phase velocity ratios ($n_{\varepsilon},n_{2\varepsilon}$) \sm{are described in Eqs.~(\ref{eq:n1}) and (\ref{eq:n2}).}

\section{\xt{WKB Method and Refraction}}\label{sec:refraction}

\pf{This appendix provides the detailed, step-by-step integrations and supporting calculations demonstrating that both the WKB approximation and fully nonlinear refraction effects — both related to the variation of the wavenumber across the breakwater — are indeed negligible, precisely as assumed in deriving the main analytical expressions in Section \ref{sec:Edensity}, specifically eqs.~(\ref{eq:energ0},\ref{eq:EnergykX2},\ref{eq:EnergykXX}).}

\begin{figure}
\centering
   \includegraphics[scale=0.47]{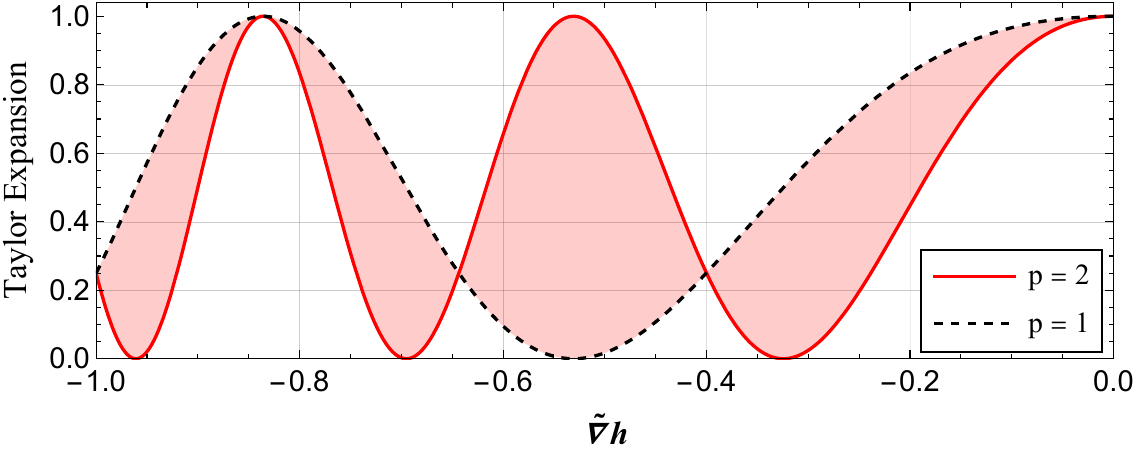}
\caption{\justifying{Taylor expansion of the WKB functions $\cos^2{[p \int k(x')dx']}$ for the special case of $x = \lambda_0$ as a function of relative slope, with $p=1,2$. The closed form expansions for the cosine arguments read $2p\pi  [ 1  - (\tilde{\nabla} h/4)   + 5(\tilde{\nabla} h)^2/12 ]$.}}
\label{fig:Group}
\end{figure}
\xt{
Now we consider the effects of the WKB approximation for handling the terms such as $\cos{(kx)}$ when we no longer deal with flat bottoms. Following eq. 4.5.6 of \citet{BMei2005} and eq. 2.314 of \citet{Dingemans1997}, we rewrite the phase as,
}
\begin{eqnarray}
\nonumber
\xt{\cos{(kx)}} &\xt{\rightarrow}& \xt{ \cos{\left[ \int_0^x k(x') dx'   \right]} \quad  \therefore}
\\
\xt{\langle   \cos^2{(kx)} \rangle_{_{\xy{\textrm{WKB}}}}} &\xt{\rightarrow}&  \xt{ \int_0^{\lambda} \cos^2{\left[ \int_{0}^x k(x') dx'   \right]}   \frac{dx}{\lambda} \quad .}
\label{eq:WKB0}
\end{eqnarray}
\xt{Taking into account that within WKB we can no longer use $\langle   \cos^2{(kx)} \rangle = 1/2$, the energy in eq.~(\ref{eq:Ekx}) is reevaluated as,}
\begin{widetext}
\begin{eqnarray}
\hspace{-0.3cm}
\nonumber
\xt{\mathscr{E}_{k0} \approx   \frac{a^2 }{2 } \frac{2k }{\sinh{(2\Lambda)}}  \Bigg[  \frac{\sinh{(2\Lambda)}+\sinh{(\frac{1}{2} k_p H_s)}}{2k}  \langle   \cos^2{(kx)} \rangle_{_{\xy{\textrm{WKB}}}}  }
+  \xt{ \left(  \frac{3 \pi \varepsilon \mathfrak{S}}{4}  \right)^2  \frac{\sinh{(4\Lambda)} + \sinh{( k_p H_s)}}{4k \sinh^6{\Lambda}}  \langle   \cos^2{(2kx)} \rangle_{_{\xy{\textrm{WKB}}}} \Bigg]  .
}
\label{eq:WKB1}
\end{eqnarray}
\end{widetext}
\xt{These modifications can be factored out of the kinetic energy in eq.~(\ref{eq:EnergykX2}), becoming}
\begin{eqnarray}
\nonumber
\hspace{-0.5cm}
\xt{
\mathscr{E}_{k0}} &\xt{=}&   \xt{ \frac{a^2}{2}  \left[ 1   +  \left(  \frac{\pi \varepsilon \mathfrak{S}}{4}  \right)^2 \chi_1   \, \frac{ \langle   \cos^2{(2kx)} \rangle_{_{\xy{\textrm{WKB}}}} }{ \langle   \cos^2{(kx)} \rangle_{_{\xy{\textrm{WKB}}}} }  \right]}
\\
&{}& \xt{ \times
\left[  1 + \frac{ \sinh{( k_p H_s)} }{ \sinh{(4\Lambda)} }    \right]   \langle   \cos^2{(kx)} \rangle_{_{\xy{\textrm{WKB}}}}     \, . }
\label{eq:WKB2}
\end{eqnarray}
\xt{Naturally, the WKB-averaged term outside the brackets will also appear in the potential energy (as well as in the numerator of $\Gamma$, see eq.~(\ref{eq:finiteampgamma2})):}
\begin{figure*}
\minipage{0.44\textwidth}
\hspace{0.0cm}
   \includegraphics[scale=0.65]{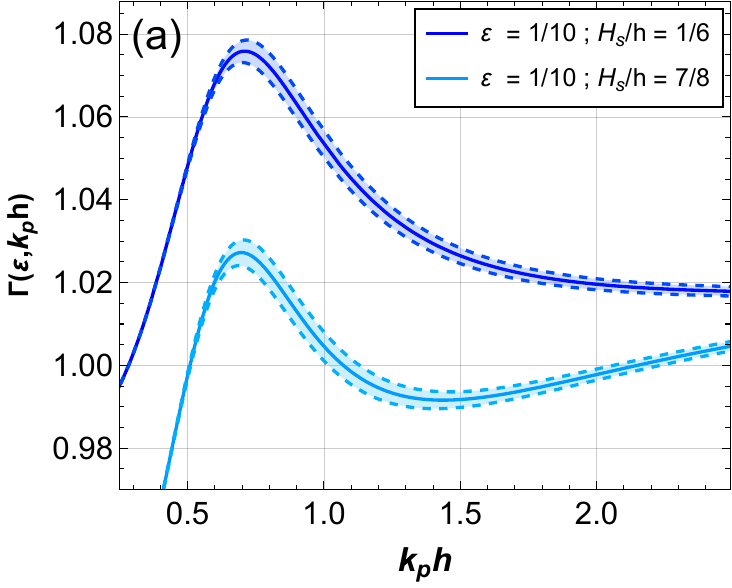}
\endminipage
\hfill
\minipage{0.51\textwidth}
   \includegraphics[scale=0.65]{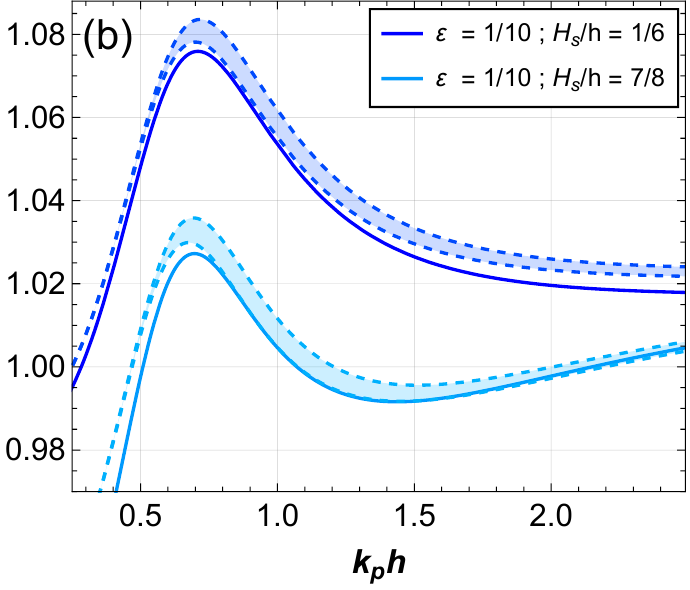}
\endminipage
\caption{\justifying{(a) Impact of the WKB-average term $\mathcal{F}$ on the inhomogeneity parameter $\Gamma$ (solid, using eq.~(\ref{eq:finiteampgamma2})) displayed by an error band corresponding to maximum and minimum of $\mathcal{F}$ (dashed lines). (b) Combined influence of WKB-averaging and rms energy on wave statistics (see eq.~(\ref{eq:EnergykXx2}) and \jfm{figure} \ref{fig:RMSEnergy}).}}
\label{fig:Group2}
\end{figure*}
\begin{eqnarray}
\hspace{-0.6cm}
\xt{
\mathscr{E}_{p} =    \frac{a^2}{2}  \left[ 1   +  \left(  \frac{\pi \varepsilon \mathfrak{S}}{4}  \right)^2 \tilde{\chi}_1   \, \frac{ \langle   \cos^2{(2kx)} \rangle_{_{\xy{\textrm{WKB}}}} }{ \langle   \cos^2{(kx)} \rangle_{_{\xy{\textrm{WKB}}}} }  \right]  \langle   \cos^2{(kx)} \rangle_{_{\xy{\textrm{WKB}}}}  
\label{eq:EnergykX3}
}
\label{eq:WKB2}
\end{eqnarray}
\xt{Therefore, following all steps of \jfm{sections} \ref{sec:bound1}-\ref{sec:bound2}, the WKB-averaged inhomogeneous parameter for steep slopes become\xy{s},}
\begin{equation}
\hspace{-0.5cm}
 \Gamma \approx \frac{ 1+   \frac{\pi^{2} \varepsilon^{2} \mathfrak{S}^{2}}{16}     \, \Tilde{\chi}_{1} \, \mathcal{F} }{  1 +  \frac{ \sinh{( k_{p}H_{s})}  }{2 \sinh{(4k_{p}h)}   }+   \frac{\pi^{2} \varepsilon^{2} \mathfrak{S}^{2}}{32}   \, \left( \Tilde{\chi}_{1}  + \left[ 1 + \frac{ \sinh{(k_{p}H_{s})}  }{ \sinh{(4k_{p}h)} } \right]  \chi_{1} \right) \, \mathcal{F} }    \, ,
\label{eq:finiteampgamma2x}
\end{equation}
with  $\mathcal{F} =  \langle   \cos^2{(2kx)} \rangle_{_{\xy{\textrm{WKB}}}}/ \langle   \cos^2{(kx)} \rangle_{_{\xy{\textrm{WKB}}}}$. Considering the model of eq.~(\ref{eq:nablalambda}), we compute the isolated WKB-average through the approximation:
\begin{equation}
\xt{
\lambda (x') \approx \lambda_0 + \frac{\tilde{\nabla} h \cdot x'}{ 2 \left( 1 + \frac{2x'}{\lambda_0} \tilde{\nabla} h \right)   } \equiv \lambda_0 \left\{ 1 +  \frac{ \frac{1}{2} \tilde{\nabla} h  \frac{ x'}{\lambda_0} }{  1 +2 \tilde{\nabla} h \, \frac{x'}{\lambda_0} } \right\} \quad ,
}
\end{equation}
\xt{Thus leading to,}
\begin{eqnarray}
\xt{
\mathcal{F}} &\xt{=}& \xt{ \frac{  \int_0^{\lambda} \cos^2{\left[  \int_{0}^x \frac{ 4\pi (2\lambda_0 + 4 \tilde{\nabla} h \, x'  ) dx'}{(2\lambda_0 + 5 \tilde{\nabla} h \, x'  ) \lambda_0}   \right]} dx }{  \int_0^{\lambda} \cos^2{\left[  \int_{0}^x \frac{ 2\pi (2\lambda_0 + 4 \tilde{\nabla} h\,  x' ) dx'}{(2\lambda_0 + 5 \tilde{\nabla} h\, x' ) \lambda_0}   \right]} dx } }
\\
\nonumber
&\xt{=}& \xt{  \frac{ \int_0^{\lambda_0} \cos^2{\left[  \frac{8\pi}{25} \left\{ 10 \frac{x}{\lambda_0} + \frac{1}{\tilde{\nabla} h} \ln{\left( 1 + \frac{5}{2} \tilde{\nabla} h \frac{x}{\lambda_0} \right)}  \right\} \right]} dx  }{ \int_0^{\lambda_0} \cos^2{\left[  \frac{4\pi}{25} \left\{ 10 \frac{x}{\lambda_0} + \frac{1}{\tilde{\nabla} h} \ln{\left( 1 + \frac{5}{2} \tilde{\nabla} h \frac{x}{\lambda_0} \right)}  \right\} \right]} dx   } \quad .
}
\end{eqnarray}
\xt{This analytical computation cannot lead to a simple closed-form. Indeed, the solution combines dozens of complex functions coupled with incomplete gamma functions. Despite the convoluted look, a simple Taylor expansion on the logarithm within brackets $\ln{\left( 1 + a x \right)} \approx a x - (1/2)a^2x^2 + (1/3)a^3x^3$ for $x \ll 1/a$ simplifies the integral, and the argument within the trigonometry function becomes $4\pi x/\lambda_0 [ 1  - (\tilde{\nabla} h/4) x/\lambda_0  + 5(\tilde{\nabla} h)^2/12   (  x/\lambda_0 )^2 + \cdots ]$. The qualitative impact of the WKB approximation (of an oscillatory behaviour around $1$) becomes clearer: \jfm{figure} \ref{fig:Group} shows that the square of the cosine of the second-order WKB function (before an average over $x$ is taken) is smaller than the linear counterpart for slopes $|\tilde{\nabla} h| \lesssim 2/5$, thus leading to $\mathcal{F} < 1$. Conversely, in the slope range of $2/5 \lesssim |\tilde{\nabla} h| \lesssim 2/3$, the second-order WKB cosine exceeds the linear counterpart, and thus $\mathcal{F} > 1$. Following the numerical integration of these functions, the WKB-averaging oscillates at most $\pm$6\% from the reference value of 1 \xy{which was assumed in \jfm{sections} \ref{sec:bound1}-\ref{sec:bound2}}. The implications of the WKB-average on eqs.~(\ref{eq:finiteampgamma2},\ref{eq:finiteampgamma2x}) are displayed in \jfm{figure} \ref{fig:Group2}\jfm{a}: maximum and minimum WKB-led variations are very small for both non-breaking and breaking waves alike, thus suggesting negligible impact on extreme wave statistics. However, these small variations that are otherwise centred at the original $\Gamma$, become slightly shifted upwards (\jfm{figure} \ref{fig:Group2}\jfm{b}) due to the small negative difference between the rms energy and the energy of the rms surface elevation, see \jfm{section} \ref{sec:bound1}.}
\begin{figure*}
\minipage{0.44\textwidth}
   \includegraphics[scale=0.76]{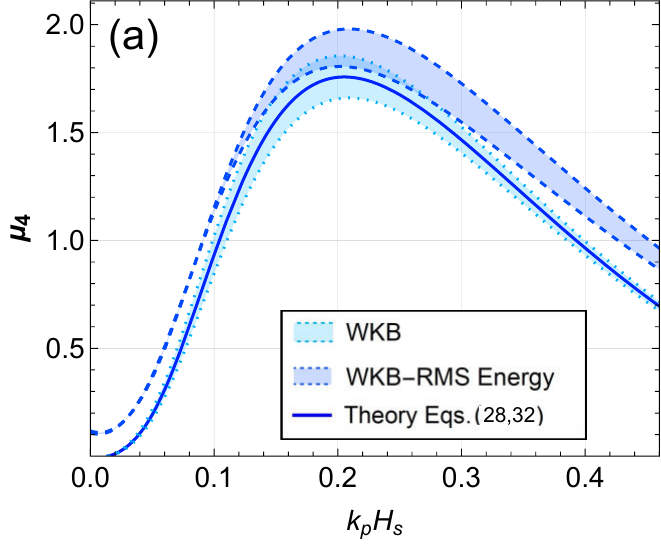}
\endminipage
\hfill
\minipage{0.55\textwidth}
   \includegraphics[scale=0.75]{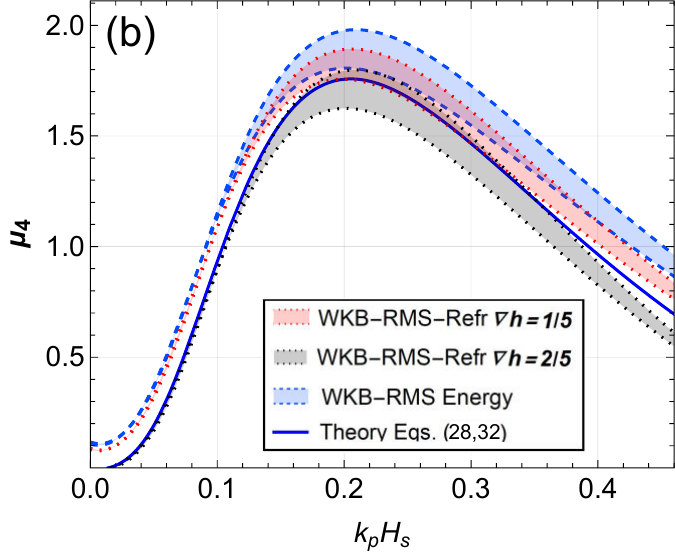}
\endminipage
   \caption{\xx{(a) Error bands of complementary effects on the excess kurtosis curve of \jfm{figure} \ref{fig:GammaApproxX2} from the WKB only (cyan) and combining with rms energy (blue). (b) Triple effect of WKB-rms-refraction terms for different bottom slopes.}}
  \label{fig:GammaWKBRefrac}
\end{figure*}

\x{The remaining \xy{complementary} effect influencing the full integration of the energy is related to the derivation of an expression for $\nabla \lambda$ for steep slopes. So far, we have assumed that we could treat this computation as if for a slowly varying bottom slope. From the mild slope equation \citep{BMei2005}, the criterion for neglecting strong refraction effects is $\nabla h / kh \ll 1$ \citep{Porter2003}. This threshold has been established since \citet{Berkhoff1973} to properly neglect the effects of $(\nabla h)^2$ and the curvature $\nabla^2 h$. However, theoretically and experimentally, in the present study, we restrict the problem to a plane slope \xy{($\nabla^2 h = 0$)}. For the experiments discussed \xy{in section \ref{sec:Exp}}, the ratio $\nabla h / kh$ \xy{ranges between} 0.07 \xy{and} 0.4 with a mean value of $\sim 0.21$, thus, our initial neglect\xy{ion} of slope-dependent refraction is reasonable. Furthermore, \citet{Booij1981,Booij1983} have shown that the mild-slope method is valid up to $\nabla h \leqslant 1/3$ despite possible concerns on refraction \citep{Lee1998}.}

\x{Nevertheless, in the pursuit of higher accuracy for the transformation of integral and other wave properties along a rapidly varying bathymetry, several models have been proposed to modify the mild-slope equations \citep{Massel1993,Chamberlain1995,Suh1997,Lee1998}. The latter are usually written as (see section 3.1 of \citet{Dingemans1997}),}
\begin{eqnarray}
\x{
-\frac{\partial^2 \phi}{\partial t^2} + \nabla \cdot (c_p c_g \nabla \phi) - (\omega^2 - k^2 c_p c_g) \phi } &\x{=}& \x{ 0 \quad ,}
\\
\x{
-\frac{\partial^2 \zeta}{\partial t^2} + \nabla \cdot (c_p c_g \nabla \zeta) - (\omega^2 - k^2 c_p c_g) \zeta } &\x{=}& \x{ 0 \quad .}
\end{eqnarray}
\x{For harmonic waves with $\zeta (x,t) = Re [\eta (x) e^{-i\omega t}]$, it can be shown \xy{that} these equations lead to \citep{Dingemans1997,BMei2005}:}
\begin{equation}
\x{
\nabla \cdot(c_p c_g \nabla \eta) + k^2 c_p c_g \eta = 0 \quad .
}
\label{eq:Ref1}
\end{equation}
\x{For long waves $(kh \ll 1/2)$, this equation can be rewritten as} \citep{BMei2005},
\begin{equation}
\x{
\nabla \cdot (h(x) \nabla \eta) + \frac{\omega^2}{g}\eta = 0 \quad .
}
\label{eq:Ref2}
\end{equation}
\x{Hence, when long waves travel in constant water depth $h$, eq.~(\ref{eq:Ref2}) reduces to the Helmholtz equation and recovers the well-known shallow water dispersion relation (see section 4.1 of \citet{BMei2005}):}
\begin{equation}
\x{
\nabla^2 \eta + k^2 \eta = 0 \quad \therefore \quad \omega^2 = k^2 gh \quad .
}
\end{equation}
\x{The results leading up to finding $\nabla \lambda$ in eq.~(\ref{eq:nablalambda}) assume that eqs.~(\ref{eq:Ref1}-\ref{eq:Ref2}) provide an accurate description for the problem. This is especially untrue for abrupt depth transitions $L/\lambda \lesssim 1$ and finite-amplitude waves. \xy{In those cases}, one may rewrite the mild-slope equation to account for significant slope-dependent evolution \citep{Massel1993,Chamberlain1995,Lee1998}:}
\begin{equation}
\x{
\nabla \cdot(c_p c_g \nabla \eta) + \Big[ k^2 c_p c_g + R_1 (\nabla h)^2 + R_2 \nabla^2 h \Big] \eta = 0 \quad ,
}
\end{equation}
\x{with coefficients $R_1 (kh)  $ and $R_2 (kh)$ to be determined from the vertical averaging of the variational principle used to obtain eq.~(\ref{eq:Ref1}) \citep{Porter2003}. For computation of these coefficients with alternative methods, see \citet{Suh1997} and \citet{Massel1993} and appendices therein. Since we restrict ourselves to the analysis of a plane bottom slope, we have $R_2 \nabla^2 h = 0$. Including the second term for the slope in the refraction equations would modify the wavelength gradient to the order of:}
\begin{equation}
\x{
\nabla \lambda \rightarrow \nabla \lambda \Big( 1 + r_1 \nabla h + \cdots \Big) \quad ,
}
\end{equation}
\x{where $r_{1}$ is the normalized version of $R_1$, $\nabla \lambda$ is already given in eq.~(\ref{eq:nablalambda}), and effectively, the wavelength gradient is proportional to $\nabla h + r_1 (\nabla h)^2$. Precise calculations for the wavenumber (or wavelength) gradient containing all orders are far from the scope of this work, and the reader may refer to eqs. 33-42 of \citet{Zhang1998} for details on all orders. Indeed, one could directly use the slope-corrected dispersion relation \citep{Zhang1998,ge2020accurate} and obtain the leading-order correction \xy{in $\nabla h $} for the wavelength gradient. Taking the full expression for $R_1$ from eq. 2.16 of \citet{Chamberlain1995}, we have $r_1 \approx -1/10$ for the typical zone relevant for wave statistics amplification (namely $kh \sim 0.5$), whereas the coefficient slowly approaches zero towards deep water (for a similar approach with slightly different coefficients, see figure 1 of \citet{Lee1998}). Accordingly, following \jfm{sections} \ref{sec:bound}-\ref{sec:bound2}, the maximum change in the energy density would now read $-(\tilde{\nabla} h/8) r_1 \nabla h = -\pi r_1 (\nabla h)^2/8\Lambda_0 \leqslant (\nabla h)^2/20$, because the linear terms are counter-balanced by $\check{\mathscr{E}}_{p2}$. Thus, the inhomogeneous parameter controlling wave statistics now reads (without WKB and rms energy effects):}
\begin{widetext}
\begin{equation}
\x{
 \Gamma \approx \frac{ 1+   \frac{\pi^{2} \varepsilon^{2} \mathfrak{S}^{2}}{16}     \, \Tilde{\chi}_{1} }{  1 + \frac{(\nabla h)^2}{20} +  \frac{ \sinh{( k_{p}H_{s})}  }{2 \sinh{(4k_{p}h)}   } \left[  1+ \frac{(\nabla h)^2}{10}  \right] +   \frac{\pi^{2} \varepsilon^{2} \mathfrak{S}^{2}}{32}   \, \left( \Tilde{\chi}_{1}  + \left[ 1 + \frac{ \sinh{(k_{p}H_{s})}  }{ \sinh{(4k_{p}h)} } \right] \left[  1+ \frac{(\nabla h)^2}{10}  \right] \chi_{1} \right)   }    \,\, .
 }
\label{eq:RefGamma}
\end{equation}
\end{widetext}
\x{To verify the impact of the combined effect of WKB-averaging and the $\langle \mathscr{E} \rangle_{\textrm{rms}}$ on extreme statistics, we plotted in \jfm{figure} \ref{fig:GammaWKBRefrac}\jfm{a} the excess kurtosis \xy{of eq.~(\ref{eq:RefGamma})} and compare it with the results of eq.~(\ref{eq:finiteampgamma2},\ref{eq:kurtApp2}), finding no significant deviations from our approach \xy{described} in \jfm{section} \ref{sec:kurtosis}. Finally, the triple combined effect that adds the refraction consideration into $\nabla \lambda$ is shown in \jfm{figure} \ref{fig:GammaWKBRefrac}\jfm{b}. \xy{It is therefore reasonable to omit} all of these corrections in the integration process. Even as these effects are important for the transformation of monochromatic waves, they seem much less relevant to extreme wave behaviour\xy{, which is the focus of the present work}.}

\bibliography{bibliography}

\end{document}